\documentclass[10pt, aps, showpacs, preprintnumbers, prd, nofootinbib, preprint]{revtex4}
\usepackage{feynmf}
\usepackage{amsmath}
\usepackage{graphicx}
\usepackage{mathrsfs}
\usepackage{stmaryrd}
\usepackage{amsfonts}
\usepackage{wrapfig}
\usepackage{subfig}
\usepackage{float}
\usepackage{bbm}
\usepackage{hyperref}
\usepackage{cleveref}
\usepackage[utf8]{inputenc}

\begin{document}
\def\sech{\mathop{\rm{sech}}\nolimits}
\def\arcsinh{\mathop{\rm{arcsinh}}\nolimits}

\title{Unified Dark Matter from a Simple Gauge Group on a Domain-Wall Brane}
\author{Benjamin D. Callen}\email{bdcallen@student.unimelb.edu.au}
\affiliation{ARC Centre of Excellence for Particle Physics at the Terascale, School of Physics, The University of Melbourne, Victoria 3010, Australia}

\begin{abstract}
 Some models of asymmetric dark matter commonly employ a gauge group structure of the form $G_{V}\times{}G_{D}$ where $G_{V}$ is the visible gauge group containing the Standard Model and $G_{D}$ is the gauge group responsible for self-interactions amongst components of dark matter. In some models, there is also an additional spontaneously broken $U(1)$ gauge symmetry coupling the visible and dark sectors at high energies. One theoretical problem is how to unify the visible and dark sectors by inducing the spontaneous breaking $G\rightarrow{}G_{V}\times{}G_{D}$ for some large gauge group $G$. In this paper, we discuss how to generate such a structure at low energies, in the context of 4+1-dimensional domain-wall brane model, by employing a generalization of the Dvali-Shifman mechanism, used to localized gauge bosons on domain walls, called the clash-of-symmetries mechanism. In one model, we describe a clash-of-symmetries domain wall solution in a theory with two scalar fields in the adjoint representation which breaks the group $SU(12)$ to two differently embedded copies of $SU(6)\times{}SU(6)\times{}U(1)$, leading to a an effective $SU(5)_{V}\times{}SU(5)_{D}\times{}U(1)_{X}$-invariant field theory on the wall. We find that fermions in the mixed representations of $SU(5)_{V}\times{}SU(5)_{D}$ do not couple to the domain wall and thus remain 5D vector-like Dirac fermions, attaining masses of order $M_{GUT}$ when we perform the breaking $SU(5)_{V}\rightarrow{}SU(3)_{c}\times{}SU(2)_{I}\times{}U(1)_{Y}$, thus being removed from the spectrum. We also outline how to build a few alternative models, one based on the group $SU(9)$, and a couple more based on non-clash-of-symmetries domain wall solutions in $SU(12)$ and $SU(10)$ models. 
\end{abstract}

\maketitle

\section{Introduction}

 Dark matter composes roughly 25 per cent of the energy content of the universe, with 70 per cent of the remaining energy content being dark energy which is responsible for the universe's expansion. Only about 5 per cent of the energy content of the universe is visible matter that is described by the particles of the Standard Model. We know that dark matter interacts with visible matter primarily through gravity, and most theories describing it postulate that it is made of a stable Weakly-Interacting Massive Particle (WIMP). Some examples of theories yielding stable WIMPs are R-parity conserving supersymmetric theories, which predict that the lightest supersymmetric particle (LSP) is a stable, neutrally charged particle, as well as theories with sterile neutrinos. There is, however, no a priori reason why dark matter has to be composed entirely of a single, stable particle; it could be that there are multiple different species of dark particles, with their own dark gauge forces, completely hidden with respect to the Standard Model particles. 

 It is also curious that, while there is more dark matter than visible matter in the universe, the disparity is not significant in terms of orders of magnitude. In fact, the dark matter mass density of the universe is only roughly five times that for visible matter,
\begin{equation}
\label{eq:darkmatterdensityvvisible}
\Omega_{DM} \simeq{} 5\Omega_{VM}.
\end{equation}
This naturally raises the question whether dark matter density is somehow related to the visible matter density at high energy scales. On the other hand, there is still the imbalance between matter and anti-matter in the visible sector which is still unaccounted for. The dominance of visible matter over anti-matter due to the number difference between baryons and anti-baryons, is characterized by the parameter $\eta{(B)}$ \cite{wmap, planck2013}, 
\begin{equation}
\label{eq:matterantimatterasymmetry} 
\eta{(B)} \equiv{} \frac{n_{B}-n_{\bar{B}}}{s} \simeq{} 10^{-10}, 
\end{equation}
where here $n_{B}$, $n_{\bar{B}}$ and $s$ are the baryon number, the anti-baryon number and the entropy densities of the universe respectively. In models of baryogenesis, this asymmetry arises from CP-violating processes as well as out-of-equilibrium dynamics. This raises the idea of a scenario where the observed ratio between visible and dark matter described in Eq.~\ref{eq:darkmatterdensityvvisible} arises fundamentally from a visible matter - dark matter asymmetry and, furthermore that this asymmetry and the matter-antimatter asymmetry are related, typically via the relation
\begin{equation}
\label{eq:matterantimattervmdm} 
n_{X}-n_{\bar{X}} \sim{} n_{B}-n_{\bar{B}}.
\end{equation}
In other words, the matter antimatter asymmetry in the visible sector leads to an asymmetry between the corresponding matter and antimatter in the dark sector. Given the above correspondence, the relative dark matter abundance  is explained if dark matter particles have masses around five times that of the proton. This scenario is called the asymmetric dark matter scenario \cite{asymmetricdarkmatterkalliaray, zurekasymmetricdm}.

 One possibility of realizing asymmetric dark matter is through grand unification, where the dark matter components are the additional components of a simple group such as the $SU(6)$ model proposed by Barr \cite{sbarrsu6dm}, or as the colors of a dark GUT $G_{D}$ in a theory based on a $G_{V}\times{}G_{D}$ gauge structure, such as the models based on $SU(5)\times{}SU(5)$ and $SO(10)\times{}SO(10)$ recently proposed by Refs.~\cite{lonsdalegrandunifieddm, lonsdaleso10xso10adm}. In the latter model, a particularly compelling model of asymmetric dark matter was made in which dark quarks form dark protons, and when the model has a certain number of dark quarks, the running of the dark gauge coupling constant induces a dark QCD scale $\Lambda_{D}$ which is of roughly the same order as $\Lambda_{QCD}$. In the model proposed by Barr, and in the $SU(6)$ and $SU(7)$ models proposed by Ma \cite{ernestmadarkgutunification}, dark matter particles are unified with visible matter inside the required multiplets of of these groups; in the case of the $SU(7)$ model of Ma, it is possible to produce an unbroken $U(1)_{D}$ group which acts solely on dark matter. Given that it is possible to generate a dark Abelian group, this raises the question of whether we can break a simple group to produce dark non-Abelian groups, leading to the $G_{V}\times{}G_{D}$ scenarios considered in Refs.~\cite{lonsdalegrandunifieddm, lonsdaleso10xso10adm}.
 
 One might consider, for example, starting with an $SU(N)$ gauge group, and then breaking it to $SU(5)_{V}\times{}SU(N-5)_{D}\times{}U(1)_{X}$, where $SU(5)_{V}$ contains the visible SM gauge groups and $SU(N-5)_{D}$ contains the dark gauge groups. If one tries to construct such a grand unified theory in ordinary 3+1D, one can easily see that one will run into some significant obstacles. As is known, in ordinary $SU(5)$ theories, the right-chiral up quark, the right-chiral electron and the quark doublet are embedded into the antisymmetric rank 2 tensor $10$ representation. This means that the most natural candidate for embedding the very same fermions in these extended GUTs is the corresponding antisymmetric rank two tensor $N(N-1)/2$ representation of $SU(N)$. Unfortunately, the same representations will contain chiral bi-quark fermions charged under representations of the form $(5, N-5)$. These mediating fermions, which are charged under both the visible and dark groups, must be made massive in some way. On top of this are the constraints that come from the requirement of anomaly cancellation in 3+1D theories with chiral fermions. Satisfying this set of constraints while not running into other problems, such as undesirable exotics, non-perturbative Yukawa interactions and a four-generation Standard Model in the visible sector, is an extremely difficult task.
 
 This seems to suggest that the generation of a $G_{V}\times{}G_{D}$ gauge theory from the spontaneous breaking of a higher simple group $G$ requires additional physics. One possibility that one might think of is to construct a grand unified theory from higher dimensions, with dimensional reduction being performed from the inclusion of, for instance, a brane. In particular, given that in odd-dimensional spacetimes chiral anomalies are absent from gauge theories, and, that in many braneworld models there is a bulk-brane mechanism called anomaly inflow \cite{callananomalyinflow, daemishaposhnikov} that cancels anomalies associated with an anomalous effective field theory, one can see that going to 4+1D spacetime with branes can resolve the problems arising in 3+1D approaches from anomalies. That leaves us to find a mechanism within braneworld models which eliminates, in particular, the unwanted bi-fundamental fermion states.
 
 One way of realizing the braneworld scenario is through the dynamical localization of fields and gravity to a domain wall \cite{rubshapdwbranes, firstpaper}. A domain wall is typically formed via a scalar field which interpolates between two discrete, disconnected vacua from negative infinity to positive infinity along some dimension. Fermions are localized by Yukawa coupling 4+1D fermionic fields to the scalar field which generates the domain wall, yielding localized 3+1D chiral zero modes \cite{jackiwrebbi}. Scalars can be localized through quartic interactions \cite{modetower}. Gravity can also be localized \cite{gremmdwgravity, adavidsondwgravity, clashofsymgravity, kehagiastamvakisgravity, slatyervolkasrsgrav, sodarsgravity, rsgravitydaviesgeorge2007}. 
 
 The localization of gauge bosons is the most difficult aspect of domain-wall brane model building, yet its conjectured solution, the Dvali-Shifman mechanism \cite{dsmech}, offers some of the most enriching and interesting parts of these types of models. To implement the Dvali-Shifman mechanism, a non-Abelian gauge group $G$ is respected and confining in the bulk, but is spontaneously broken to some subgroup $H$ in the interior of the domain-wall brane. The confining bulk will then act as a dual superconductor, repelling the `electric' field lines of H (or H-field lines) from the bulk. If a test charge is placed on the wall, the H-field lines will simply diverge out through the world volume of the domain wall. If a test charge is placed in the bulk, the H-field lines will form a flux string onto the domain wall and then diverge, behaving as if it were actually placed on the wall. In this way, gauge bosons are localized without violating gauge charge universality \cite{dubrub}. Given the requirement of a large gauge group $G$ being spontaneously broken to a subgroup $H$ in order to implement the Dvali-Shifman mechanism, as well as the need for $H$ to contain the Standard Model gauge fields, this obviously motivates interesting models based on grand unification in 4+1D, such as the minimal choice $G=SU(5)$ and $H=SU(3)\times{}SU(2)\times{}U(1)$ \cite{firstpaper} or, alternatively, a non-minimal choice such as $G=SO(10)$ \cite{jayneso10paper}. Also, in the minimal choice, one finds that the profiles for the various fermions and scalars transforming under the SM gauge group are split, leading to resolution of the the fermion mass hierarchy problem and colored-Higgs induced proton decay \cite{su5branemassfittingpaper, su5a4braneworldcallen}. 
 
 Furthermore, the Dvali-Shifman mechanism has an interesting extension called the clash-of-symmetries (CoS)) mechanism \cite{o10kinks, abeliankinkscos, clashofsymmetries, e6domainwallpaper, pogosianvachaspaticos, vachaspaticos2, pogosianvachaspaticos3}. Here, rather than just leave $G$ unbroken, we now give the scalar field (or fields) which generates the domain wall a gauge charge, and give it a potential such that the scalar field has a disconnected vacuum manifold, whose path-connected components are homeomorphic to the coset $G/H$. This means that on one side of the domain wall, $G$ is broken to $H$ while on the other side of the wall, $G$ is broken generally to an isomorphic but differently embedded subgroup $H'$. Due to $H'$ not being exactly the same copy of $H$, this leads to further symmetry breaking to $H\cap{}H'$ in the interior of the wall. Gauge bosons of $H\cap{}H'$, whether Abelian or non-Abelian, are then localized if the corresponding generators originate entirely from the non-Abelian subgroups of $H$ and $H'$. An interesting model based on $E_{6}$ was constructed in Ref.~\cite{e6domainwallpaper} using the Clash-of-Symmetries mechanism, in which $H$ and $H'$ are isomorphic to $SO(10)\times{}U(1)$, leading to $H\cap{}H' = SU(5)\times{}U(1)\times{}U(1)$, with the $SU(5)$ subgroup being localized. The same reference gave a treatment for dynamical localization of fermions in the same model, and given that the scalar field generating the domain wall is now charged under the gauge group, the localization of the various $H\cap{}H'$-covariant components of the fermions depend non-trivially on how they couple to the kink. This leads to the interesting property that for some given sign of the Yukawa coupling to the kink, some of the $H\cap{}H'$-covariant fermionic components attain localized left-chiral zero modes, some will attain right-chiral zero modes, and some components can be \emph{completely decoupled} from the domain wall.
 
 In this paper, we will show that this last interesting property of fermion localization in the context of a Clash-of-Symmetries domain wall can be exploited to eliminate the troublesome fermionic mediators which arise in attempting to generate a GUT which leads to both visible and dark gauge sectors after symmetry breaking. In particular, we choose our gauge group to be $SU(12)$, and we generate a series of Clash-of-Symmetries domain-wall solutions in 4+1D from a scalar field theory with two scalars transforming under the adjoint $143$ representation, with a potential which is invariant under a $\mathbb{Z}_{2}$-symmetry which interchanges the two scalar fields. We will show that in a special parameter regime, a Clash-of-Symmetries domain wall which has an $SU(5)_{V}\times{}SU(5)_{D}\times{}U(1)_{X}$ gauge group localized to its world volume can be made to be the most energetically stable of the solutions. Upon coupling fermions in the fundamental $12$ and rank-two antisymmetric $66$ representations, we find in particular that the potentially troublesome $(5, 5)$ bi-fundamental fermion in the $66$ is completely decoupled from the domain-wall brane. This means that this bi-fundamental fermion remains a 4+1D \emph{Dirac} fermion and thus vector-like, and when we include an additional adjoint scalar field which induces the usual breaking $SU(5)_{V}\rightarrow{}SU(3)\times{}SU(2)\times{}U(1)$, the SM-covariant components of this bi-fundamental fermion will attain masses of order $M_{GUT}$ in the interior of the domain wall, removing them entirely from the low energy 3+1D spectrum on the wall. It turns out that other troublesome components, such as the additional $SU(5)_{V}$ and $SU(5)_{D}$ quintets in the $66$, the singlet components in the $12$, and the lone singlet in the $66$, are either semi-delocalized or fully delocalized, and/or attain only massive modes (either through a 4+1D mass through the effective coupling to the kink, or after breaking $U(1)_{X}$). The only components which attain localized chiral zero modes are the $(5, 1)$ and $(1, 5)$ components in the $12$ as well as the $(10, 1)$ and $(1, 10)$ components of the $66$. Interestingly, for given signs of the Yukawa couplings, if the $(5, 1)$ component in the $12$ or the $(10, 1)$ component in the $66$, which transform solely under $SU(5)_{V}\times{}U(1)_{X}$, attain a chiral zero modes of a given chirality (either left or right), the corresponding dark multiplets for $SU(5)_{D}$, the $(1, 5)$ component in the $12$ and the $(1, 10)$ component in the $66$, attain chiral zero modes of the \emph{opposite} chirality. This is very interesting as it means that if we break $SU(5)_{V}\times{}SU(5)_{D}$ symmetrically, this leads directly to the mirror matter scenario \cite{footvolkasorigmirrormatter, footlewvolkas2, leeyangmm, kobzarevmm, pavsicmm}, which can be thought of as a realization of asymmetric dark matter \cite{footvolkasmirroradm1, footvolkasmirroradm2, asymmetricdarkmatterkalliaray}. We also have the option to break $SU(5)_{V}\times{}SU(5)_{D}$ asymmetrically, leading to the kind of scenarios described in Refs.~\cite{lonsdalegrandunifieddm, lonsdaleso10xso10adm}. At the very least, we have a 3+1D effective field theory in which the particle content contains a left-chiral $\overline{5}$ and a left-chiral $10$ under $SU(5)_{V}$ in the visible sector, and a right chiral $\overline{5}$ and a right-chiral $10$ under $SU(5)_{D}$ in the dark sector. At low energies, after appropriate breaking of $SU(5)_{V}$ to the Standard Model as well as the breaking of $U(1)_{X}$, these sectors have no mediators and are completely sequestrated. Scalars can also be localized, yielding Higgs potentials for both the visible and dark sectors.
 
  We also present some alternative models which generate hidden sectors, including an $SU(9)$ model in which the localized gauge group is $SU(5)_{V}\times{}SU(2)\times{}U(1)$, and a model based on the non-CoS domain wall in the $SU(12)$ model. In the $SU(9)$ model, we again have two adjoint scalar fields which generate the domain wall, and these scalars break $SU(9)$ to differently embedded copies of $SU(6)\times{}SU(3)\times{}U(1)$, which overlap to yield a localized $SU(5)_{V}\times{}SU(2)_{D}\times{}U(1)_{X'}$ on the wall. It turns out that if we choose two copies of the fundamental $9$ representation and one copy of the totally antisymmetric rank three $84$ representation, we attain the desired particle content without mediators at low energies. With the model based on the non-CoS kink in $SU(12)$, the gauge group respected on the wall is $H\cap{}H' = SU(6)_{V}\times{}SU(6)_{D}\times{}U(1)$. The $SU(6)$ subgroups are then broken with additional scalar fields to $SU(5)$ (or $SU(5)\times{}U(1)$) subgroups, leading to the localization of an $SU(5)_{V}\times{}SU(5)_{D}$-invariant theory by the original Dvali-Shifman mechanism. Just as before, the undesired mediators are eliminated from the spectrum in the same way. The cost of using the non-CoS domain wall is additional scalar fields as well as some additional fermionic particle content, since we have more localized $SU(5)_{V}$ and $SU(5)_{D}$ quintets than we need. We show in the same subsection that this non-CoS domain wall scenario can be refined and simplified by using an $SU(10)$ model, in which the gauge group is broken to the same $SU(5)_{V}\times{}SU(5)_{D}\times{}U(1)$ subgroup and, subsequently, the visible $SU(5)_{V}$ group is broken directly to the Standard Model gauge group.
  
  In the next section, we go into further detail as to why 3+1D unification of visible and dark gauge sectors is difficult. We give the best examples of 3+1D GUTs that the author invented which leads to a $G_{V}\times{}G_{D}$ structure, namely a model based on $SU(7)$ which is broken to $SU(5)_{V}\times{}SU(2)_{D}\times{}U(1)$, and another based on $SU(9)$ being broken to $SU(5)_{V}\times{}SU(4)_{D}\times{}U(1)$. These models turn out to have highly undesirable features, including four-generation Standard Models in the visible sector as well as fermionic mediators attaining their masses from electroweak symmetry breaking, both of which lead to non-perturbative Yukawa interactions. In Sec.~\ref{sec:cosmechanism}, we give a short treatment of domain walls, the Dvali-Shifman mechanism and the Clash-of-Symmetries mechanism. In Sec.~\ref{sec:solution}, we describe the scalar potential with two adjoint scalar fields and find the CoS solutions for several parameter choices. In Sec.~\ref{sec:fermionlocalization}, we deal with fermion localization and describe how the fermionic mediators are eliminated. Section \ref{sec:scalarlocalization} describes scalar localization. In Sec.~\ref{sec:alternativemodels} we give some nice alternative models: a sketch for a Clash-of-Symmetries model based on $SU(9)$ is given in Sec.~\ref{subsec:su9model}, and two more models based on non-Clash-of-Symmetries domain walls in $SU(12)$ and $SU(10)$ gauge theories are given in Sec.~\ref{subsec:noncossolution}. Section \ref{sec:conclusion} is our conclusion.

\section{The Difficulty of attaining $G_{V}\times{}G_{D}$ from Grand Unification in 3+1D} 
\label{sec:whynot3+1DGUT}

 In this section, we discuss in detail why ordinary $3+1D$ GUTs are unpromising candidates for the unification of the visible Standard Model gauge forces with a hidden gauge sector which includes non-Abelian interactions. We mainly do this in the context of unification for $SU(N)$, but there are some reasons we will give at the end of this section as to why $SO(N)$ unifications are not promising either. Given the $SU(6)$ model proposed by Barr \cite{sbarrsu6dm} in which dark matter arises as a sixth color, and the $SU(7)$ model proposed by Ma \cite{ernestmadarkgutunification}, in which a dark $U(1)$ interaction is generated, it is natural to ask whether a similar unification theory can generate gauge interactions in the dark matter sector which are non-Abelian. Consider breaking an $SU(N)$ gauge theory to $SU(5)_{V}\times{}SU(N-5)_{D}\times{}U(1)$ with an adjoint scalar field. The decomposition of the fundamental representation in terms of representations of $SU(5)_{V}\times{}SU(N-5)_{D}\times{}U(1)$ is
 \begin{equation}
  \label{eq:sunfundamental}
  N = (5, 1, N-5)\oplus{}(1, N-5, -5).
 \end{equation}
Naturally, this is the representation to include the Standard Model fermions which are embedded in a quintet in ordinary $SU(5)$ grand unification, and the $(1, N-5, -5)$ component is identified as a dark quark. However, we also need to include the fermions which are embedded in the decuplet representation of $SU(5)$. The decuplet representation is a rank two antisymmetric representation, and thus the natural and minimal candidate representation to embed these fermions in an $SU(N)$ theory is the corresponding rank two antisymmetric $N(N-1)/2$ representation. The decomposition of the $N(N-1)/2$ representation can be deduced by multiplying the $N$ representation with itself and then taking the antisymmetric products. The result is 
\begin{equation}
 \label{eq:sunantisymmetric}
 \frac{N(N-1)}{2} = \big(10, 1, 2(N-5)\big)\oplus{}\big(5, N-5, N-10\big)\oplus{}\big(1, \frac{(N-5)(N-6)}{2}, -10\big).
\end{equation}
From this we see that we get not only components which transform under the rank 2 antisymmetric representations for the visible and dark gauge groups, but also an undesirable bi-fundamental state, which is the $(5, N-5, N-10)$ component. This bi-fundamental fermion is chiral just like the rest of the fermions in these representations, and also needs to be made massive. This introduces the problem of choosing a number of representations such that the chiral fermions which are charged under both $SU(5)_{V}$ and $SU(N-5)_{D}$, the fermionic mediators, will all attain masses after electroweak symmetry breaking or, preferably, the breaking of a subgroup of $SU(N-5)_{D}$. This is on top of the usual chiral anomaly cancellation constraint for 3+1D GUTs.

 We have found a couple of models in which the fermionic mediators all attain masses after electroweak symmetry breaking. The first model is based on $SU(7)$, which is broken to $SU(5)_{V}\times{}SU(2)_{D}\times{}U(1)$. In our construction of these models, we restricted ourselves to totally antisymmetric representations, as the anomalies coming from symmetric representations are larger and grow faster with rank, as well as leading to more potentially undesirable components. The combination of left-chiral fermionic representations that we choose for the $SU(7)$ model is the anomaly free combination $\overline{7}\oplus{}21\oplus{}\overline{35}$. Under $SU(5)_{V}\times{}SU(2)_{D}\times{}U(1)$, these three representations decompose as
 \begin{equation}
 \begin{aligned}
  \label{eq:su7reps}
  \overline{7} &= (\overline{5}, 1, -2)\oplus{}(1, 2, +5), \\
  21 &= (10, 1, +4)\oplus{}(5, 2, -3)\oplus{}(1, 1, -10), \\
  \overline{35} &= (10, 1, -6)\oplus{}(\overline{10}, 2, +1)\oplus{}(\overline{5}, 1, +8)
  \end{aligned}
 \end{equation}
In this particular scenario, given we want to preserve a non-Abelian group in the dark sector, we don't have the option of breaking $SU(2)_{D}$, so we must make the all the fermions massive through electroweak symmetry breaking. In ordinary $SU(5)$ unification, the electroweak Higgs doublet is usually embedded into either the $\overline{5}$ representation. If we embed this anti-quintet into the $\overline{7}$ representation, we see that we have the following possible invariant Yukawa interactions (which are assumed to be either of the form $\overline{(\psi^{R_{1}}_{L})^{c}}\psi^{R_{2}}_{L}\phi^{R_{3}}$ or $\overline{(\psi^{R_{1}}_{L})^{c}}\psi^{R_{2}}_{L}(\phi^{R_{3}})^{*}$) in the theory:
\begin{equation}
\label{eq:su7bifundamentalmass1}
\overline{35}_{F}\times{}21_{F}\times{}7_{S}\supset{}(\overline{10}, 2, +1)_{F}\times{}(5, 2, -3)_{F}\times{}(5, 1, +2)_{S}\supset{}(1, 1, 0),
\end{equation}
\begin{equation}
 \label{eq:su7bifundamentalmass2}
\overline{35}_{F}\times{}\overline{35}_{F}\times{}\overline{7}_{S}\supset{}(\overline{10}, 2, +1)_{F}\times{}(\overline{10}, 2, +1)_{F}\times{}(\overline{5}, 1, -2)_{S}\supset{}(1, 1, 0),
\end{equation}
and
\begin{equation}
 \label{eq:su7bifundamentalmass3}
\overline{7}_{F}\times{}21_{F}\times{}\overline{7}_{S}\supset{}(1, 2, +5)_{F}\times{}(5, 2, -3)_{F}\times{}(\overline{5}, 1, -2)_{S}\supset{}(1, 1, 0),
\end{equation}
where $F$ denotes a fermionic component and $S$ denotes a scalar component. Given these all contain singlets, they generate mass terms. The interaction in Eq.~\ref{eq:su7bifundamentalmass1} generates masses for the electron-like and down quark-like components of the $(5, 2, -3)$ and $(\overline{10}, 2, +1)$ fermionic mediators after electroweak symmetry breaking. The interaction in Eq.~\ref{eq:su7bifundamentalmass2} generates a masses for the up quark-like components of the $(\overline{10}, 2, +1)$ fermionic mediator. Finally, the interaction in Eq.~\ref{eq:su7bifundamentalmass3} generates a mass between the left-chiral neutrino-like component of the $(5, 2, -3)$ mediator and the dark quark $(1, 2, +5)$ doublet. Thus, all the fermionic mediators attain masses. However, the interaction in Eq.~\ref{eq:su7bifundamentalmass3} contains
\begin{equation}
 \label{eq:su7firstgendownelectronmass}
\overline{7}_{F}\times{}21_{F}\times{}\overline{7}_{S}\supset{}(\overline{5}, 1, -2)_{F}\times{}(10, 1, +4)_{F}\times{}(\overline{5}, 1, -2)_{S}\supset{}(1, 1, 0),
\end{equation}
which generates down-quark and electron masses for the generation of visible SM fermions coming from the $(\overline{5}, 1, -2)$ and $(10, 1, +4)$ components, which implies that the Dirac mass formed between the $\nu_L$-like component of the $(5, 2, -3)$ state and the dark $(1, 2, +5)$ quark is of order the MeV scale. This is clearly in opposition to experiment since the state formed from these components would couple to the $W$ and $Z$ bosons, if we choose the $(\overline{5}, 1, -2)$ and $(10, 1, +4)$ components to generate the first generation of visible fermions. With only a Higgs doublet coming from the $\overline{7}$, the only term that generates masses for the up quark components of the visible $SU(5)_{V}$ decuplets is
\begin{equation}
\label{eq:tenpletmass1} 
 (10, 1, -6)_{F}\times{}(10, 1, +4)_{F}\times{}(5, 1, +2)_{S}\subset{}\overline{35}_{F}\times{}21_{F}\times{}7_{S}.
\end{equation}
Unless we introduce additional Higgs multiplets which can yield a Yukawa interaction which produces a second independent mass term amongst the $(10, 1, -6)$ and $(10, 1, +4)$ multiplets, we will have massless up quarks. Fortunately, the $35$ representation contains a $(5, 1, -8)$ component, so if we introduce a $35$ scalar, we attain the Yukawa interaction
\begin{equation}
 \label{eq:secondsu7higgsdoubletyukawa}
 21_{F}\times{}21_{F}\times{}35_{S}\supset{}(10, 1, +4)_{F}\times{}(10, 1, +4)_{F}\times{}(5, 1, -8)_{S}\supset{}(1, 1, 0),
\end{equation}
which yields a mass term for the up quark component in the  $(10, 1, +4)$ fermion. We could also choose an appropriate representation coming from the tensor product $\overline{35}\times{}\overline{35}$ which contains a $(5, 1, +12)$ component, which can generate a mass term for the up quark in the $(10, 1, -6)$ fermion. 

 Having produced mass terms for all the fermions, we need to break the $U(1)$ subgroup. We can simply do this with a scalar in the $21$ representation, because this representation contains a $(1, 1, -10)$ component, which can break the $U(1)$ group when it attains a VEV. When the $(1, 1, -10)$ condenses, it also yields a Majorana mass term for the dark $(1, 2, +5)$ quark doublet from the interaction
 \begin{equation}
  \label{eq:darkquarkmajmass}
  \overline{7}_{F}\times{}\overline{7}_{F}\times{}21_{S}\supset{}(1, 2, +5)_{F}\times{}(1, 2, +5)_{F}\times{}(1, 1, -10)_{S}\supset{}(1, 1, 0).
 \end{equation}

 We have constructed a model in which all the fermionic mediators and also the visible and dark fermions in the combination $\overline{7}\oplus{}21\oplus{}\overline{35}$ attain masses after electroweak symmetry breaking. This means that in a situation where the relevant Yukawa coupling constants are natural, the fermionic mediators will attain masses which are of order the electroweak scale. Given that the LHC has so far failed to detect such exotics at the TeV scale, this is undesirable, implying that the associated Yukawa coupling constants must enter the non-perturbative regime. Furthermore, we can see that one generation of the $\overline{7}\oplus{}21\oplus{}\overline{35}$ combination yields \emph{two} visible Standard Model generations, implying that the minimal model based on this group theoretic structure will contain a four-generation SM. The recent results from the LHC also put strong constraints on a fourth generation \cite{eberhardt4gensmlhc, pdg2014}. 
 
 In light of the numerous undesirable properties of the $SU(7)$ model, one may think of extending to a higher gauge group so that we could possibly make the mediators massive through symmetry breaking in the dark sector rather than the visible sector. The simplest model that the author found which could possibly lead to this outcome is based on $SU(9)$. Unfortunately, this also does not work. 
 
 The $SU(9)$ model that the author formulated is based on the initial symmetry breaking pattern $SU(9)\rightarrow{}SU(5)_{V}\times{}SU(4)_{D}\times{}U(1)$ with an adjoint. Also, we choose each generation of left-chiral fermions to consist of the combination $\overline{9}\oplus{}36\oplus{}\overline{84}\oplus{}126$ of representations of $SU(9)$. These $SU(9)$ representations decompose under $SU(5)_{V}\times{}SU(4)_{D}\times{}U(1)$ as
 \begin{equation}
 \begin{aligned}
  \label{eq:su9admcombo}
  \overline{9} &= (\overline{5}, 1, -4)\oplus{}(1, \overline{4}, +5), \\
  36 &= (10, 1, +8)\oplus{}(5, 4, -1)\oplus{}(1, 6, -10), \\
  \overline{84} &= (10, 1, -12)\oplus{}(\overline{10}, \overline{4}, -3)\oplus{}(\overline{5}, 6, +6)\oplus{}(1, 4, +15), \\
  126 &= (\overline{5}, 1, +16)\oplus{}(\overline{10}, 4, +7)\oplus{}(10, 6, -2)\oplus{}(5, \overline{4}, -11)\oplus{}(1, 1, -20).
  \end{aligned}
 \end{equation}
  Now, one may think of making the fermionic mediators massive through symmetry breaking in the dark sector. The most obvious symmetry breaking pattern to consider is the breaking $SU(4)_{D}\rightarrow{}SU(3)_{D}$ with one of the various dark quartets embeded in the representations given in Eq.~\ref{eq:su9admcombo}. If we introduce a scalar in the $9$ representation, and use the $(1, 4, -5)$ component to break $SU(4)_{D}\rightarrow{}SU(3)_{D}$, we see that we get the following mass-generating terms:
  \begin{equation}
   \label{eq:su93684bar9}
   36_{F}\times{}\overline{84}_{F}\times{}9_{S}\supset{}(5, 4, -1)_{F}\times{}(\overline{5}, 6, +6)_{F}\times{}(1, 4, -5)_{S}\supset{}(5, 3)_{F}\times{}(\overline{5}, \overline{3})_{F}\times{}(1, 1)_{S}\supset{}(1, 1),
  \end{equation}
  \begin{equation}
   \label{eq:su984bar1269bar1}
   \overline{84}_{F}\times{}126_{F}\times{}\overline{9}_{S}\supset{}(\overline{5}, 6, +6)_{F}\times{}(5, \overline{4}, -11)_{F}\times{}(1, \overline{4}, +5)\supset{}(\overline{5}, 3)_{F}\times{}(5, \overline{3})_{F}\times{}(1, 1)_{S}\supset{}(1, 1),
  \end{equation}
\begin{equation}
   \label{eq:su984bar1269bar2}
   \overline{84}_{F}\times{}126_{F}\times{}\overline{9}_{S}\supset{}(\overline{10}, \overline{4}, -3)_{F}\times{}(10, 6, -2)_{F}\times{}(1, \overline{4}, +5)\supset{}(\overline{10}, \overline{3})_{F}\times{}(10, 3)_{F}\times{}(1, 1)_{S}\supset{}(1, 1),
  \end{equation}
  and
\begin{equation}
 \label{eq:su91261269}
 126_{F}\times{}126_{F}\times{}9_{S}\supset{}(\overline{10}, 4, +7)_{F}\times{}(10, 6, -2)_{F}\times{}(1, 4, -5)_{S}\supset{}(\overline{10}, 3)_{F}\times{}(10, \overline{3})_{F}\times{}(1, 1)_{S}\supset{}(1, 1).
\end{equation}
Here, we have suppressed the $U(1)$ charge in the representations under $SU(5)_{V}\times{}SU(3)_{D}$. All the above interactions imply that the mediators with non-trivial charges under both $SU(5)_{V}$ and $SU(3)_{D}$ attain masses of order the breaking scale of $SU(4)_{D}$. Unfortunately, the leftover $SU(3)_{D}$-singlet components which are charged under $SU(5)_{V}$ also go on to attain masses from the breaking of $SU(4)_{D}$ coming from the following interactions:
\begin{equation}
 \label{eq:su99bar369bar}
 \overline{9}_{F}\times{}36_{F}\times{}\overline{9}_{S}\supset{}(\overline{5}, 1, -4)_{F}\times{}(5, 4, -1)_{F}\times{}(1, \overline{4}, +5)_{S}\supset{}(\overline{5}, 1)_{F}\times{}(5, 1)_{F}\times{}(1, 1)_{S}\supset{}(1, 1),
\end{equation}
\begin{equation}
 \label{eq:su93684bar92}
 36_{F}\times{}\overline{84}_{F}\times{}9_{S}\supset{}(10, 1, +8)_{F}\times{}(\overline{10}, \overline{4}, -3)_{F}\times{}(1, 4, -5)_{S}\supset{}(10, 1)_{F}\times{}(\overline{10}, 1)_{F}\times{}(1, 1)_{S}\supset{}(1, 1),
\end{equation}
\begin{equation}
 \label{eq:su984bar1269bar3}
 \overline{84}_{F}\times{}126_{F}\times{}\overline{9}_{S}\supset{}(10, 1, -12)_{F}\times{}(\overline{10}, 4, +7)_{F}\times{}(1, \overline{4}, +5)_{S}\supset{}(10, 1)_{F}\times{}(\overline{10}, 1)_{F}\times{}(1, 1)_{S}\supset{}(1, 1),
\end{equation}
and 
\begin{equation}
 \label{eq:su912612692}
 126_{F}\times{}126_{F}\times{}9_{S}\supset{}(\overline{5}, 1, +16)_{F}\times{}(5, \overline{4}, -11)_{F}\times{}(1, 4, -5)_{S}\supset{}(\overline{5}, 1)_{F}\times{}(5, 1)_{F}\times{}(1, 1)_{S}\supset{}(1, 1).
\end{equation}

 Hence, all the visible fermions attain masses at the $SU(4)_{D}$ breaking scale, which we wish to be above the electroweak scale. Like the $SU(7)$ model, we can also make the mediators massive through electroweak symmetry breaking via a combination of the various Higgs quintets embedded in the representations in Eq.~\ref{eq:su9admcombo}. Again, it turns out that we have many of the same problems: non-perturbative coupling constants, a four-generation Standard Model at low energies, and a complicated Higgs sector.

  Given the troubles that we have encountered with $SU(N)$ groups, one might consider $SO(N)$ gauge theories instead. Given that we need complex representations to embed the SM fermions, we would need to choose $N = 4n+2$. Naturally, one would try a breaking pattern of the form $SO(4n+2)\rightarrow{}SO(10)_{V}\times{}SO(4n-8)_{D}$. Given that all the SM fermions naturally fit into the $16$ spinor representation of $SO(10)_{V}$, the natural representation to consider embedding the SM fermions along with dark quarks is the spinor representation of $SO(4n+2)$, which has a dimension of $2^{2n}$. The problem is that the $2^{2n}$ spinor representation typically decomposes completely into components which are charged under both $SO(10)_{V}$ and $SO(4n-8)_{D}$. For example, for $SO(18)$, the spinor $256$ representation decomposes under $SO(10)_{V}\times{}SO(8)_{D}$ as 
  \begin{equation}
   \label{eq:so18spinor}
   256 = (16, 8)\oplus{}(\overline{16}, 8'),
  \end{equation}
 where here the $8$ and $8'$ denote the two different, complex 8-dimensional spinor representations of $SO(8)_{D}$. Hence, both components in the $256$ are mixed fermions. Having experimented with many special orthogonal groups and combinations of their representations, this seems to be a generic trait that is difficult to overcome. Hence, in the context of $3+1D$ GUTs, $SO(N)$ theories do not show much promise either.

  We have not disproven that a satisfying $3+1D$ GUT yielding a non-Abelian dark gauge group can be constructed. However, the above examples seem to highlight the major difficulties in constructing such a theory. The large numbers of Higgs fields required to induce the various breakings, as well as the complicated representations required for the fermions to satisfy the constraints coming from anomaly cancellation and to ensure that the fermionic mediators can attain masses, make the types of models described above very undesirable. This seems to suggest that we need to consider additional physics to efficiently eliminate the fermionic mediators from the spectrum and to perhaps reduce the number of constraints on the theory. One may consider adding an extra dimension and localizing the desired fields on a domain-wall brane. Going to $4+1D$ automatically eliminates the constraints coming from anomalies, and, as it turns out in the context of a Clash-of-Symmetries domain wall, presents a way to make the fermionic mediators attain masses of order the GUT scale. We now turn our attention to domain-wall brane models and, as we will show later, a desirable Clash-of-Symmetries domain-wall brane model based on the gauge group $SU(12)$ in $4+1D$ which resolves many of the problems found in the $3+1D$ constructions in this section can be constructed.

\section{Domain Walls, the Dvali-Shifman Mechanism, and the Clash-of-Symmetries Mechanism}
\label{sec:cosmechanism}
 
 Domain walls are topological defects in which the boundary conditions for a scalar field(s) at positive and negative infinity along some spatial dimension are mapped to two discrete, degenerate and disconnected vacua for that (those) scalar field(s). Their topological stability is ensured due to the fact that they are mappings which belong to non-trivial homotopy classes of $\pi_{0}(M)$, where $M$ is the moduli space of the theory, unlike standard homogeneous vacuum states which belong to the trivial class. The simplest scalar field theory supporting topologically stable domain-wall solutions is a $\mathbb{Z}_{2}$-symmetric quartic scalar field theory for a single scalar field $\eta$ with a tachyonic mass, which may be written

 \begin{equation}
 \label{eq:dwscalarpotential}  
 V(\eta) = \frac{1}{4}\lambda{}(\eta^2-v^2)^2.
 \end{equation}

\begin{figure}[h]
 \includegraphics[scale=0.3]{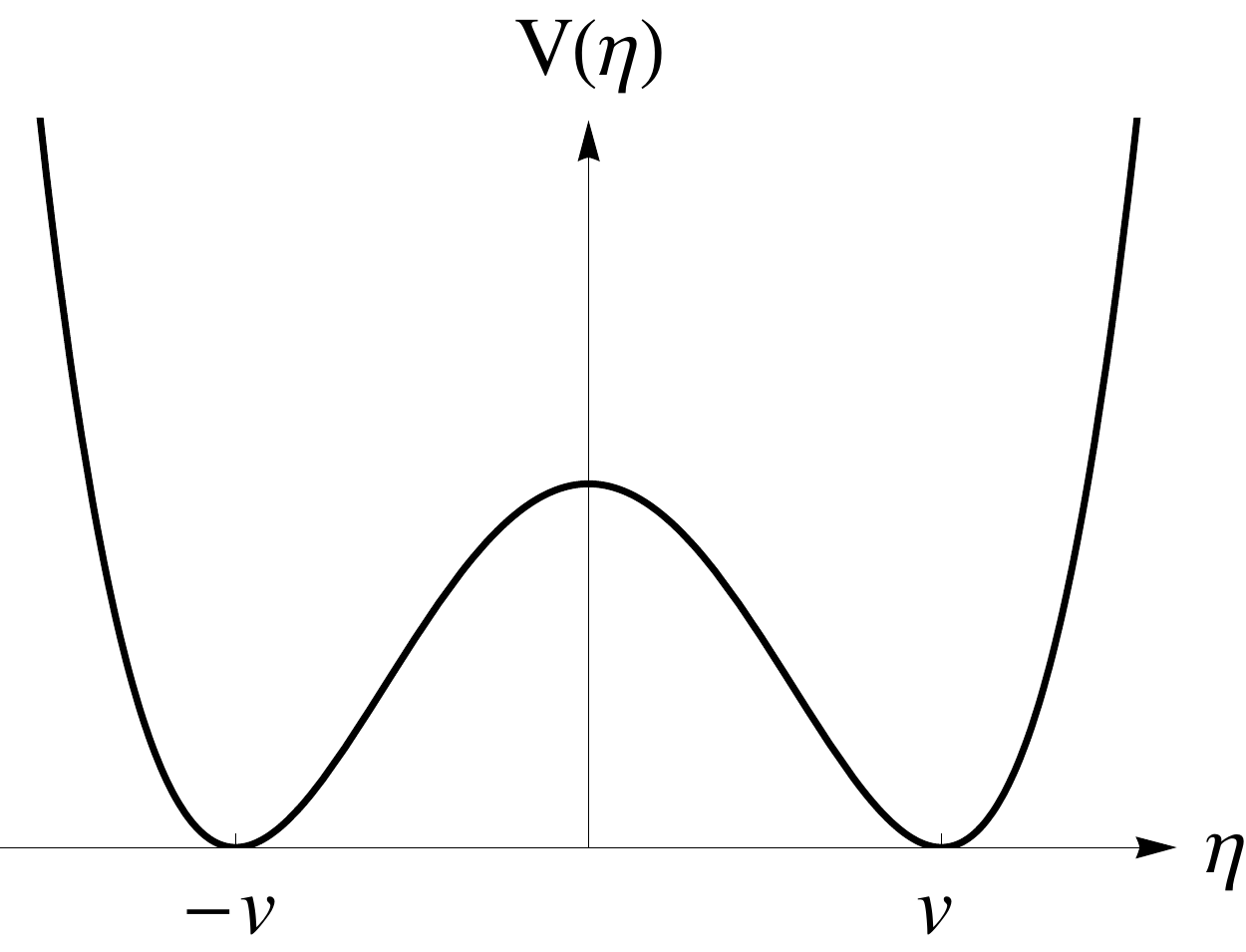}
\caption{A plot of the potential $V(\eta)$ from Eq.~\ref{eq:dwscalarpotential}.}
\label{fig:dwpotential}
\end{figure}

\begin{figure}[h]
 \includegraphics[scale=0.3]{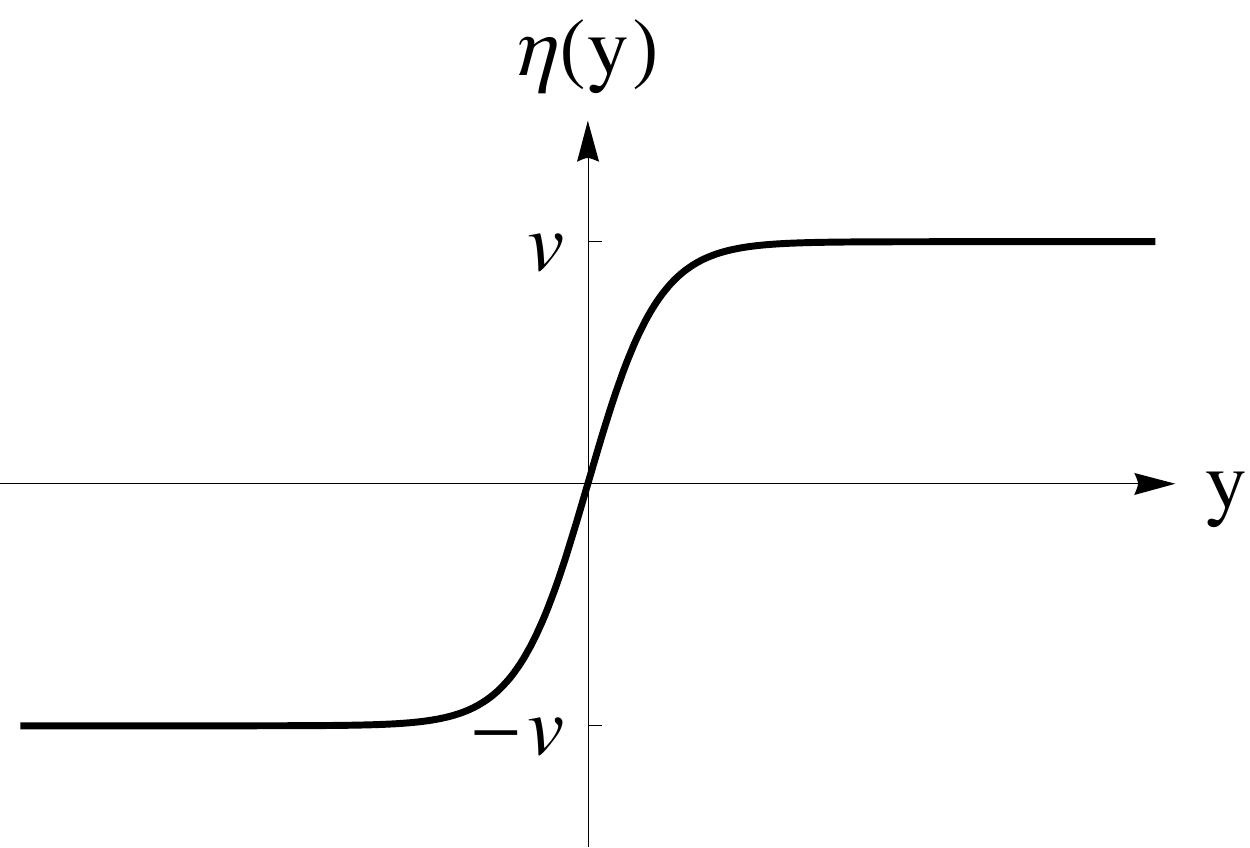}
\caption{A plot of the domain-wall solution for $\eta$ given by Eq.~\ref{eq:dwsolution}.}
\label{fig:dwsolution}
\end{figure}

 This potential has two discrete, degenerate vacua $\eta=-v$ and $\eta=+v$. One can show that 
\begin{equation}
\label{eq:dwsolution}
 \eta{(y)} = v\tanh{(ky)},
\end{equation}
 where $k^2 = \lambda{}v^2/2$, is a solution to the Euler-Lagrange equations with the scalar potential in Eq.~\ref{eq:dwscalarpotential}. This solution also satisfies the boundary conditions 
\begin{equation}
\begin{gathered}
\label{eq:dqboundaryconditions} 
\eta{(y\rightarrow{}-\infty{})} = -v, \\
\eta{(y\rightarrow{}+\infty{})} = +v,
\end{gathered}
\end{equation}
and is thus a domain-wall solution. A plot of the potential in terms of $\eta$ is given in Fig.~\ref{fig:dwpotential} and a plot of the solution in Eq.~\ref{eq:dwsolution} is given in Fig.~\ref{fig:dwsolution}.

 Having formed a domain wall, the next goal is to localize the various particle content required to formulate a realistic domain-wall brane localized effective field theory which contains the Standard Model. This involves the introduction of various dynamical localization mechanisms for fermions, scalars, gravitons and gauge bosons. We will not deal with the localization of gravity at all in this paper, but one can show that the localization of gravitons is possible \cite{gremmdwgravity, adavidsondwgravity, clashofsymgravity, kehagiastamvakisgravity, slatyervolkasrsgrav, sodarsgravity, rsgravitydaviesgeorge2007}. Fermionic chiral zero modes can be localized by Yukawa interactions to the domain wall as first shown in Ref.~\cite{jackiwrebbi}. Scalar modes can be localized via quartic interactions \cite{rsgravitydaviesgeorge2007}. We will deal with localization of fermions and scalars in the context of the model proposed in this paper in later sections. We will now turn to the localization of gauge bosons.

 Gauge bosons are the most difficult species of particle to localize to a domain wall. They cannot be localized to the domain wall through direct cubic or quartic couplings to it because the development of a zero mode profile for a gauge boson will in general mean that gauge charge universality in non-Abelian theories is lost \cite{dubrub}, because the effective gauge couplings for the different fermions and scalars depend on overlap integrals of the profiles of the particles involved. The only known, plausible mechanism which is conjectured to localize gauge bosons and retain gauge charge universality is the Dvali-Shifman mechanism \cite{dsmech}. This mechanism involves breaking a gauge group $G$ to a subgroup $H$ in the interior of the domain wall and then localizing the gauge bosons of $H$ through the confinement dynamics of $G$ in the bulk. To help illustrate this, we will consider the original $SU(2)$-model that Dvali and Shifman considered. 

 Consider taking the original model with a singlet scalar $\eta$ described by the potential in Eq.~\ref{eq:dwscalarpotential} and now adding to it another scalar field $\chi$ in the adjoint representation of $SU(2)$. Under the discrete $\mathbb{Z}_{2}$ symmetry, $\eta{}\rightarrow{}-\eta$ and $\chi{}\rightarrow{}-\chi{}$. The potential of this new theory is given by
\begin{equation}
\begin{aligned}
\label{eq:dvalishifmanmodel}
V(\eta, \chi) &= \frac{1}{4}\lambda_{\eta}(\eta^2-v^2)^2+\lambda_{\eta\chi}(\eta^2-v^2)\rm{Tr}[\chi^2]+\mu^2_{\chi}\rm{Tr}[\chi^2] \\
              &+\lambda_{\chi}\rm{Tr}[\chi^2]^2.
\end{aligned}
\end{equation}
To ensure that the above potential is bound from below and to ensure stable domain-wall solutions, we impose the parameter conditions
\begin{equation}
\label{eq:boundednessconditions}
\lambda_{\eta}>0, \qquad{} \lambda_{\chi}>0, \qquad \lambda_{\eta\chi}v^2>\mu^2_{\chi}>0.
\end{equation}
In this region of parameter space, the global minima are $\eta = \pm{}v$, $\chi = 0$. In the middle of the wall, the field $\chi$ develops a tachyonic mass and should condense. Without loss of generality, we choose the component proportional to the isospin operator $I = diag(-1, +1)$, which we call $\chi_{1}$, to condense with the other components set to zero. If we impose the additional special parameter choice
\begin{equation}
\label{eq:singlekinklumpconditions}
2\mu^2_{\chi}(\lambda_{\eta\chi}-\lambda_{\chi})+(\lambda_{\eta}\lambda_{\chi}-\lambda^2_{\eta\chi})v^2=0,
\end{equation}
one finds that 
\begin{equation}
\begin{aligned}
\label{eq:singlekinklumpsolution}
\eta(y)&=v\tanh{(ky)}, \\
\chi_{1}(y)&=A\sech{(ky)},
\end{aligned}
\end{equation}
where $k^2=\mu^2_{\chi}$ and $A^2=\frac{\lambda_{\eta\chi}v^2-2\mu^2_{\chi}}{\lambda_{\chi}}$, is a solution to the Euler-Lagrange equations satisfying the boundary conditions $\eta{(y\rightarrow{}\pm{}\infty)} = \pm{}v$ and $\chi{(y\rightarrow{}\pm{}\infty)} = 0$. For this solution, $\eta$ still generates the domain-wall kink while $\chi_{1}$ forms what we call a \emph{lump} which condenses to a non-zero vacuum expectation value in the interior of the wall. In parameter regions outside that implied by Eq.~\ref{eq:singlekinklumpconditions} (but within those of Eq.~\ref{eq:boundednessconditions}), there will still exist a solution in which $\eta$ generates a kink and $\chi_{1}$ generates a lump, although the solution will have to be solved numerically.

 In the interior of the wall, $\chi$ attains a non-zero vacuum expectation value which induces the breaking $SU(2)\rightarrow{}U(1)$. In the bulk, $\chi$ asymptotes to zero, leaving $SU(2)$ unbroken and confining there. Suppose a  $U(1)$ test charge is placed on the wall. In the bulk, $SU(2)$ is unbroken and confining which implies, under the dual superconductor picture of confinement first proposed by 't Hooft and Mandelstam \cite{thooftdualsuperconductor, mandelstamdualsuperconductor}, that the bulk behaves as a dual superconductor. It follows that the electric field lines emanating from the test charge will be repelled from the bulk by the dual Meissner effect and will diverge outwards parallel to the 3D world volume of the domain-wall brane. Now imagine that the test charge is placed in the bulk. The electric field lines will still be repelled from the bulk and they will form a flux string which diverges out onto the wall, so that the charge behaves as if it really were on the wall. In this way, the couplings of localized fermion and scalar modes, which are charged under the remaining $U(1)$ theory, to the $U(1)$ photon will be independent of where these modes are localized. The Dvali-Shifman mechanism as proposed generalizes this simple $SU(2)\rightarrow{}U(1)$ toy model to the case where the gauge symmetry $G$ respected is a larger non-Abelian group and the subgroup $H$ to which it is broken on the wall is a non-Abelian semi-simple group. Extrapolating the results for the test charge in the case described above to the case where $H$ is non-Abelian, this means that the couplings of localized quarks to gluons is independent of the quark profiles, preserving gauge charge universality as well as localizing the gluons. Note that the Dvali-Shifman idea depends on whether 5D Yangs-Mills gauge theories are confining. Although not absolutely proven, we have good numerical evidence that in 4+1D, SU(2) \cite{5dconfinement} and SU(5) \cite{damiengeorgephd} gauge theories are in fact confining. 

 The Dvali-Shifman mechanism is an attractive mechanism for the localization of gauge bosons as far as the building of domain-wall brane models is concerned. One may ask whether there are ways to extend this mechanism. In particular, given that the Dvali-Shifman mechanism requires two scalar fields which condense, one to form a kink and one to form a lump, one can ask whether it is possible to achieve the same dynamics with a single scalar field in some representation of a gauge group $G$. The first idea which naturally comes to mind is to reassign the kink-forming field $\eta$ from the gauge singlet representation to a non-trivial representation of $G$. If we ensure that the discrete $\mathbb{Z}_{2}$ symmetry is \emph{outside} the gauge group $G$, then, instead of just the two discrete vacua we had in the potential of Eq.~\ref{eq:dwscalarpotential}, we now have two disconnected vacuum \emph{manifolds}. If we choose parameters such that the most stable breaking pattern is from $G$ to a subgroup $H$, both these vacuum manifolds are then (individually) diffeomorphic to the coset manifold $G/H$. One can then think of forming domain wall solutions which interpolate between the two disconnected manifolds, which opens the possibility of breaking $G$ to two differently embedded copies of $H$ on each side of the domain wall, one of which we will call $H'$. Because these isomorphic subgroups are not exactly the same, there has to be further breaking in the core of the defect to the overlap of these groups, $H\cap{}H'$. The idea is then that if $H$ and $H'$ (or subgroups of them) are non-Abelian and confining in the bulk, then smaller subgroups can be localized to the domain wall interior by Dvali-Shifman dynamics. This version of the Dvali-Shifman mechanism is called the Clash-of-Symmetries (CoS) mechanism. Many attempts at forming either realistic models or simply just domain walls from the CoS mechanism exist in the literature Refs.~\cite{o10kinks, abeliankinkscos, clashofsymmetries, e6domainwallpaper, intersectingclashofsym, pogosianvachaspaticos, vachaspaticos2, pogosianvachaspaticos3}. A thorough exploration of the group theoretic aspects underlying it are given in Ref.~\cite{clashofsymgrouptheorypaper}.

 We now turn to the requirements of localization for non-Abelian and Abelian subgroups of $H\cap{}H'$. In general, both $H$ and $H'$ are semi-simple and may be written as
\begin{equation}
  \begin{aligned}
  \label{eq:semisimple}
  H &= N_{1}\times{}N_{2}\times{}N_{3}\times{}...\times{}N_{k-1}\times{}N_{k}\times{}U(1)_{Q_{1}}\times{}U(1)_{Q_{2}}\times{}U(1)_{Q_{3}}...U(1)_{Q_{l-1}}\times{}U(1)_{Q_{l}}, \\
  H' &= N'_{1}\times{}N'_{2}\times{}N'_{3}\times{}...\times{}N'_{k-1}\times{}N'_{k}\times{}U(1)_{Q'_{1}}\times{}U(1)_{Q'_{2}}\times{}U(1)_{Q'_{3}}...U(1)_{Q'_{l-1}}\times{}U(1)_{Q'_{l}},
  \end{aligned}
  \end{equation}
  where the $N_{i}$ and $N'_{i}$ denote the non-Abelian factor groups and the $Q_{i}$ and $Q'_{i}$ denote the generators of the Abelian factor groups belonging to $H$ and $H'$ respectively. 
  Since, $H$ and $H'$ are semi-simple, $H\cap{}H'$ is also semi-simple. We will denote its non-Abelian factor groups as $n_{i}$ and the generators of its Abelian factor groups as $q_{i}$ and write
  \begin{equation}
  \label{eq:intersectingsemisimple}
  H\cap{}H' = n_{1}\times{}n_{2}\times{}n_{3}\times{}...\times{}n_{r-1}\times{}n_{r}\times{}U(1)_{q_{1}}\times{}U(1)_{q_{2}}\times{}U(1)_{q_{3}}\times{}...\times{}U(1)_{q_{s-1}}\times{}U(1)_{q_{s}}.
  \end{equation}

 We will deal with localization of non-Abelian groups first. Given the Dvali-Shifman mechanism relies on confinement dynamics, it follows that for a gauge group to be localized to the domain wall, it must lie inside a larger non-Abelian group in the bulk. Given that that on one side of the wall, a non-Abelian subgroup $n_{i}$ of $H\cap{}H'$ will lie inside a non-Abelian factor $N_{a}$ of $H$, while on the other it will be a subgroup of a non-Abelian factor $N'_{b}$ of $H'$, it follows that to be fully localized $n_{i}$ must be a proper subgroup of both $N_{a}$ and $N'_{b}$
  \begin{equation}
   \begin{gathered}
  \label{eq:nonabelianlocclashofsym}
  n_{i}\subset{}N_{a} \; \mathrm{and} \; n_{i}\subset{}N'_{b}.
  \end{gathered}
   \end{equation}
   If, on the other hand, $n_{i}$ is precisely equal to one of these groups, then on one side of the bulk is will be free to propogate and thus semi-delocalized, and if, in the rare case, $n_{i}=N_{a}=N'_{b}$, then $n_{i}$ will be fully delocalized. 
   
   Localization of Abelian gauge bosons is slightly more complex, but similar. In general, the Abelian generators $q_{i}$ of $H\cap{}H'$ must, to be respected at the level of \emph{symmetries} on the wall, be able to be written as linear combinations of generators of $H$ and $H'$ independently. Obviously, the Abelian generators in $H$ and $H'$, $Q_{i}$ and $Q'_{i}$, can contribute to these respective linear combinations, but there are also leftover generators inside the non-Abelian factors $N_{a}$ and $N'_{b}$, which we call $T_{i}$ and $T'_{i}$ respectively, which are outside the non-Abelian factors $n_{i}$ of $H\cap{}H'$. Hence the condition for $U(1)_{q_{i}}$ to be a symmetry inside $H\cap{}H'$ respected on the domain-wall brane is for the generator $q_{i}$ to satisfy
   \begin{equation}
  \begin{aligned}
  \label{eq:abeliangeneratorsymmetrycondition}
  q_{i} &= \sum^{l}_{j=1}\alpha^{i}_{j}Q_{j}+\sum^{m}_{j=1}\beta^{i}_{j}T_{j}, \\
        &= \sum^{l}_{j=1}\alpha'^{i}_{j}Q'_{j}+\sum^{m}_{j=1}\beta'^{i}_{j}T'_{i}.
  \end{aligned}
  \end{equation}
  The conditions for full \emph{localization} of an Abelian gauge boson are more stringent. Just like the localized non-Abelian gauge bosons, the Abelian gauge bosons must lie completely inside non-Abelian groups wherever they may propagate through the bulk. Furthermore, the photons corresponding to the respective Abelian generators of $H$ and $H'$, $Q_{i}$ and $Q'_{i}$,  are able to propagate through the halves of the bulk in which they are respected, and if they contribute to the linear combination for $q_{i}$, there is a chance that the photon associated with $q_{i}$ will leak into the bulk. The consequence is that for the photon of $q_{i}$ to be fully localized to the domain wall, it must only be a linear combination of the $T_{i}$ and $T'_{i}$ generators of $H$ and $H'$ respectively, satisfying
  \begin{equation}
  \label{eq:abeliangeneratorlocalizationcondition}
  q_{i} = \sum^{m}_{j=1}\beta^{i}_{j}T_{j} = \sum^{m}_{j=1}\beta'^{i}_{j}T'_{j}, \qquad{} \alpha^{i}_{j} = \alpha'^{i}_{j}  = 0\; \forall{} \; j.\\
  \end{equation}
  If any of the $\alpha^{i}_{j}$ are non-zero while all of the $\alpha'^{i}_{j}$ are zero, then the photon is semi-delocalized and able to propagate into the $H$-respecting side of the bulk, but not the $H'$-respecting side, and vice versa if some $\alpha'^{i}_{j}$ are non-zero and all the $\alpha^{i}_{j}$ are zero. If there exist both some non-zero $\alpha^{i}_{j}$ and some non-zero $\alpha'^{i}_{j}$, the photon can leak into both sides of the bulk and is thus fully delocalized. 
  
  In this section, we have explained the formation of domain walls, the Dvali-Shifman mechanism for gauge boson localization and ended with the clash-of-symmetries mechanism. There have been attempts to form viable domain-wall brane models with a localized Standard Model based on $SO(10)$ \cite{o10kinks} and $E_{6}$ \cite{e6domainwallpaper} using the clash-of-symmetries mechanism. A slightly altered version of this mechanism can be used to localize gauge fields on to the intersection of two domain-wall branes in $5+1D$ spacetime, assuming both $4+1D$ and $5+1D$ Yang-Mills theories are confining \cite{intersectingclashofsym}. In this paper, we will exploit this mechanism in an $SU(12)$-invariant theory to generate a localized $SU(5)_{V}\times{}SU(5)_{D}\times{}U(1)_{X}$ gauge theory, where $SU(5)_{V}$ contains the gauge groups of the visible Standard Model sector and $SU(5)_{D}$ contains the gauge groups of a dark matter hidden sector, and $U(1)_{X}$ is an Abelian gauge group coupling the two sectors (and thus must be broken spontaneously by adding further Higgs fields). Furthermore, we will show that troublesome fermionic and scalar mediators which are charged under both $SU(5)_{V}$ and $SU(5)_{D}$ are eliminated from the spectrum, leading to sufficient sequestration of the visible and hidden sectors. In particular, we will show that the mixed $(5, 5)$ fermion in the rank two antisymmetric representation of $SU(12)$ is completely decoupled from the wall, meaning it remains 5D and vector-like and will thus attain a mass of order $M_{GUT}$ on the brane, once we break $SU(5)_{V}$ to the Standard Model.

\section{A Localized $SU(5)_{V}\times{}SU(5)_{D}\times{}U(1)_{X}$-Invariant Effective Action from a Clash-of-Symmetries Domain Wall in a 4+1D $SU(12)\times{}\mathbb{Z}_{2}$ Scalar Field Theory}
\label{sec:solution}

 In this section, we will describe a CoS domain-wall solution which yields a localized $SU(5)_{V}\times{}SU(5)_{D}\times{}U(1)_{X}$ gauge theory from a 4+1D $SU(12)$ theory. To achieve this, we break $SU(12)$ to two differently embedded copies of $SU(6)\times{}SU(6)\times{}U(1)$ on each side of the wall. One may first consider achieving this with a single adjoint scalar field $\eta$ which transforms under a discrete $\mathbb{Z}_{2}$-symmetry as $\eta{}\rightarrow{}-\eta{}$. The $\mathbb{Z}_{2}$-symmetric scalar potential for this field is
\begin{equation}
\label{eq:singleadjointpotential} 
V(\eta) = -\mu^{2}Tr[\eta^2]+\lambda_{1}(Tr[\eta^2])^2+\lambda_{2}Tr[\eta^4].
\end{equation}
 The global minima of the potential have been well studied \cite{lingfongli74}. For $\lambda_{2}<0$, this potential induces the breaking $SU(12)\rightarrow{}SU(11)\times{}U(1)$, and, for $\lambda_{2}>0$, the most stable symmetry breaking pattern is $SU(12)\rightarrow{}SU(6)\times{}SU(6)\times{}U(1)$. Thus, we desire that $\lambda_{2}>0$. However, given that, in a convenient choice of basis, $\langle{}\eta{}\rangle{}\propto{}diag(-1, -1, -1, -1, -1, -1, +1, +1, +1, +1, +1, +1)$, one can see that for $U=i\sigma_{1}\otimes{}\mathbbm{1}_{6\times{}6}\in{}SU(12)$ (where here $\sigma_{1}$ is the first Pauli matrix), $U^{\dagger}\langle{}\eta{}\rangle{}U = -\langle{}\eta{}\rangle{}$. This means that $\langle{}\eta{}\rangle{}$ is related to $-\langle{}\eta{}\rangle{}$ by a gauge transformation, and thus the vacuum manifold is in fact \emph{connected} and contains only a single component diffeomorphic to $G/H = SU(12)/SU(6)\times{}SU(6)\times{}U(1)$. This means that there will not exist any stable domain wall solutions since we require a disconnected vacuum manifold.

 The simplest way to resolve the connectedness problem of the single adjoint $\mathbb{Z}_{2}$-symmetric scalar potential is to use a scalar field theory with two independent adjoint scalar fields. We denote these two adjoint scalar fields as $\eta$ and $\chi$, and, instead of a discrete reflection symmetry, we utilize the interchange symmetry $\eta{}\rightarrow{}\chi{}$, $\chi{}\rightarrow{}\eta{}$ as our discrete $\mathbb{Z}_{2}$ symmetry. The most general potential for this system may be written
\begin{equation}
 \label{eq:twoadjointpotentialfull}
 V(\eta, \chi) = V(\eta)+V(\chi)+I(\eta, \chi),
\end{equation}
where
\begin{equation}
\begin{aligned}
\label{eq:twoadjointpotentialparts}
V(\eta) &= -\mu^{2}Tr[\eta^2]-\frac{1}{3}cTr[\eta^3]+\lambda_{1}(Tr[\eta^2])^2+\lambda_{2}Tr[\eta^4], \\
V(\chi) &= -\mu^{2}Tr[\chi^2]-\frac{1}{3}cTr[\chi^3]+\lambda_{1}(Tr[\chi^2])^2+\lambda_{2}Tr[\chi^4], \\
I(\eta, \chi) &= 2\delta^2Tr[\eta{}\chi]+dTr[\eta^2\chi]+dTr[\eta{}\chi^2]+l_{1}Tr[\eta^2]Tr[\chi^2]+l_{2}Tr[\eta^2\chi^2]+l_{3}(Tr[\eta{}\chi])^2 \\
              &+l_{4}Tr[\eta{}\chi{}\eta{}\chi{}]+l_{5}Tr[\eta^2]Tr[\eta{}\chi]+l_{5}Tr[\eta{}\chi]Tr[\chi^2]+l_{6}Tr[\eta^3\chi]+l_{6}Tr[\eta{}\chi^3].
\end{aligned}
\end{equation}
 The single-field potentials $V(\eta)$ and $V(\chi)$ are simply the single-adjoint scalar potential with the cubic invariant, while $I(\eta, \chi)$ is the interaction potential containing all the terms which couple $\eta$ and $\chi$ non-trivially. The determination of the global minima for the two-adjoint scalar Higgs potential is obviously much more complicated. An analysis of the most general potential (without the discrete symmetry that we have imposed) was first given by Wu \cite{dandiwusymbreaking}. Nevertheless, we can present an argument for the existence of domain wall solutions and an argument for the existence of a region of parameter space in which the desired solution with $SU(5)_{V}\times{}SU(5)_{D}\times{}U(1)_{X}$ localized to the wall is the most stable domain wall solution. 
 
  The existence of domain wall solutions is ensured by the disconnectedness of the vacuum manifold. If we choose parameters such that the potential is bound from below, $V(\eta, \chi)$ must have at least one global minimum. If this minimum is given by $\eta = A$, $\chi = B$, then by the interchange symmetry $\eta = B$, $\chi = A$ is also a global minimum. Given that both the fields are adjoint fields, there is a connected component of the vacuum manifold described by $G/H = \lbrace{}(\eta, \chi) = (U^{\dagger}AU, U^{\dagger}BU); U\in{}G\rbrace{}$ and another component described by $(G/H)_{\eta{}\leftrightarrow{}\chi{}} = \lbrace{}(\eta, \chi) = (U^{\dagger}BU, U^{\dagger}AU); U\in{}G\rbrace{}$. Given that $\eta$ and $\chi$ are independent fields, the interchange symmetry is by construction outside $G$, ensuring that $G/H$ and $(G/H)_{\eta{}\leftrightarrow{}\chi{}}$ are disconnected.
  
  To argue for the existence of a parameter space yielding the desired CoS solution, it will greatly help us if we make some choices which simplify the analysis a great amount. There are two things which make the analysis rather simple: choosing parameters such that the vacua of the respective disconnected components of the vacuum manifold are of the form $\eta\neq{}0$, $\chi=0$ and $\eta=0$, $\chi\neq{}0$, and, choosing parameters such that the solutions for which $[\eta, \chi] \neq 0$, at any point, will clearly be non-minimal. Consider the analogous potential with two gauge singlets, $\phi$ and $\varphi$, with the interchange symmetry $\phi{}\leftrightarrow{}\varphi{}$,
  \begin{equation}
   \label{eq:twosingletinterchangepotential}
   V(\phi, \varphi) = -\frac{1}{2}M^2\phi^2-\frac{1}{3}a\phi^3+\frac{1}{4}F\phi^{4}-\frac{1}{2}M^2\varphi^2-\frac{1}{3}a\varphi^3+\frac{1}{4}F\varphi^{4}+N^2\phi\varphi+\frac{1}{2}L\phi^2\varphi^2+g\phi^3\varphi.
  \end{equation}
 Suppose we turn off the interactions involving odd powers of $\phi$ and $\varphi$ for now by setting the $N=a=g=0$. The potential in this case has a few additional reflection symmetries, $\phi\rightarrow{}-\phi$, $\varphi\rightarrow{}\varphi$ and $\phi\rightarrow{}\phi$, $\varphi\rightarrow{}-\varphi$. A quick calculation of the stationary points shows that, other than $\phi=0$, $\varphi=0$, which is always a maximum, there are stationary points at $\phi = \pm{}\sqrt{M^2/F}$, $\varphi = 0$ and $\phi = 0$, $\varphi = \pm{}\sqrt{M^2/F}$, and at $\phi = \pm{}\sqrt{M^2/(F+L)}$, $\varphi = \pm{}\sqrt{M^2/(F+L)}$. The values of the potential for these respective vacua are $V(\pm{}\sqrt{M^2/F}, 0) = V(0, \pm{}\sqrt{M^2/F}) = -M^4/4F$ and $V(\pm{}\sqrt{M^2/F}, \pm{}\sqrt{M^2/F}) = -M^4/2(F+L)$. In the region $L>F$, the stationary points $\phi = \pm{}\sqrt{M^2/F}$, $\varphi = 0$, and $\phi = 0$, $\varphi = \pm{}\sqrt{M^2/F}$ are degenerate global minima while the stationary points $\phi = \pm{}\sqrt{M^2/(F+L)}$, $\varphi = \pm{}\sqrt{M^2/(F+L)}$ are saddle points. In the region $L<F$, the points $\phi = \pm{}\sqrt{M^2/(F+L)}$, $\varphi = \pm{}\sqrt{M^2/(F+L)}$ are the global minima and $\phi = \pm{}\sqrt{M^2/F}$, $\varphi = 0$ and $\phi = 0$, $\varphi = \pm{}\sqrt{M^2/F}$ are the saddle points. For $F=L$, the symmetry of the potential is enhanced to $SO(2)$. 
 
 Note that for the above potential for two singlet fields that choosing the coupling constant for the $\phi^2\varphi^2$ interaction to be larger than that for the $\phi^4$ and $\varphi^4$ self-interactions, the minima are such that only one of the fields develops a non-zero vacuum expectation value. Similarly, at least when we leave the interactions with odd powers of $\eta$ and $\chi$ switched off, the minima of the potential in Eq.~\ref{eq:twoadjointpotentialparts} should be of the form $\eta \neq{} 0$, $\chi = 0$ and $\eta = 0$, $\chi \neq{} 0$ if some of the coupling constants involving the products of quadratic powers of $\eta$ and $\chi$, $l_{1}$, $l_{2}$, $l_{3}$ and $l_{4}$, are made sufficiently positive and larger compared to the quartic self-couplings $\lambda_{1}$ and $\lambda_{2}$. In particular, along any direction $\eta = v_{1}A$, $\chi = v_{2}B$, where $A$ and $B$ are generators, we need the effective coupling for $v^{2}_{1}v^2_{2}$ to be larger than that for $v^4_{1,2}$. This can be easily done by making $l_{1}$ sufficiently large, since $Tr[\eta^2]Tr[\chi^2]$ is independent of the vacuum alignment. 
 
 In choosing conditions such that the minima are of the form $\eta \neq{} 0$, $\chi = 0$ and $\eta = 0$, $\chi \neq{} 0$, $I(\eta, \chi)$ vanishes and becomes positive if we deviate from the minima. This implies that for the $\eta \neq{} 0$, $\chi = 0$ minima, $\eta$ must exist at the minimum of the single adjoint Higgs potential $V(\eta)$ (and likewise, by the interchange symmetry, $\chi$ must exist at a minimum of $V(\chi)$ for the $\eta = 0$, $\chi \neq{} 0$ minima). This means that the symmetry breaking patterns under such conditions reduce to those of the single adjoint Higgs potential, for which the minimal breaking patterns are well known \cite{lingfongli74, rueggscalarpot}. Setting $c=0$ and choosing $\lambda_{2}>0$, $\eta$ in the $\eta \neq{} 0$, $\chi = 0$ minimum, and $\chi$ in the $\eta = 0$, $\chi \neq{} 0$ minimum, will attain vacuum expectation values which break $SU(12)$ to $SU(6)\times{}SU(6)\times{}U(1)$. 
 
 A domain-wall solution can then be obtained by looking for a solution which interpolates from the $\eta \neq{} 0$, $\chi = 0$ minimum as $y\rightarrow{}-\infty$ to the $\eta = 0$, $\chi \neq{} 0$ as $y\rightarrow{}+\infty$. In the parameter regime we have chosen, this means that with this solution we have two domains in which $SU(12)$ is broken to $SU(6)\times{}SU(6)\times{}U(1)$ subgroups which need not be exactly the same. In other words, in the domain in which the solution converges to the $\eta \neq{} 0$, $\chi = 0$ minimum, $SU(12)$ is broken to the embedding $H_{1} = SU(6)_{1}\times{}SU(6)_{2}\times{}U(1)_{A}$, while in the domain in which the solution converges to the $\eta = 0$, $\chi \neq{} 0$ minimum, $SU(12)$ is broken to a potentially different embedding $H_{2} = SU(6)_{3}\times{}SU(6)_{4}\times{}U(1)_{B}$. In the interior of the domain wall, the symmetry is broken to $H_{1}\cap{}H_{2}$.
 
 To analyze what $H_{1}\cap{}H_{2}$ should be, we need to look at the vacua attained by $\eta$ and $\chi$ at $y=-\infty$ and $y=+\infty$ respectively. Without loss of generality, we can choose $\eta$ to attain the VEV pattern
 \begin{equation}
  \label{eq:etaminusinfinity}
  \eta{(y\rightarrow{}-\infty)} = vA,
 \end{equation}
where 
\begin{equation}
 \label{eq:amatrix}
 A = \begin{pmatrix} -\mathbbm{1}_{6\times{}6} & 0 \\
                     0 & +\mathbbm{1}_{6\times{}6}
     \end{pmatrix}.
\end{equation}
In general, at $y=+\infty$, $\chi$ will in general attain a VEV of the form $\chi(y\rightarrow{}+\infty) = vB$, where
\begin{equation}
 \label{eqbmatrixgeneral}
 B = U^{\dagger}AU,
\end{equation}
where $U$ is some unitary rotation matrix. In the general case, $A$ and $B$ will not commute. To make the upcoming analysis much simpler, we will further restrict the parameter space such that we ensure that $A$ and $B$ commute and are simultaneously diagonalizable, and, more generally, that $\eta{(y)}$ and $\chi{(y)}$ commute along the entire extra dimension. If we rewrite $I(\eta, \chi)$ in terms of $[\eta, \chi]$ and $\{\eta, \chi\}$, it turns out there is only one term which depends non-trivially on the commutator; that term is precisely
\begin{equation}
 \label{eq:commutatorterm}
 \frac{1}{4}(l_{4}-l_{2})Tr([\eta, \chi]^2).
\end{equation}
Given that $\eta$ and $\chi$ are real fields, and that the commutator $[\eta, \chi]$ is anti-hermitian and thus has complex eigenvalues, it follows that $[\eta, \chi]^2$ is a negative definite operator and that the trace, $Tr([\eta, \chi]^2)$ should always yield a negative number. Hence, to ensure $[\eta, \chi]$ along the domain wall solution, we need to make the difference $l_{2}-l_{4}$ sufficiently positive. We will always assume this is the case. This means that $B$ will be simultaneously diagonalizable with $A$ and, in general, may be written in the form 
\begin{equation}
 \label{eq:bmatrixdiagonalized}
 B = \begin{pmatrix} +\mathbbm{1}_{(6-m)\times{}(6-m)} & 0 & 0 & 0 \\
                     0 & -\mathbbm{1}_{m\times{}m} & 0 & 0 \\
                     0 & 0 & +\mathbbm{1}_{m\times{}m} & 0 \\
                     0 & 0 & 0 & -\mathbbm{1}_{(6-m)\times{}(6-m)} \end{pmatrix},
\end{equation}
where here $m$ is an integer between zero and six. In the cases that $m=0$ and $m=6$, we see that $B=-A$ and $B=+A$ respectively. This means that $H_{1}=H_{2}$ in these cases and that they are non-CoS domain walls. For $m=0$ and $m=6$, the symmetry respected in the interior of the wall is simply the same $SU(6)\times{}SU(6)\times{}U(1)_{A}$ subgroup respected in the bulk. Otherwise, in the case that $1\leq{}m\leq{}5$, we can see by inspection of $A$ and $B$ that the symmetry in the interior of the wall can be written
\begin{equation}
 \label{eq:cossym}
 H_{1}\cap_{}H_{2} = SU(6-m)_{1}\times{}SU(m)_{1}\times{}SU(6-m)_{2}\times{}SU(m)_{2}\times{}U(1)_{X_{m}}\times{}U(1)_{A}\times{}U(1)_{B}.
\end{equation}
Here, $X_{m}$ is a generator which is a sum of leftover generators from the original $SU(6)$ subgroups, namely $T_{1m}$, $T_{2m}$, $T_{3m}$ and $T_{4m}$ from $SU(6)_{1}$, $SU(6)_{2}$, $SU(6)_{3}$ and $SU(6)_{4}$ respectively. We choose to write these generators as
\begin{equation}
 \label{eq:t1su61}
 T_{1m} = \begin{pmatrix} +m\mathbbm{1}_{(6-m)\times{}(6-m)} & 0 & 0 \\
                         0 & -(6-m)\mathbbm{1}_{m\times{}m} & 0 \\
                         0 & 0 & 0_{6\times{}6} \end{pmatrix},
\end{equation}
\begin{equation}
 \label{eq:t2su62}
 T_{2m} = \begin{pmatrix} 0_{6\times{}6} & 0 & 0 \\
                         0 & -(6-m)\mathbbm{1}_{m\times{}m} & 0 \\
                         0 & 0 & +m\mathbbm{1}_{(6-m)\times{}(6-m)} \end{pmatrix},
\end{equation}
\begin{equation}
 \label{eq:t3su63}
 T_{3m} = \begin{pmatrix} +m\mathbbm{1}_{(6-m)\times{}(6-m)} & 0 & 0 & 0\\
                         0 & 0_{m\times{}m} & 0 & 0 \\
                         0 & 0 & -(6-m)\mathbbm{1}_{m\times{}m} & 0 \\
                         0 & 0 & 0 & 0_{(6-m)\times{}(6-m)} \end{pmatrix},
\end{equation}
and
\begin{equation}
 \label{eq:t4su64}
 T_{4m} = \begin{pmatrix} 0_{(6-m)\times{}(6-m)} & 0 & 0 & 0\\
                         0 & -(6-m)\mathbbm{1}_{m\times{}m} & 0 & 0 \\
                         0 & 0 & 0_{m\times{}m} & 0 \\
                         0 & 0 & 0 & +m\mathbbm{1}_{(6-m)\times{}(6-m)} \end{pmatrix}.
\end{equation}
Putting this together, we see that 
\begin{equation}
\begin{aligned}
 \label{eq:xgenerator}
 X_{m} &= T_{1m}+T_{2m} \\
   &= T_{3m}+T_{4m} \\
   &= \begin{pmatrix} +m\mathbbm{1}_{(6-m)\times{}(6-m)} & 0 & 0 & 0\\
                         0 & -(6-m)\mathbbm{1}_{m\times{}m} & 0 & 0 \\
                         0 & 0 & -(6-m)\mathbbm{1}_{m\times{}m} & 0 \\
                         0 & 0 & 0 & +m\mathbbm{1}_{(6-m)\times{}(6-m)} \end{pmatrix},
 \end{aligned}
\end{equation}
satisfies not only Eq.~\ref{eq:abeliangeneratorsymmetrycondition} but also the condition of Eq.~\ref{eq:abeliangeneratorlocalizationcondition} and is thus fully localized on the domain wall. Likewise, given $1\leq{}m\leq{}5$, all the $SU(m)_{1, 2}$ and $SU(6-m)_{1, 2}$ subgroups are smaller than their parent $SU(6)$ subgroups on both sides of the wall and are thus localized. Given that $U(1)_{A}$ and $U(1)_{B}$ are unbroken in their respective domains, their photons are semi-delocalized. Hence, each of the solutions for $m$ between one and five leads to the localization of gauge bosons associated with a $SU(6-m)_{1}\times{}SU(m)_{1}\times{}SU(6-m)_{2}\times{}SU(m)_{2}\times{}U(1)_{X_{m}}$ subgroup on the domain wall. Obviously, we are most interested in the $m=1$ and $m=5$ CoS solutions as they lead to a localized $SU(5)\times{}SU(5)\times{}U(1)$ gauge theory. 

 We have outlined some of the parameter region and types of boundary conditions needed for the domain-wall solutions of interest. To calculate a domain-wall solution we also need to solve the Euler-Lagrange equations. Having chosen conditions such that $\eta$ and $\chi$ will be diagonal along the whole extra dimension described by the coordinate $y$, we may write $\eta{(y)} = diag(a_{1}(y), a_{2}(y), ..., a_{12}(y))$ and $\chi = diag(b_{1}(y), b_{2}(y), ..., b_{12}(y))$. In terms of $\eta$ and $\chi$, noting that the kinetic terms for these fields are given by
 \begin{equation}
  \label{eq:etachikinetic}
  K(\eta, \chi) = \frac{1}{2}Tr[\partial^{\mu}\eta{}\partial_{\mu}\eta]+\frac{1}{2}Tr[\partial^{\mu}\chi{}\partial_{\mu}\chi],
 \end{equation}
the Euler-Lagrange equations resulting from the potential in Eq.~\ref{eq:twoadjointpotentialparts} are given by
 \begin{equation}
 \begin{aligned}
  \label{eq:eulerlagrangetwoadjointscalarpotential}
  &\Box{}\eta-2\mu^2\eta-c\eta^2+4\lambda_{1}Tr(\eta^2)\eta+4\lambda_{2}\eta^3+2\delta^{2}\chi+d\eta{}\chi{}+d\chi{}\eta{}+d\chi^2+2l_{1}Tr(\chi^2)\eta+l_{2}\eta{}\chi^{2}+l_{2}\chi^{2}\eta \\
            &+2l_{3}Tr(\eta\chi)\chi+2l_{4}\chi\eta\chi+2l_{5}Tr(\eta\chi)\eta+l_{5}Tr(\eta^2)\chi+l_{5}Tr(\chi^2)\chi+l_{6}\eta^2\chi+l_{6}\eta\chi\eta+l_{6}\chi{}\eta^2+l_{6}\chi^3 = 0, \\
  &\Box{}\chi-2\mu^2\chi-c\chi^2+4\lambda_{1}Tr(\chi^2)\chi+4\lambda_{2}\chi^3+2\delta^{2}\eta+d\eta{}\chi{}+d\chi{}\eta{}+d\eta^2+2l_{1}Tr(\eta^2)\chi+l_{2}\chi{}\eta^{2}+l_{2}\eta^{2}\chi \\
            &+2l_{3}Tr(\eta\chi)\eta+2l_{4}\eta\chi\eta+2l_{5}Tr(\eta\chi)\chi+l_{5}Tr(\chi^2)\eta+l_{5}Tr(\eta^2)\eta+l_{6}\chi^2\eta+l_{6}\chi\eta\chi+l_{6}\eta{}\chi^2+l_{6}\eta^3 = 0.           
  \end{aligned}
 \end{equation}
Under the conditions we have chosen, in terms of $a_{i}(y)$ and $b_{i}(y)$, these equations simply reduce to
\begin{equation}
 \begin{aligned}
  \label{eq:eulerlagrangeab}
  &-\frac{d^{2}a_{i}}{dy^2}-2\mu^2a_{i}-ca^{2}_{i}+4\lambda_{1}(\sum^{12}_{j=1}a^2_{j})a_{i}+4\lambda_{2}a^{3}_{i}+2\delta^{2}b_{i}+2da_{i}b_{i}+db^{2}_{i}+2l_{1}(\sum^{12}_{j=1}b^2_{j})a_{i}+2(l_{2}+l_{4})b^{2}_{i}a_{i} \\
  &+2l_{3}(\sum^{12}_{j=1}a_{j}b_{j})b_{i}+2l_{5}(\sum^{12}_{j=1}a_{j}b_{j})a_{i}+l_{5}(\sum^{12}_{j=1}a^2_{j})b_{i}+l_{5}(\sum^{12}_{j=1}b^2_{j})b_{i}+3l_{6}a^2_{i}b_{i}+l_{6}b^{3}_{i} = 0, \\
  &-\frac{d^{2}b_{i}}{dy^2}-2\mu^2b_{i}-cb^{2}_{i}+4\lambda_{1}(\sum^{12}_{j=1}b^2_{j})b_{i}+4\lambda_{2}b^{3}_{i}+2\delta^{2}a_{i}+2da_{i}b_{i}+da^{2}_{i}+2l_{1}(\sum^{12}_{j=1}a^2_{j})b_{i}+2(l_{2}+l_{4})a^{2}_{i}b_{i} \\
  &+2l_{3}(\sum^{12}_{j=1}a_{j}b_{j})a_{i}+2l_{5}(\sum^{12}_{j=1}a_{j}b_{j})b_{i}+l_{5}(\sum^{12}_{j=1}b^2_{j})a_{i}+l_{5}(\sum^{12}_{j=1}a^2_{j})a_{i}+3l_{6}b^2_{i}a_{i}+l_{6}a^{3}_{i} = 0.
 \end{aligned}
\end{equation}

 Note in the above equation implies that the coupled Euler-Lagrange equation for each pair $(a_{i}, b_{i})$ is the same, independent of the index $i$. This means, similarly to other clash-of-symmetries models \cite{e6domainwallpaper, o10kinks}, that the solutions are determined entirely from the boundary conditions. If one looks at the boundary conditions at infinity, namely $\eta(y\rightarrow{}-\infty) = vA$, $\chi(y\rightarrow{}-\infty) = 0$ and $\eta(y\rightarrow{}+\infty) = 0$, $\chi(y\rightarrow{}+\infty) = vB$, one notices that for some $i$, $a_{i}(y\rightarrow{}-\infty)$ and $b_{i}(y\rightarrow{}+\infty)$ have the same sign (either both $-v$ or $+v$), and for other pairs they have the opposite sign (either $(-v, +v)$ or $(+v, -v)$). If we further impose the symmetry $\eta{}\rightarrow{}-\eta$, $\chi{}\rightarrow{}-\chi{}$, which eliminates the cubic terms from the potential (ie. $c=d=0$), then the underlying equations describing components in which both $a_{i}(y\rightarrow{}-\infty)=b_{i}(y\rightarrow{}+\infty)=-v$ and $a_{j}(y\rightarrow{}-\infty)=b_{j}(y\rightarrow{}+\infty)=+v$ are the same. Likewise, the equations describing the components for which $a_{i}(y\rightarrow{}-\infty)=-b_{i}(y\rightarrow{}+\infty)=-v$ and $a_{j}(y\rightarrow{}-\infty)=-b_{j}(y\rightarrow{}+\infty)=+v$ are also the same. That means, when we set $c=d=0$, we can think of the solution for $\eta$ and $\chi$ as being of the form
 \begin{equation}
 \label{eq:gensolutionetachi}
 \begin{aligned}
  \eta{(y)} &= \begin{pmatrix} +\eta_{-}(y)\mathbbm{1}_{(6-m)\times{}(6-m)} & 0 & 0 & 0 \\
                         0 & +\eta_{+}(y)\mathbbm{1}_{m\times{}m} & 0 & 0 \\
                         0 & 0 & -\eta_{+}(y)\mathbbm{1}_{m\times{}m} & 0 \\
                         0 & 0 & 0 & -\eta_{-}(y)\mathbbm{1}_{(6-m)\times{}(6-m)} \end{pmatrix}, \\
 \chi{(y)} &= \begin{pmatrix} +\chi_{-}(y)\mathbbm{1}_{(6-m)\times{}(6-m)} & 0 & 0 & 0 \\
                         0 & +\chi_{+}(y)\mathbbm{1}_{m\times{}m} & 0 & 0 \\
                         0 & 0 & -\chi_{+}(y)\mathbbm{1}_{m\times{}m} & 0 \\
                         0 & 0 & 0 & -\chi_{-}(y)\mathbbm{1}_{(6-m)\times{}(6-m)} \end{pmatrix}, 
 \end{aligned}                        
 \end{equation}
where $\eta_{\pm{}}$ and $\chi_{\pm{}}$ satisfy the boundary conditions 
\begin{equation}
\label{eq:etachiminusbcs}
 \begin{aligned}
  \eta_{-}(y\rightarrow{}-\infty) &= -v, \quad{} \eta_{-}(y\rightarrow{}+\infty) = 0, \\
  \chi_{-}(y\rightarrow{}-\infty) &= 0, \quad{} \chi_{-}(y\rightarrow{}+\infty) = +v, 
 \end{aligned}
\end{equation}
and
\begin{equation}
\label{eq:etachimplusbcs}
 \begin{aligned}
  \eta_{+}(y\rightarrow{}-\infty) &= -v, \quad{} \eta_{+}(y\rightarrow{}+\infty) = 0, \\
  \chi_{+}(y\rightarrow{}-\infty) &= 0, \quad{} \chi_{+}(y\rightarrow{}+\infty) = -v. 
 \end{aligned}
\end{equation}
This means that Eq.~\ref{eq:eulerlagrangeab} now simplifies to 
\begin{equation}
 \label{eq:eulerlagrange+-}
 \begin{aligned}
  &-\frac{d^{2}\eta_{\pm{}}}{dy^2}-2\mu^2\eta_{\pm{}}+4\lambda_{1}((12-2m)\eta^2_{-}+2m\eta^2_{+})\eta_{\pm{}}+4\lambda_{2}\eta^{3}_{\pm{}}+2\delta^{2}\chi_{\pm{}}+2l_{1}((12-2m)\chi^{2}_{-}+2m\chi^{2}_{+})\eta_{\pm{}}+2(l_{2}+l_{4})\chi^{2}_{\pm{}}\eta_{\pm{}} \\
  &+2l_{3}((12-2m)\eta_{-}\chi_{-}+2m\eta_{+}\chi_{+})\chi_{\pm{}}+2l_{5}((12-2m)\eta_{-}\chi_{-}+2m\eta_{+}\chi_{+})\eta_{\pm{}}+l_{5}((12-2m)\eta^2_{-}+2m\eta^2_{+})\chi_{\pm{}} \\
  &+l_{5}((12-2m)\chi^{2}_{-}+2m\chi^{2}_{+})\chi_{\pm{}}+3l_{6}\eta^2_{\pm{}}\chi_{\pm{}}+l_{6}\chi^{3}_{\pm{}} = 0, \\
  &-\frac{d^{2}\chi_{\pm{}}}{dy^2}-2\mu^{2}\chi_{\pm{}}+4\lambda_{1}((12-2m)\chi^{2}_{-}+2m\chi^{2}_{+})\chi_{\pm{}}+4\lambda_{2}\chi^{3}_{\pm{}}+2\delta^{2}\eta_{\pm{}}+2l_{1}((12-2m)\eta^2_{-}+2m\eta^2_{+})\chi_{\pm{}}+2(l_{2}+l_{4})\eta^{2}_{\pm{}}\chi_{\pm{}} \\
  &+2l_{3}((12-2m)\eta_{-}\chi_{-}+2m\eta_{+}\chi_{+})\eta_{\pm{}}+2l_{5}((12-2m)\eta_{-}\chi_{-}+2m\eta_{+}\chi_{+})\chi_{\pm{}}+l_{5}((12-2m)\chi^{2}_{-}+2m\chi^{2}_{+})\eta_{\pm{}} \\
  &+l_{5}((12-2m)\eta^2_{-}+2m\eta^2_{+})\eta_{\pm{}}+3l_{6}\chi^2_{\pm{}}\eta_{\pm{}}+l_{6}\eta^{3}_{\pm{}} = 0.
 \end{aligned}
\end{equation}

 To analyze the stability of the solutions for the various values of $m$ under the simplifying assumptions and conditions that we have made, we need to solve Eq.~\ref{eq:eulerlagrange+-} numerically. We first do this initially with $\delta$, $l_{5}$ and $l_{6}$ set to zero. To calculate solutions numerically, we must work in non-dimensionalized coordinates, and in this particular case we do this by non-dimensionalizing the variables and parameters of the theory in terms of an arbitrary energy scale $k$. This means that we calculated the solutions numerically with respect to the non-dimensionalized coordinate $\tilde{y} = ky$, and the masses and coupling constants in terms of dimensionless numbers multiplied by the powers of $k$ corresponding to their mass dimensions. 
 
 We calculated solutions to Eq.~\ref{eq:eulerlagrange+-} using the relaxation technique on a mesh with 2001 grid points, with the domain of $\tilde{y}$ truncated to $(-10, 10)$, and then calculated the energy density of the solutions for various values of $m$. The relaxation technique we used was a higher order technique, in which for most points we used not only the point $\tilde{y}_{i}$ and its neighbors $\tilde{y}_{i\pm{}1}$, but also the next nearest neighbors $\tilde{y}_{i\pm{}2}$ to approximate the second order derivatives of the functions generating the domain walls at $\tilde{y}_{i}$. This approximation of the second order derivative is accurate to $O(\epsilon^{4})$, where $\epsilon$ is the mesh spacing, and this means that when we apply this to the relaxation technique, the functions can be evaluated to an accuracy of $O(\epsilon^{6})$. For the $i=1$ and $i=1999$ points, the points which are neighbors to the points on the boundaries of the domain, we used a different combination of points, ranging from $\tilde{y}_{i-1}$ to $\tilde{y}_{i+4}$ for $i=1$, and from $\tilde{y}_{i-4}$ to $\tilde{y}_{i+1}$ for $i=1999$, to generate the same accuracy.
 
 We first did this for the parameter choice
 
\begin{equation}
\begin{aligned}
\label{eq:nonCoschoice}
\mu^{2} &=  2.0k^2, \\
\lambda_{1} &= \frac{1.0}{k}, \\
\lambda_{2} &= \frac{1.0}{k}, \\
l_{1} &= \frac{8.0}{k}, \\
l_{2} &= \frac{7.0}{k}, \\
l_{3} &= -\frac{3.0}{k}, \\
l_{4} &= -\frac{2.0}{k}, \\
l_{5} &= 0.0, \\
l_{6} &= 0.0.
\end{aligned}
\end{equation}
With this parameter choice, we found that the energy densities, which we denote $\epsilon{(m)}$, for each of the choices of $m$ from zero to six were respectively
\begin{equation}
\label{eq:nonCoschoiceenergydensities}
\begin{aligned}
\epsilon{(m = 0)} &= 0.806759k, \\ 
\epsilon{(m = 1)} &= 0.855328k, \\ 
\epsilon{(m = 2)} &= 0.907039k, \\ 
\epsilon{(m = 3)} &= 0.947975k, \\ 
\epsilon{(m = 4)} &= 0.907039k, \\ 
\epsilon{(m = 5)} &= 0.855328k, \\ 
\epsilon{(m = 6)} &= 0.806759k. \\ 
\end{aligned}
\end{equation}
Hence, for the above choice, the non-Clash-of-Symmetries domain wall solutions, the $m=0$ and $m=6$ solutions, are the most stable and are degenerate, while the $SU(3)\times{}SU(3)\times{}SU(3)\times{}SU(3)\times{}U(1)$ generating wall corresponding to the choice $m=3$ is the least stable. This is not unexpected given that for the choice in Eq.~\ref{eq:nonCoschoice}, the coupling constant $l_{3}$ which corresponds to the $[Tr(\eta{}\chi{})]^2$ interaction, is chosen to be negative. The term $[Tr(\eta{}\chi{})]^2$ is maximized when $m=0$ or $m=6$, since in this case the term is roughly proportional to $[Tr(AB)]^2 = [\pm{}Tr(A^2)]^2 = 1/4$ near $\tilde{y} = 0$, yielding a negative contribution to the energy density for negative $l_3$, while for the $m=3$ solution, $Tr(AB) = 0$, and hence the $[Tr(\eta{}\chi{})]^2$ interaction does not contribute to its energy density.

 Conversely, if we choose $l_3$ to be positive, we expect that the $m=3$ solution will be the most stable, and the $m=0$ and $m=6$ solutions will be the most unstable. For the parameter choice 
\begin{equation}
\label{eq:maximalCoschoice}
\begin{aligned}
\mu^{2} &= 2.0k^2, \\
\lambda_{1} &= \frac{1.0}{k}, \\
\lambda_{2} &= \frac{1.0}{k}, \\
l_{1} &= \frac{8.0}{k}, \\
l_{2} &= \frac{7.0}{k}, \\
l_{3} &= \frac{6.0}{k}, \\
l_{4} &= -\frac{2.0}{k}, \\
l_{5} &= 0.0, \\
l_{6} &= 0.0.
\end{aligned}
\end{equation}
we find that the energy densities are
\begin{equation}
\begin{aligned}
\label{eq:maximalCoschoiceenergydensities}
\epsilon{(m = 0)} = 1.077678k, \\
\epsilon{(m = 1)} = 0.985394k, \\ 
\epsilon{(m = 2)} = 0.956093k, \\
\epsilon{(m = 3)} = 0.947975k, \\
\epsilon{(m = 4)} = 0.956093k, \\
\epsilon{(m = 5)} = 0.985394k, \\
\epsilon{(m = 6)} = 1.077678k. 
\end{aligned}
\end{equation}
Indeed, the $SU(3)\times{}SU(3)\times{}SU(3)\times{}SU(3)\times{}U(1)$ solution is the most stable for this choice, with the solutions with $m$ decreasing or increasing from $m=3$ getting progressively more unstable, leaving the $m=0$ and $m=6$ non-CoS domain walls with the highest energy density and the least amount of stability.

 With the above two choices, the outcome has been that either the non-CoS solutions ($m=0$ and $m=6$) or the $m=3$ solution are the most stable. Howevever, the solutions we would like to make the most stable are the ones generating a localized $SU(5)_{V}\times{}SU(5)_{D}\times{}U(1)_{X}$ subgroup, namely one of the $m=1$ or $m=5$ solutions. In the previous parameter choices, we turned off the interactions which were proportional to odd powers of either $\eta$ or $\chi$ (or both) as well as the mixed $Tr(\eta{}\chi{})$ mass term. In this area of parameter space, there is an enhanced symmetry, with $\eta{}\rightarrow{}\chi{}$, $\chi{}\rightarrow{}-\eta$ and $\eta{}\rightarrow{}-\eta{}$, $\chi{}\rightarrow{}-\chi{}$ being symmetries of the potential in Eq.~\ref{eq:twoadjointpotentialparts}. The second of these is what eliminates the cubic interactions and is what allows us to write the form of the solutions for $\eta$ and $\chi$ solely in terms of four functions: $\eta_{-}$, $\chi_{-}$, $\eta_{+}$ and $\chi_{+}$.
 If the second of these symmetries is broken by allowing cubic interactions, then eight functions are required: the functions which take a component along the diagonal of $\eta$ from $-v$ to zero and the corresponding component of $\chi$ from zero to $+v$ are not exactly the negative of the functions which take a component along the diagonal of $\eta$ from $+v$ to zero and the corresponding component of $\chi$ from zero to $-v$. Hence, for simplicity of analysis, we will keep the second of these symmetries and keep the cubic interactions set to zero. 
  The first of the symmetries mentioned in the above paragraph, $\eta{}\rightarrow{}\chi{}$, $\chi{}\rightarrow{}-\eta$, is the one which arises by setting the quartic interactions proportional to odd powers of both $\eta$ and $\chi$ as well as the mixed mass $Tr[\eta{}\chi{}]$ term to zero. It is the one that is responsible for the degeneracy in energy density between solutions with $m = n$ and $m = 6-n$. We will break this symmetry by turning on these terms, thus breaking the degeneracies between solutions with $m = n$ and $m = 6-n$ for $n = 0, 1, 2$. If $l_{3}$ is negative, the energy density of one of the non-CoS domain walls will be raised but the other lowered, and terms like $Tr(\eta{}\chi^3)$ are still maximized in magnitude for the non-CoS solutions, so in this case one of the non-CoS domain walls will be the most stable. Hence, the parameter region we are interested in is one where $l_{3}$ is positive.
  
  In the following analysis, we fix $\delta^2$ to be
  \begin{equation}
  \label{eq:deltafixing}
  \delta^2 = -\frac{12l_{5}+l_{6}}{48\lambda_{1}+4\lambda_{2}}\mu^{2}. 
  \end{equation}
  We make this fixing purely for computational convenience, since it ensures that the minima remain of the form $\eta{}\neq{}0$, $\chi = 0$ and $\eta{} = 0$, $\chi{}\neq{}0$. The main worry if we break this fixing is whether the global minima still generate the same symmetry breaking patterns, since they will have both $\eta$ and $\chi$ non-zero. There is reason to believe that this is the case, since if we perturb away from the fixing in Eq.~\ref{eq:deltafixing}, we can analyze what happens to the minima by doing a resultant perturbation from $\eta{}\neq{}0$, $\chi = 0$ (and vice versa). Consider a minimum of the form $\eta = vA$, $\chi = \epsilon{}C$, where $C$ is a generator, and $\epsilon$ is a small number resulting from perturbing away from the condition in Eq.~\ref{eq:deltafixing}. Then, to first order in $\epsilon$, only the mixed mass term $Tr[\eta{}\chi{}]$ and the $Tr[\eta^{3}\chi{}]$ and $Tr[\eta^{2}]Tr[\eta{}\chi{}]$ interactions contribute to the perturbation in energy of the minima. Given that $A^2$ is proportional to the identity, the contributions from all these terms to the perturbation in energy are proportional to $Tr(AC)$. Hence, the symmetry breaking pattern that will result will be the one which extremizes $Tr(AC)$ such that the perturbation in energy is minimal, which will correspond to the case that $Tr(AC)$ is maximally positive if $\epsilon$ is a negative, or to the case that $Tr(AC)$ is maximally negative if $\epsilon{}$ is positive. These cases happen respectively if $C$ is either totally aligned or totally anti-aligned with $A$: in other words, if $C = \pm{}A$. This means that, at least for a small perturbation, the symmetry breaking patterns of the minima remain the same.

\begin{figure}[H]
\begin{center}
\includegraphics[scale=1.0]{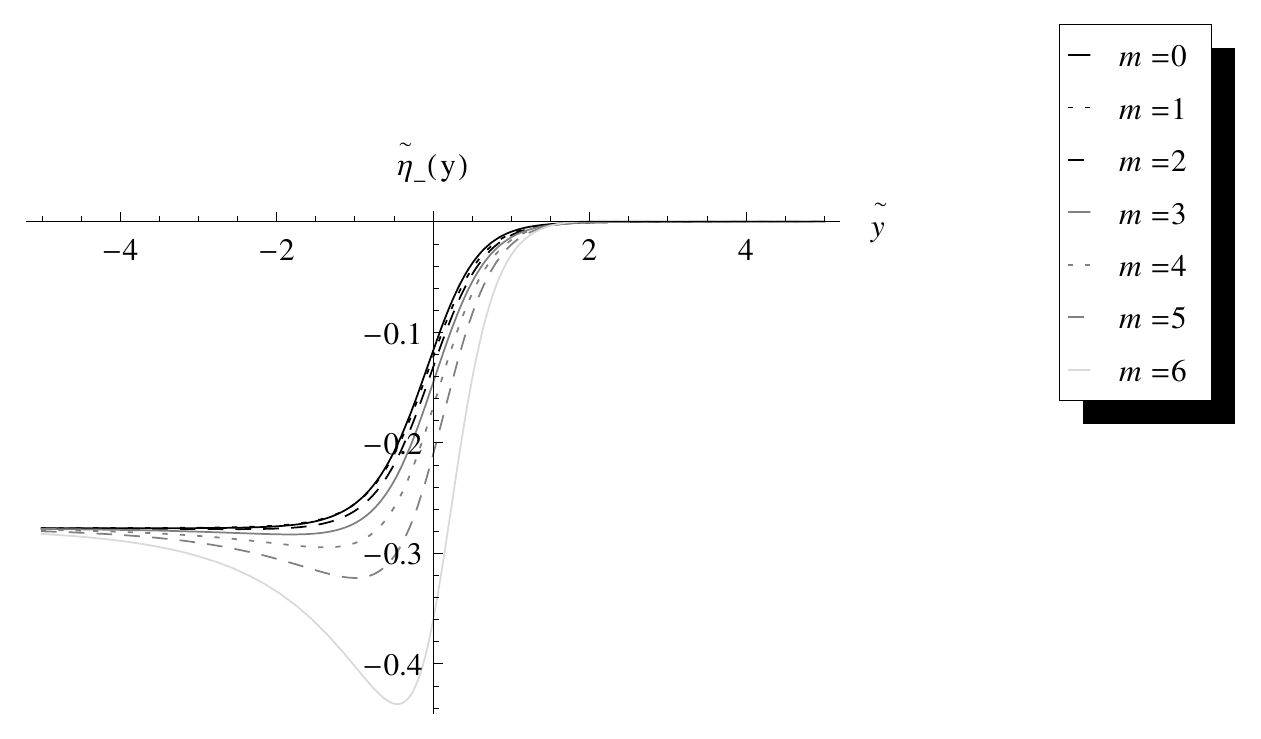}
\caption{A plot of the solutions for $\tilde{\eta}_{-}$ for $0\leq{}m\leq{}5$ for the parameter choice in Eq.~\ref{eq:su5xsu5Coschoice} subject to the constraint in Eq.~\ref{eq:deltafixing}.}
\label{fig:etaminus3plots} 
\end{center}
\end{figure}

\begin{figure}[H]
\begin{center}
\includegraphics[scale=1.0]{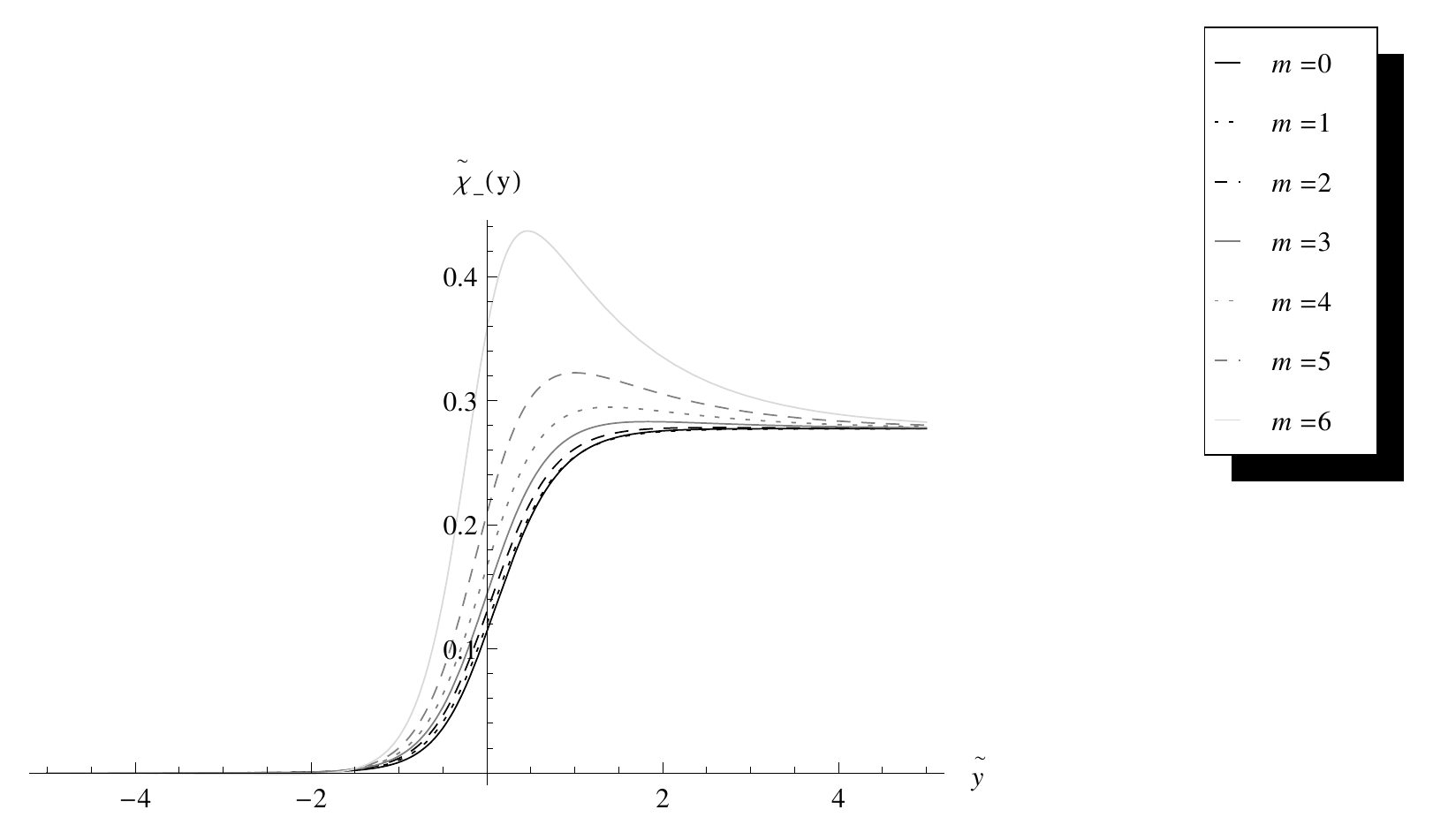}
\caption{A plot of the solutions for $\tilde{\chi}_{-}$ for $0\leq{}m\leq{}5$ for the parameter choice in Eq.~\ref{eq:su5xsu5Coschoice} subject to the constraint in Eq.~\ref{eq:deltafixing}.}
\label{fig:chiminus3plots} 
\end{center}
\end{figure}

\begin{figure}[H]
\begin{center}
\includegraphics[scale=1.0]{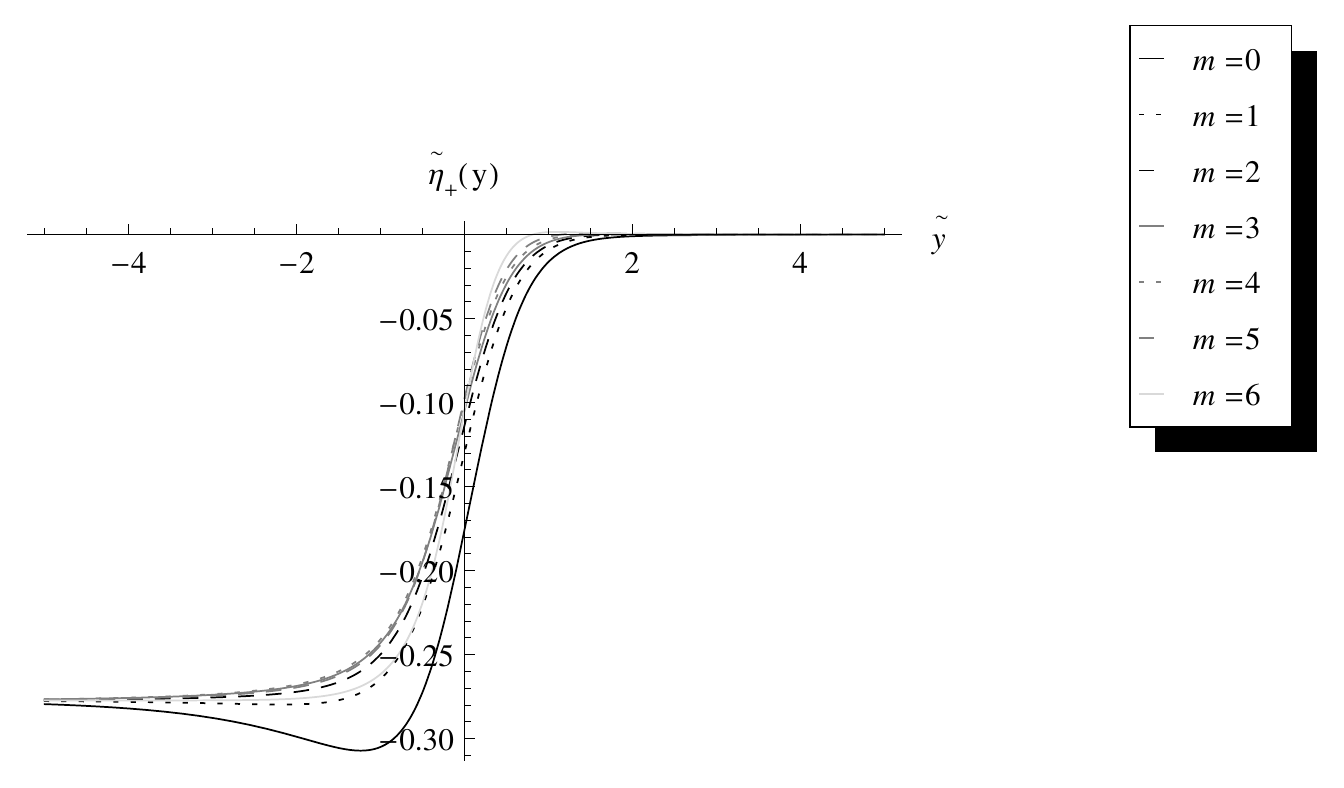}
\caption{A plot of the solutions for $\tilde{\eta}_{+}$ for $1\leq{}m\leq{}6$ for the parameter choice in Eq.~\ref{eq:su5xsu5Coschoice} subject to the constraint in Eq.~\ref{eq:deltafixing}.}
\label{fig:etaplus3plots} 
\end{center}
\end{figure}

\begin{figure}[H]
\begin{center}
\includegraphics[scale=1.0]{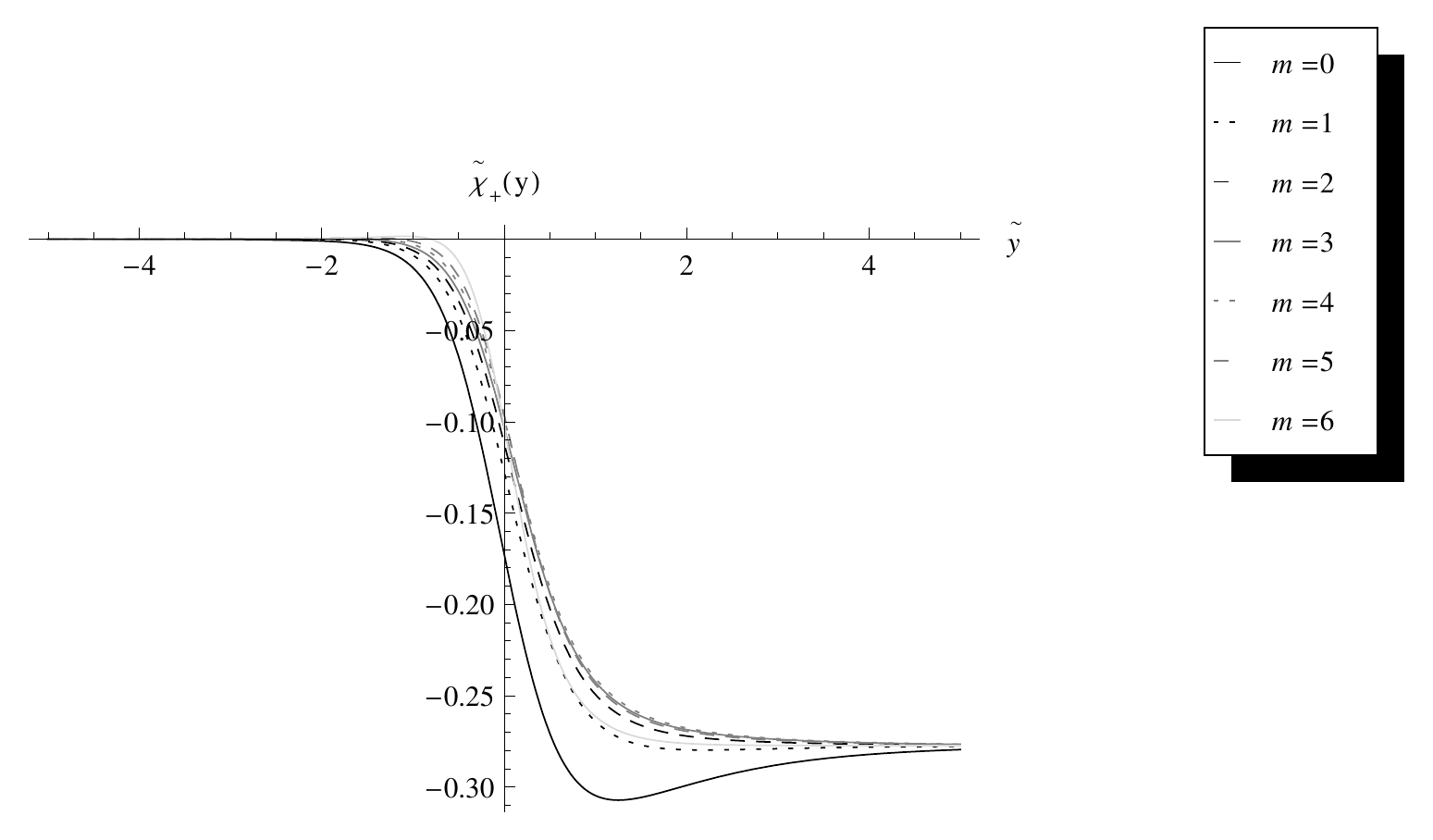}
\caption{A plot of the solutions for $\tilde{\chi}_{+}$ for $1\leq{}m\leq{}6$ for the parameter choice in Eq.~\ref{eq:su5xsu5Coschoice} subject to the constraint in Eq.~\ref{eq:deltafixing}.}
\label{fig:chiplus3plots} 
\end{center}
\end{figure}

 Given the assumptions we have made, we indeed find that it is possible to make one of the desired $m=1$ or $m=5$ solutions the most energetically stable. For the parameter choice 
\begin{equation}
\label{eq:su5xsu5Coschoice}
\begin{aligned}
\mu^{2} &= 2.0k^2, \\
\lambda_{1} &= \frac{1.0}{k}, \\
\lambda_{2} &= \frac{1.0}{k}, \\
l_{1} &= \frac{8.0}{k}, \\
l_{2} &= \frac{7.0}{k}, \\
l_{3} &= \frac{6.0}{k}, \\
l_{4} &= -\frac{2.0}{k}, \\
l_{5} &= -\frac{2.2}{k}, \\
l_{6} &= -\frac{2.0}{k},
\end{aligned}
\end{equation}
along with the condition used in Eq.~\ref{eq:deltafixing}, we find that the resultant energy densities are
\begin{equation}
\label{eq:su5xsu5Coschoiceenergydensities}
\begin{aligned}
\epsilon{(m = 0)} = 0.893757k, \\
\epsilon{(m = 1)} = 0.883037k, \\ 
\epsilon{(m = 2)} = 0.891332k, \\
\epsilon{(m = 3)} = 0.913322k, \\
\epsilon{(m = 4)} = 0.952365k, \\
\epsilon{(m = 5)} = 1.023219k, \\
\epsilon{(m = 6)} = 1.220998k. 
\end{aligned}
\end{equation}
The graphs of $\tilde{\eta}_{-} = \eta_{-}k^{-3/2}$, $\tilde{\chi}_{-} = \chi_{-}k^{-3/2}$, $\tilde{\eta}_{+} = \eta_{+}k^{-3/2}$ and $\tilde{\chi}_{+} = \chi_{+}k^{-3/2}$ for the various choices of $m$ for this parameter choice are shown in Figs.~\ref{fig:etaminus3plots}, \ref{fig:chiminus3plots}, \ref{fig:etaplus3plots}, and \ref{fig:chiplus3plots}.

 To conclude this section, we have successfully generated a parameter choice for which one of the domain walls leading to a localized $SU(5)_{V}\times{}SU(5)_{D}\times{}U(1)_{X}$ gauge group is the most energetically stable. The next step is to show that fermions and scalars can be localized in an acceptable way, which we shall show in the following two sections. One worry about the solutions resulting from the parameter choice in Eq.~\ref{eq:su5xsu5Coschoice} is that the energy density of the desired $m=1$ solution only differs from the $m=0$ and $m=2$ solutions by about one percent. It is thus plausible that the desired construction could be unstable when we account for quantum corrections. We leave the analysis of these corrections to later work. In the section after those dealing with fermion and scalar localization, we detail several alternative models.

\section{Fermion Localization and the Elimination of Fermionic Mediators from the Spectrum of the 3+1D Effective Field Theory}
\label{sec:fermionlocalization}

 In this section, we will show how to couple fermions to the $SU(5)_{V}\times{}SU(5)_{D}\times{}U(1)_{X}$-generating domain-wall solutions described in the previous section. From the last section, depending on the parameter region we choose, there are two options which generate the desired localized group: the $m=1$ solution, for which $\eta$ and $\chi$ can be described as being composed of five copies of $\pm{}(\eta_{-}, \chi_{-})$ and one of $\pm{}(\eta_{+}, \chi_{+})$, and the $m=5$ solution, which can be described by five copies of $\pm{}(\eta_{+}, \chi_{+})$ and one of $\pm{}(\eta_{-}, \chi_{-})$. 
 
  Normally, when coupling a 4+1D fermion field $\Psi$ to a scalar field $\phi$, which transforms under a discrete $\mathbb{Z}_{2}$ reflection symmetry $\phi{}\rightarrow{}-\phi{}$ and which generates a domain wall, the only acceptable Yukawa coupling one can write down is $h\overline{\Psi}\Psi{}\phi$. For this interaction, the reflection symmetry is extended so that $\overline{\Psi}\Psi{}\rightarrow{}-\overline{\Psi}\Psi{}$ (which can be achieved by taking $\Psi\rightarrow{}i\Gamma^{5}\Psi$), and, depending on the sign of the coupling constant $h$, this interaction leads to an effective $y$-dependent mass term which is either a kink or an anti-kink, leading respectively to either a localized, massless left-chiral or right-chiral zero mode. These type of fermionic chiral zero modes are generally the candidates for embedding the Standard Model fermions. In the case of the CoS domain walls from the two-field model with an interchange symmetry described in the previous section, the various $SU(5)_{V}\times{}SU(5)_{D}\times{}U(1)_{X}$-covariant components embedded in the $SU(12)$ multiplets have different localization properties, and, furthermore, we have two different types of Yukawa coupling. 
  
  The two ways of Yukawa coupling a fermion to $\eta$ and $\chi$ which respect the interchange symmetry $\eta{}\leftrightarrow{}\chi$ as well as $SU(12)$ can be described as follows. Let $\Psi_{R}$ be a fermion in some non-trivial representation $R$ of $SU(12)$. Then we can either couple $\Psi_{R}$ to $\eta$ and $\chi$ as 
  \begin{equation}
  \label{eq:yukawacoupling1} 
   h(\overline{\Psi_{R}}\eta{}\Psi_{R})_{1}+h(\overline{\Psi_{R}}\chi{}\Psi_{R})_{1},
  \end{equation}
  with $\Psi_{R}$ invariant under the discrete interchange symmetry, or, secondly, as
  \begin{equation}
  \label{eq:yukawacoupling2} 
  h'(\overline{\Psi_{R}}\eta{}\Psi_{R})_{1}-h'(\overline{\Psi_{R}}\chi{}\Psi_{R})_{1},
  \end{equation}
  with $\Psi\rightarrow{}i\Gamma^{5}\Psi$ under the discrete interchange symmetry. Here, in both equations, the $1$ subscript denotes taking the gauge singlet component of the $\overline{R}\times{}Adjoint\times{}R$ structure which arises in Yukawa couplings between $\Psi_{R}$, $\eta$ and $\chi$. 

  The $m=1$ solution is effectively a domain wall between $\eta = vA$, $\chi = 0$ at negative infinity and $\eta = 0$ $\chi = vB$ at positive infinity. It is therefore helpful to notice the charges of the various $SU(5)_{V}\times{}SU(5)_{D}\times{}U(1)_{X}$ components of the $12$ and $66$ representations under $A$ and $B$. Under $SU(5)_{V}\times{}SU(5)_{D}\times{}U(1)_{X}\times{}U(1)_{A}\times{}U(1)_{B}$, the fundamental $12$ representation breaks down as
  \begin{equation}
   \label{eq:fundamental12charges}
   12 = (5, 1, +1, -1, +1)\oplus{}(1, 5, +1, +1, -1)\oplus{}(1, 1, -5, -1, -1)\oplus{}(1, 1, -5, +1, +1),
  \end{equation}
 and the rank two anti-symmetric $66$ representation breaks down as
 \begin{equation}
  \label{eq:rank2antisymmetric66charges}
  \begin{gathered}
  66 = (10, 1, +2, -2, +2)\oplus{}(1, 10, +2, +2, -2)\oplus{}(5, 5, +2, 0, 0)\oplus{}(5, 1, -4, -2, 0) \\
     \oplus{}(5, 1, -4, 0, +2)\oplus{}(1, 5, -4, 0, -2)\oplus{}(1, 5, -4, +2, 0)\oplus{}(1, 1, -10, 0 , 0).
  \end{gathered}
 \end{equation}
 Let's now consider applying the coupling in Eq.~\ref{eq:yukawacoupling1} to the components of the fundamental. We shall label the $(5, 1, +1, -1, +1)$ component by $\Psi^{5V}$, the $(1, 5, +1, +1, -1)$ component by $\Psi^{5D}$, the $(1, 1, -5, -1, -1)$ component by $\Psi^{--}$ and the $(1, 1, -5, +1, +1)$ component by $\Psi^{++}$. Generally, if the coupling of a fermion to a domain wall is represented by the $\tilde{y}$-dependent mass term $W(\tilde{y})$, or, in other words, that the resultant $4+1D$ Dirac equation is given by 
\begin{equation}
\label{eq:4+1DDiracfromsuperpotential}
i\Gamma^{M}\partial_{M}\Psi - W(y)\Psi = 0,
\end{equation}
then if we expand $\Psi$ as a tower of left- and right-chiral modes of the form
\begin{equation}
 \label{eq:towerofmodes}
 \Psi{(x, \tilde{y})} = \sum_{m} f^{m}_{L}(\tilde{y})\psi^{m}_{L}(x)+ f^{R}_{L}(\tilde{y})\psi^{m}_{R}(x),
\end{equation}
where the $3+1D$ modes $\psi_{L,R}$ satisfy the $3+1D$ Dirac equation
\begin{equation}
 \label{eq:3+1Ddirac}
 i\gamma^{\mu}\psi_{L,R} = m\psi_{R,L},
\end{equation}
then the profiles $f^{m}_{L,R}$ satisfy Schr\'{o}dinger equations with the potentials
\begin{equation}
\label{eq:leftSEpotential} 
V_{L}(\tilde{y}) = W(\tilde{y})^2-W'(\tilde{y}),
\end{equation}
for the left-chiral modes, and 
\begin{equation}
\label{eq:leftSEpotential} 
V_{R}(\tilde{y}) = W(\tilde{y})^2+W'(\tilde{y}),
\end{equation}
for the right-chiral modes. Note that given the above potentials are of the form of those that arise in supersymmetric quantum mechanics, $W(\tilde{y})$ can be thought of as a superpotential.

\begin{figure}[H]
\begin{center}
\includegraphics[scale=0.9]{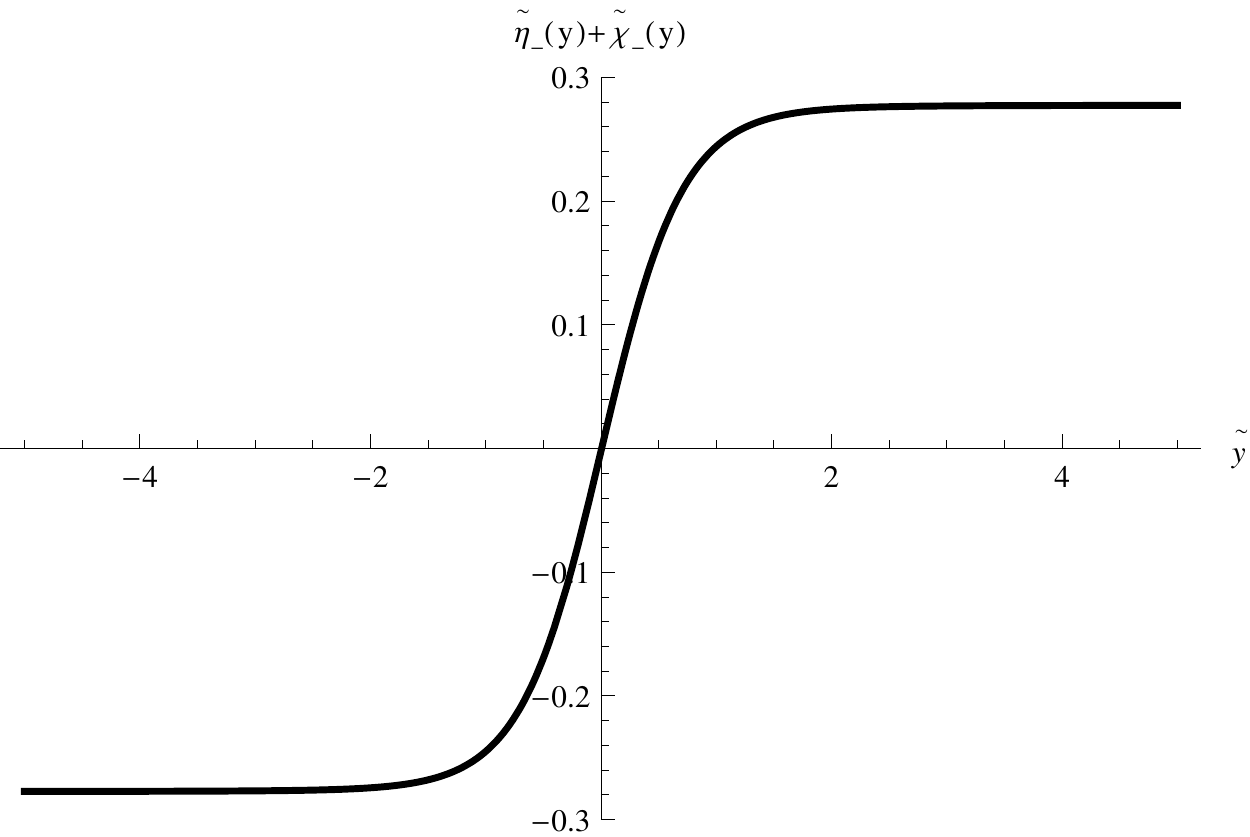}
\caption{A plot of the solutions for $\tilde{\eta}_{-}+\tilde{\chi}_{-}$ for $m=1$ for the parameter choice in Eq.~\ref{eq:su5xsu5Coschoice} subject to the constraint in Eq.~\ref{eq:deltafixing}.}
\label{fig:etaminuspluschiminus13} 
\end{center}
\end{figure}

 Applying the interaction of Eq.~\ref{eq:yukawacoupling1} to the components of the fundamental $12$ representation, we attain the superpotentials
 \begin{equation}
 \label{eq:visiblequintetsuperpot}
  W^{5V}(\tilde{y}) = h[\eta_{-}(\tilde{y})+\chi_{-}(\tilde{y})],
 \end{equation}
for the visible quintet,
 \begin{equation}
 \label{eq:darkquintetsuperpot}
  W^{5D}(\tilde{y}) = -h[\eta_{-}(\tilde{y})+\chi_{-}(\tilde{y})] = -W^{5V}(\tilde{y}),
 \end{equation}
for the dark quintet, 
\begin{equation}
 \label{eq:mmsingletsuperpot}
  W^{--}(\tilde{y}) = h[\eta_{+}(\tilde{y})+\chi_{+}(\tilde{y})],
 \end{equation}
for the $\Psi^{--}$ singlet component, and
\begin{equation}
 \label{eq:ppsingletsuperpot}
  W^{++}(\tilde{y}) = -h[\eta_{+}(\tilde{y})+\chi_{+}(\tilde{y})] = -W^{--}(\tilde{y}).
 \end{equation}
To know if we will end up with chiral zero modes for the visible and dark quintets, we need to know the form of $\eta_{-}+\chi_{-}$. This should be kink like as $\eta_{-}\rightarrow{}-v$, $\chi_{-}\rightarrow{}0$ as $\tilde{y}\rightarrow{}-\infty$ (or $-10$, in our truncation), and $\eta_{-}\rightarrow{}0$, $\chi_{-}\rightarrow{}+v$ as $\tilde{y}\rightarrow{}+\infty$, which means that $\eta_{-}+\chi_{-}\rightarrow{}\pm{}v$ as $\tilde{y}\rightarrow{}\pm{}\infty{}$. Indeed, as the plot in Fig.~\ref{fig:etaminuspluschiminus13} of $\eta_{-}+\chi_{-}$ for the $m=1$ solution for the parameter choice of Eq.~\ref{eq:su5xsu5Coschoice} shows, it is indeed kink-like. This means that standard result of Jackiw and Rebbi \cite{jackiwrebbi} for chiral zero modes holds for the superpotential $W^{5V}(\tilde{y})$, and we will attain a single left-chiral zero mode for the visible quintet if $h$ is positive, or a single right-chiral zero mode if $h$ is negative. Interestingly, due to the relative minus sign in Eq.~\ref{eq:darkquintetsuperpot}, which is due to the visible and dark quintets having the opposite charges under $A$ and $B$, the spectra for the left- and right-chiral modes for the dark quintet is flipped with respect to that for the visible quintet. This means that for $h>0$, we will attain a right-chiral zero mode for the dark quintet, or a left-chiral zero mode if $h<0$. Thus, given that that zero modes for the visible and dark quintets have opposite chirality, this suggests the possibility of reproducing a mirror matter theory on the domain-wall brane. 

\begin{figure}[H]
\begin{center}
\includegraphics[scale=0.8]{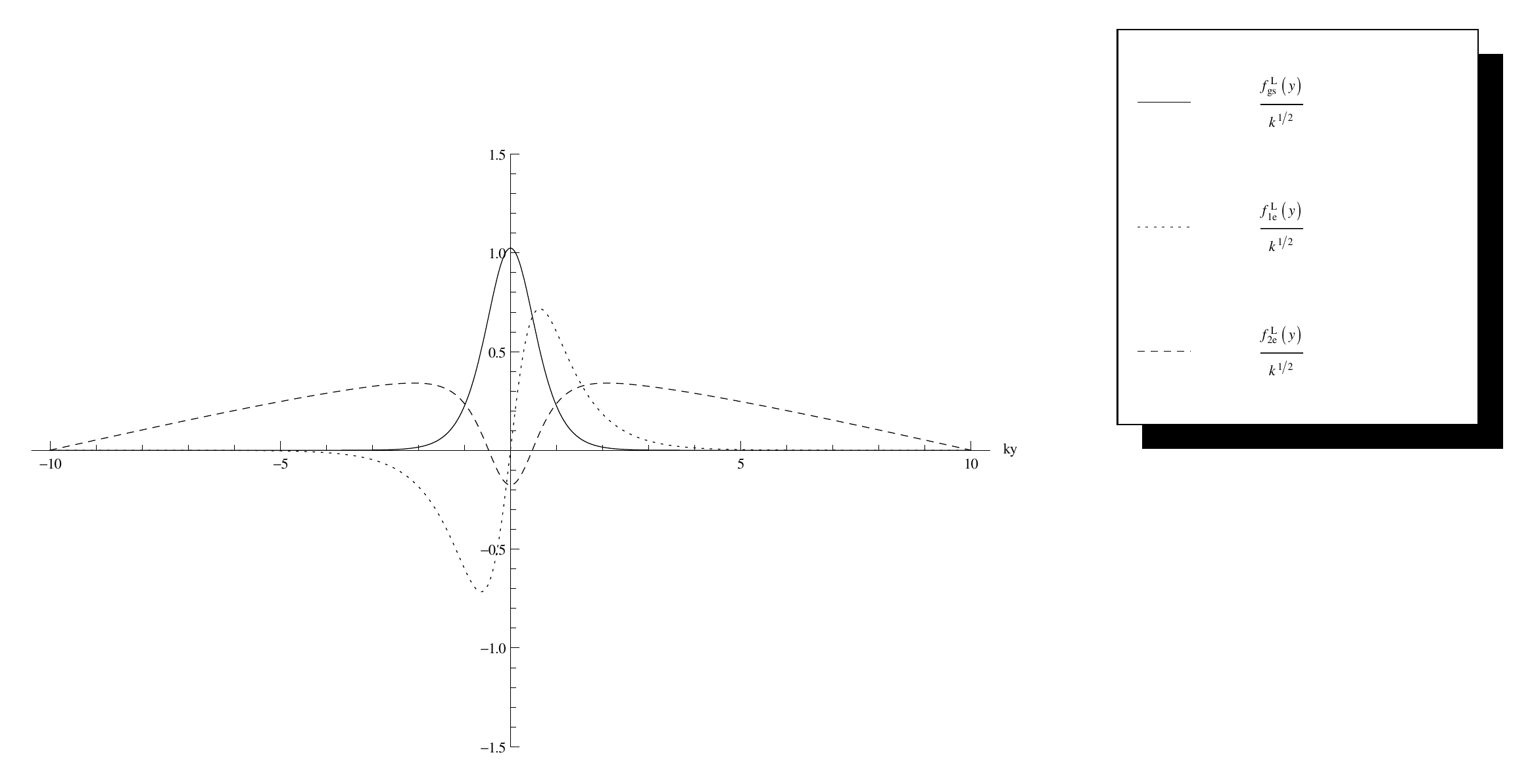}
\caption{A plot of the first three left-chiral (right-chiral) modes, including the zero mode, of the visible (dark) quintet for $\tilde{h}=10$.}
\label{fig:leftchiralh10} 
\end{center}
\end{figure}

\begin{figure}[H]
\begin{center}
\includegraphics[scale=0.9]{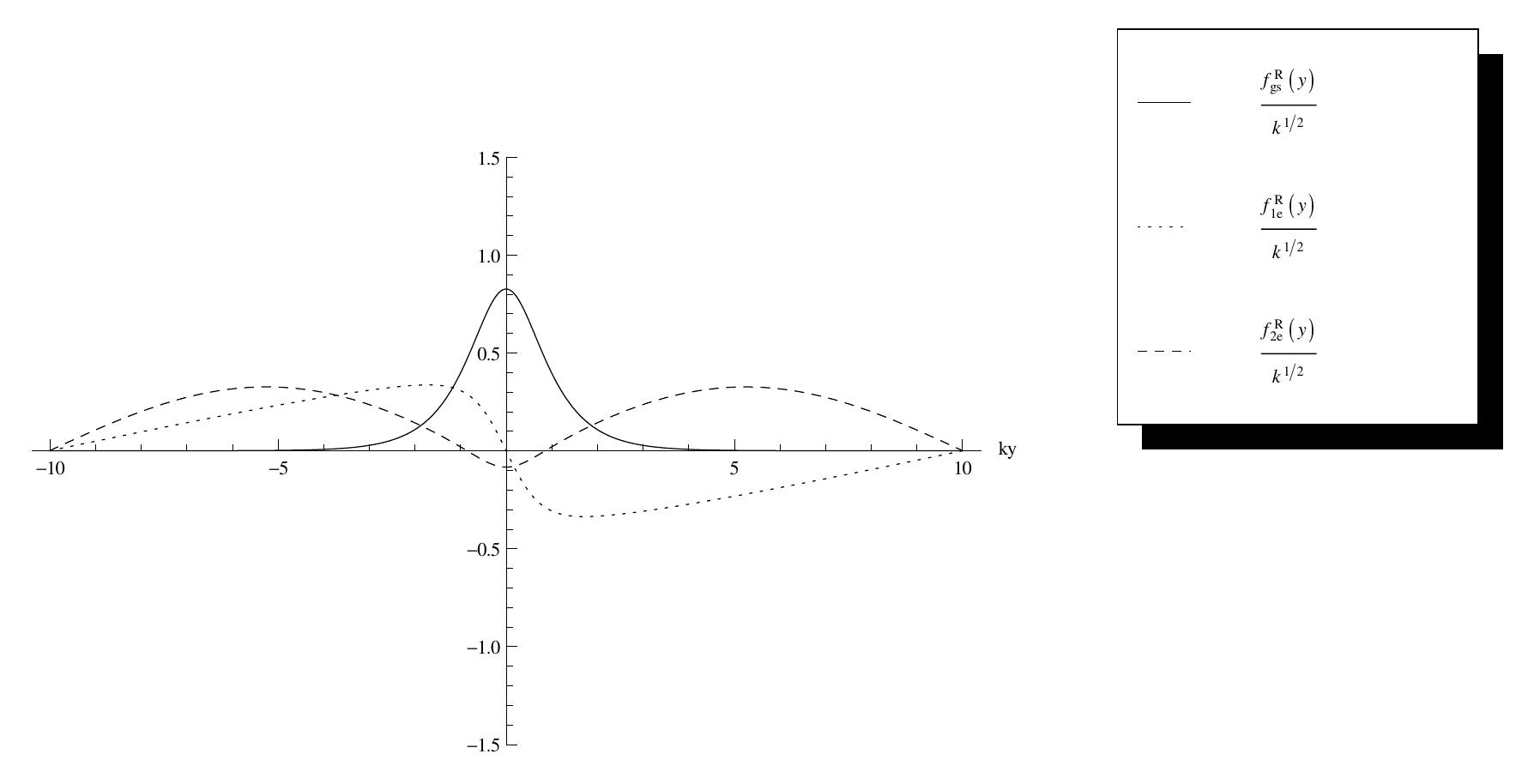}
\caption{A plot of the first three right-chiral (left-chiral) modes of the visible (dark) quintet for $\tilde{h}=10$.}
\label{fig:rightchiralh10} 
\end{center}
\end{figure}

\begin{figure}[H]
\begin{center}
\includegraphics[scale=0.9]{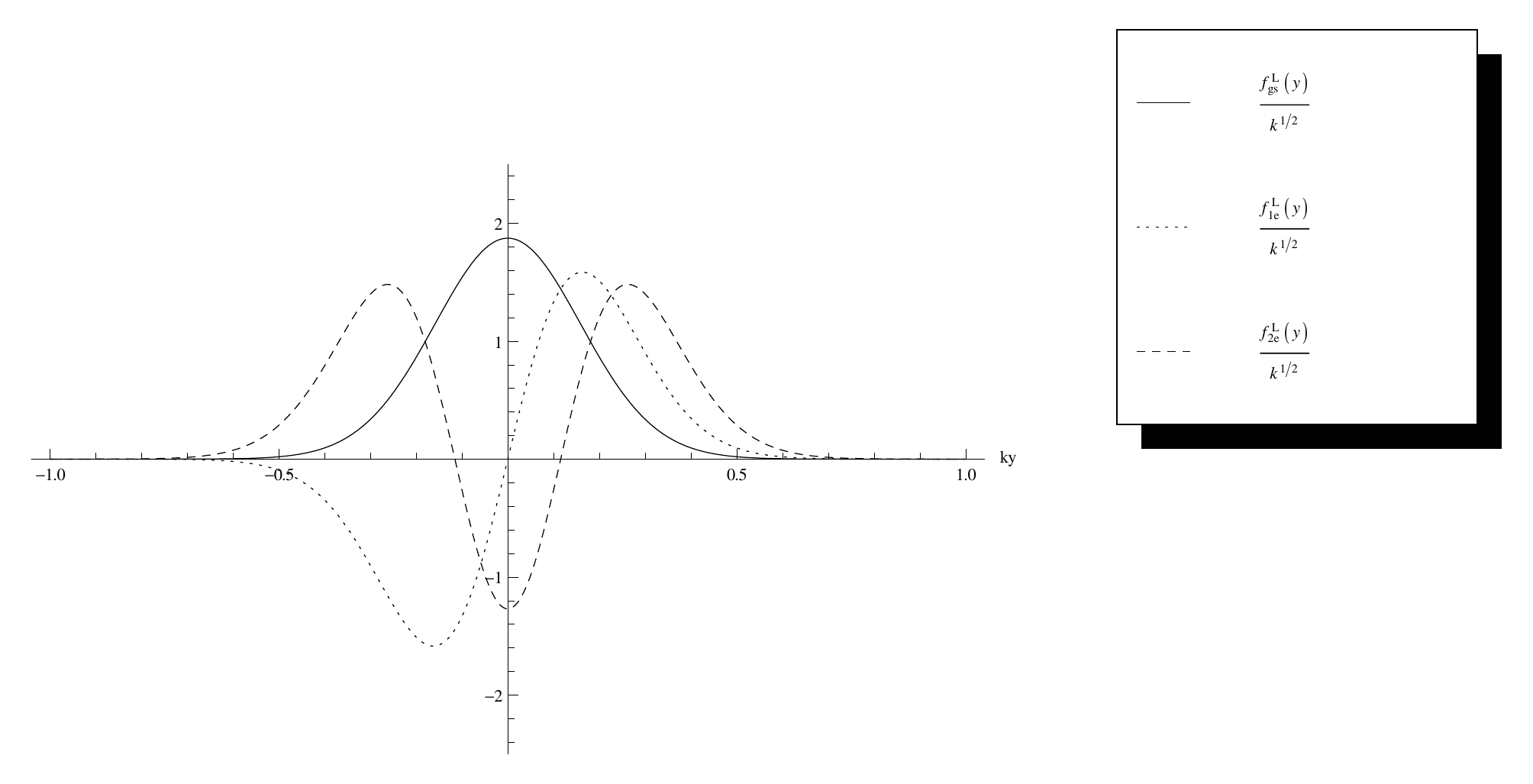}
\caption{A plot of the first three left-chiral (right-chiral) modes, including the zero mode, of the visible (dark) quintet for $\tilde{h}=100$.}
\label{fig:leftchiralh100} 
\end{center}
\end{figure}

\begin{figure}[H]
\begin{center}
\includegraphics[scale=0.9]{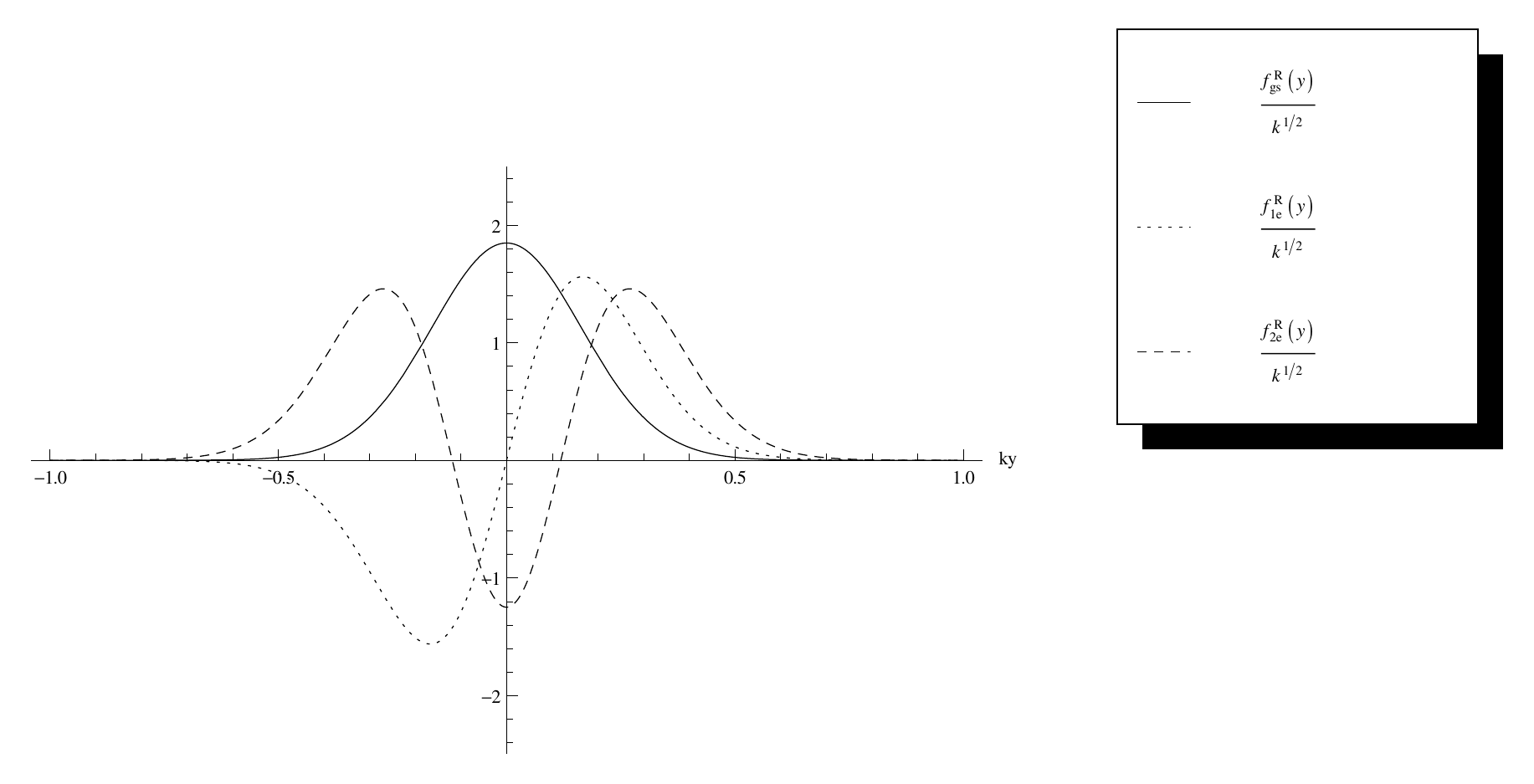}
\caption{A plot of the first three right-chiral (left-chiral) modes of the visible (dark) quintet for $\tilde{h}=100$.}
\label{fig:rightchiralh100} 
\end{center}
\end{figure}

\begin{figure}[H]
\begin{center}
\includegraphics[scale=0.9]{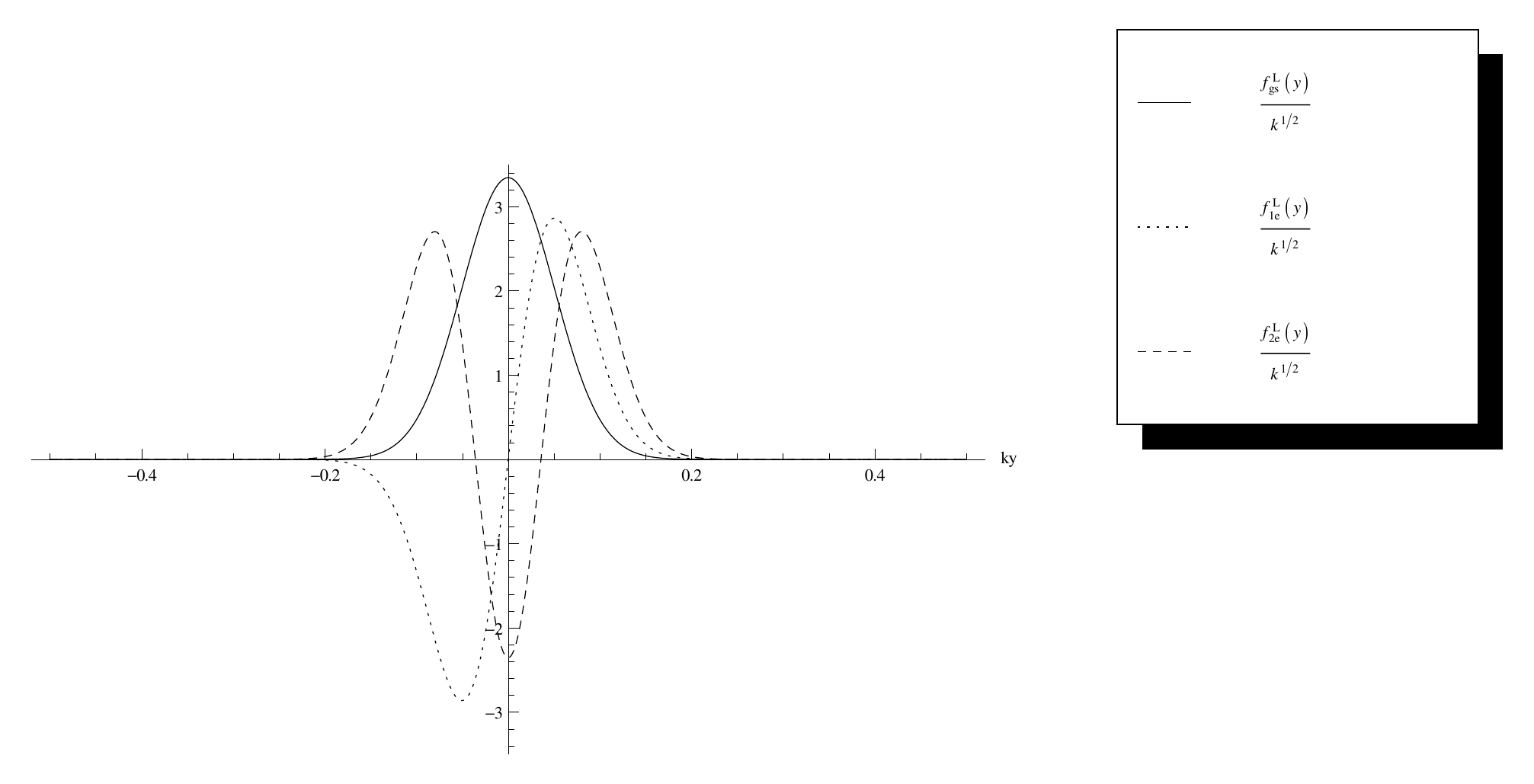}
\caption{A plot of the first three left-chiral (right-chiral) modes, including the zero mode, of the visible (dark) quintet for $\tilde{h}=1000$.}
\label{fig:leftchiralh1000} 
\end{center}
\end{figure}

\begin{figure}[H]
\begin{center}
\includegraphics[scale=0.9]{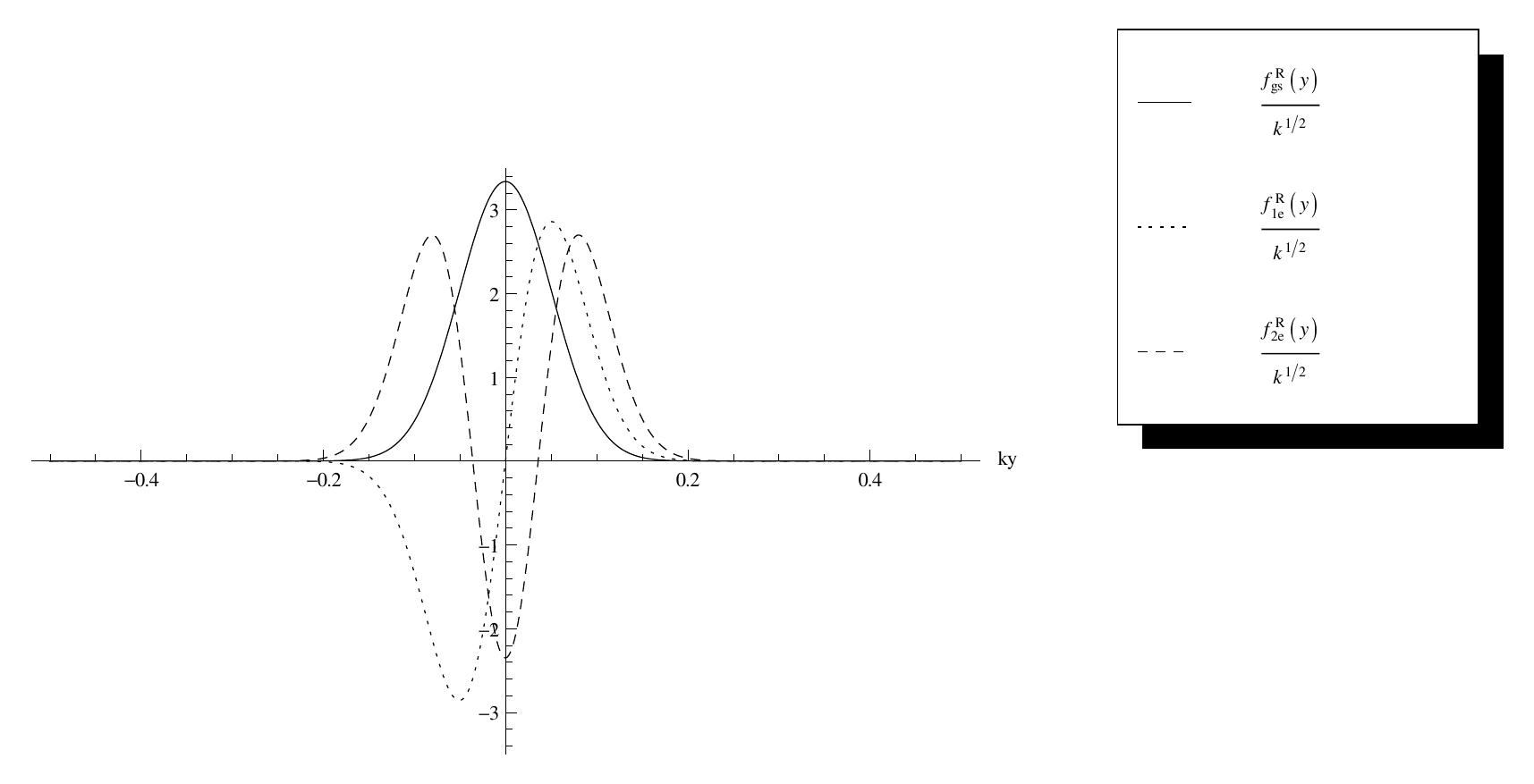}
\caption{A plot of the first three right-chiral (left-chiral) modes of the visible (dark) quintet for $\tilde{h}=1000$.}
\label{fig:rightchiralh1000} 
\end{center}
\end{figure}

 To calculate some of the profiles for the modes of the visible and dark quintets, we again use dimensionless variables. We firstly define the dimensionless Yukawa coupling by 
 \begin{equation}
 \tilde{h} = \frac{hv}{k},
\end{equation}
and the non-dimensionalized profiles by 
\begin{equation}
 \label{eq:nondimfermprofiles}
 \tilde{f}_{L,R}(\tilde{y}) = k^{-\frac{1}{2}}f_{L,R}(\tilde{y}),
\end{equation}
and we utilize the same dimensionless coordinate $\tilde{y}$ from the previous section. We solve the relevant differential equations on the same mesh that we used before, with the domain of $\tilde{y}$ truncated to $(-10, 10)$ and split into 2000 intervals, and thus we solve for the profile functions on 2001 mesh points. We solve for the profile functions in the usual way, by defining $f(\tilde{y}_{i}) = 0$ for $i=0$ and $i=2000$ (here $\tilde{y}_{0} = -10$, $\tilde{y}_{2000} = +10$), and then writing the Hamiltonian operator for the relevant Schr\"{o}dinger equations in terms of the $f(\tilde{y}_{i})$, with the second order derivative in $\tilde{y}$ of a profile $f$ at $\tilde{y}_{i}$ calculated in terms of $f$ computed at the adjacent points, turning the Schr\"{o}dinger equation into an eigenvalue/eigenvector problem for a symmetric matrix on a 2001-dimensional vector space, with the components of the eigenvectors in this space being the values of the eigenfunction $f$ at the various mesh points $\tilde{y}_{i}$. We calculate all the derivatives 
in the Hamiltonian to sixth order in the mesh spacing, which we will call here $\epsilon$. This means that we calculate the kinetic term as well as the derivative of the superpotential $W$ in terms of the relevant functions evaluated not only at $\tilde{y}_{i-1}$ and $\tilde{y}_{i+1}$, but also at $\tilde{y}_{i-2}$, $\tilde{y}_{i+2}$ and $\tilde{y}_{i-3}$ and $\tilde{y}_{i+3}$. Because the derivative of the superpotential will involve dividing known functions, $\eta_{\pm{}}$ and $\chi_{\pm{}}$, which are known to $O(\epsilon^{6})$ in the mesh spacing, this term, and thus the whole Hamiltonian operator, is known to $O(\epsilon^{5})$. All of this means that instead of having a Hamiltonian which is a symmetric tridiagonal matrix, we end up with a Hamiltonian which is symmetric, septa-diagonal matrix. This makes things a bit more complicated and slower in terms of computation but is nevertheless doable: we first convert the septa-diagonal matrix to a tridiagonal matrix via a series of Householder transformations, calculate the eigenvalues and eigenvectors of the tridiagonal matrix, and then transform back to the original basis to get the eigenvectors of the septa-diagonal matrix. We then produced plots for the ground state, the first and second excited states of both the left- and right-chiral towers for each of the components of the fundamental, for the choices $\tilde{h} = 10$, $\tilde{h} = 100$, and $\tilde{h} = 1000$, which are shown in Figs.~\ref{fig:leftchiralh10}, \ref{fig:rightchiralh10}, \ref{fig:leftchiralh100}, \ref{fig:rightchiralh100}, \ref{fig:leftchiralh1000} and \ref{fig:rightchiralh1000}.

For $\tilde{h} = 10$, the squared masses of the first few localized, left-chiral modes, which we label $m^{2}_{L,gs}$, $m^{2}_{L,1e}$ and $m^{2}_{L,2e}$, for the visible quintet are
\begin{equation}
\begin{aligned}
 \label{eq:masssqrleft12visibleh10}
 m^{2}_{L,gs} &= 0, \\
 m^{2}_{L,1e} &= 5.8010k^2, \\
 m^{2}_{L,2e} &= 7.7023k^2.
 \end{aligned}
\end{equation}
Similarly, those for the first few right-chiral modes 
\begin{equation}
\begin{aligned}
 \label{eq:masssqrright12visibleh10}
 m^{2}_{R,gs} &= 5.8010k^2, \\
 m^{2}_{R,1e} &= 7.7005k^2, \\
 m^{2}_{R,2e} &= 7.8002k^2.
 \end{aligned}
\end{equation}
The squared masses for the first few left- and right-chiral modes for the dark quintet are, as implied previously, the same as those just above for the visible quintet but with the chiralities reversed.

For $\tilde{h} = 100$, the squared masses of the first few localized chiral modes of the visible quintet are 
\begin{equation}
\begin{aligned}
 \label{eq:masssqrleft12visibleh100}
 m^{2}_{L,gs} &= 0, \\
 m^{2}_{L,1e} &= 76.4038k^2, \\
 m^{2}_{L,2e} &= 148.6622k^2.
 \end{aligned}
\end{equation}
and
\begin{equation}
\begin{aligned}
 \label{eq:masssqrright12visibleh100}
 m^{2}_{R,gs} &= 76.4038k^2, \\
 m^{2}_{R,1e} &= 148.6622k^2, \\
 m^{2}_{R,2e} &= 216.7872k^2.
 \end{aligned}
\end{equation}

For $\tilde{h} = 1000$, the squared masses of the first few localized chiral modes of the visible quintet are 
\begin{equation}
\begin{aligned}
 \label{eq:masssqrleft12visibleh1000}
 m^{2}_{L,gs} &= 0, \\
 m^{2}_{L,1e} &= 782.7132k^2, \\
 m^{2}_{L,2e} &= 1561.2653k^2.
 \end{aligned}
\end{equation}
and
\begin{equation}
\begin{aligned}
 \label{eq:eq:masssqrright12visibleh1000}
 m^{2}_{R,gs} &= 782.7153k^2, \\
 m^{2}_{R,1e} &= 1561.2736k^2, \\
 m^{2}_{R,2e} &= 2335.6715k^2.
 \end{aligned}
\end{equation}

\begin{figure}[H]
\begin{center}
\includegraphics[scale=0.9]{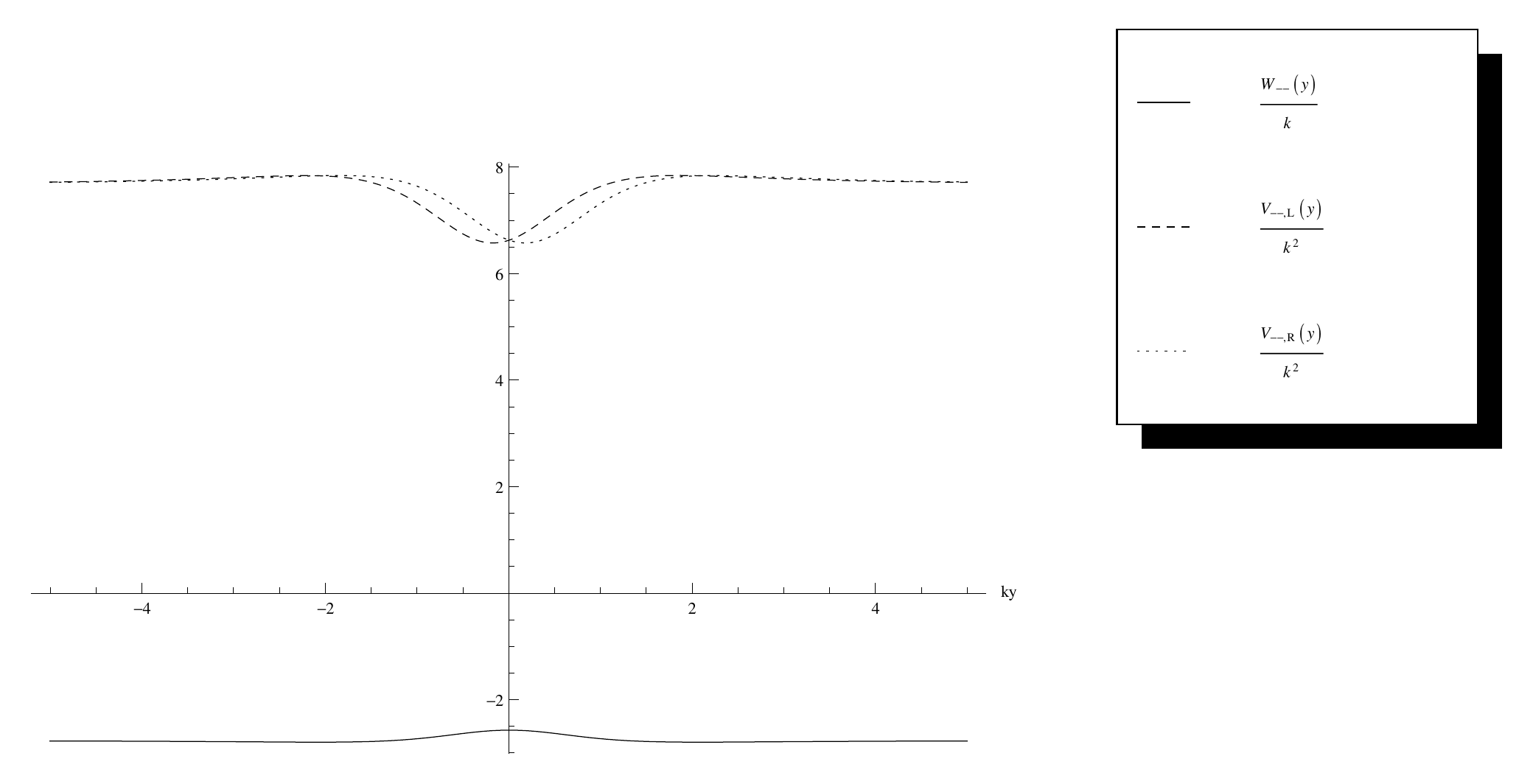}
\caption{Plots of $W^{--}(\tilde{y})/k$, $V^{--}_{L}(\tilde{y})/k^2$ ($V^{++}_{R}(\tilde{y})/k^2$), and $V^{--}_{L}(\tilde{y})/k^2$ ($V^{++}_{L}(\tilde{y})/k^2$) for $\tilde{h}=10$.}
\label{fig:fermionsingletpotentialplot} 
\end{center}
\end{figure}

 Now we repeat the analysis for the singlet components of the fundamental. These experience a superpotential proportional to $\eta_{+}+\chi_{+}$. Given that $\eta_{+}\rightarrow{}-v$, $\chi_{+}\rightarrow{}0$ as $\tilde{y}\rightarrow{}-\infty$, and $\eta_{+}\rightarrow{}0$, $\chi_{+}\rightarrow{}-v$ as $\tilde{y}\rightarrow{}+\infty$, it obvious that $\eta_{+}+\chi_{+}$ is not kink-like, and, given it approaches the same non-zero, constant value at both positive and negative infinity, there will not exist a normalizable profile for either a left- or right-chiral zero mode. Considering that, in most instances, $\eta_{+}$ can be approximated by something proportional to $M(1-\tanh{(k_{+}y)})/2$ and $\chi_{+}$ can be approximated by something proportional to $M(1+\tanh{(k_{+}y)})/2$, for some mass scale $M$ and inverse wall width $k_{+}$, we anticipate that $\eta_{+}+\chi_{+}$ will behave to a first order approximation as a simple 5D bulk mass $M$, and that only massive modes will exist. This is in fact that case, and it is easy to see this from plotting the potentials $V^{--}_{L, R}$ ($V^{++}_{R, L}$) that arise from the superpotential $W^{--}(\tilde{y})$ ($W^{++}(\tilde{y})$), which are shown along with the superpotential in Fig.~\ref{fig:fermionsingletpotentialplot} for $\tilde{h} = 10$. We can see clearly that despite the existence of small wells near the centre of the wall, the potentials are positive definite and never drop below about $6.5k^2$. Thus, in this case at least, only massive modes exist. We find that this property holds for $\tilde{h} = 100$ and $\tilde{h} = 1000$, although the wells are a bit deeper, supporting more localized massive modes.

For $\tilde{h} = 10$, the squared masses of the first few left-chiral (right-chiral) modes of the $\Psi^{--}$ ($\Psi^{++}$) singlet are 
\begin{equation}
\begin{aligned}
 \label{eq:masssqrleft12visible}
 m^{2}_{L,gs} &= 7.3984k^2, \\
 m^{2}_{L,1e} &= 7.8209k^2, \\
 m^{2}_{L,2e} &= 7.8542k^2.
 \end{aligned}
\end{equation}
Similarly, those for the first few right-chiral (left-chiral)
\begin{equation}
\begin{aligned}
 \label{eq:masssqrleft12visible}
 m^{2}_{R,gs} &= 7.3984k^2, \\
 m^{2}_{R,1e} &= 7.8209k^2, \\
 m^{2}_{R,2e} &= 7.8542k^2.
 \end{aligned}
\end{equation}
For $\tilde{h} = 100$, the squared masses of the first few localized chiral modes for the singlets are 
\begin{equation}
\begin{aligned}
 \label{eq:masssqrleft12visible}
 m^{2}_{L,gs} &= 674.2393k^2, \\
 m^{2}_{L,1e} &= 696.7534k^2, \\
 m^{2}_{L,2e} &= 717.2189k^2.
 \end{aligned}
\end{equation}
and
\begin{equation}
\begin{aligned}
 \label{eq:masssqrleft12visible}
 m^{2}_{R,gs} &= 674.2393k^2, \\
 m^{2}_{R,1e} &= 696.7534k^2, \\
 m^{2}_{R,2e} &= 717.2189k^2.
 \end{aligned}
\end{equation}

\begin{figure}[H]
\begin{center}
\includegraphics[scale=0.8]{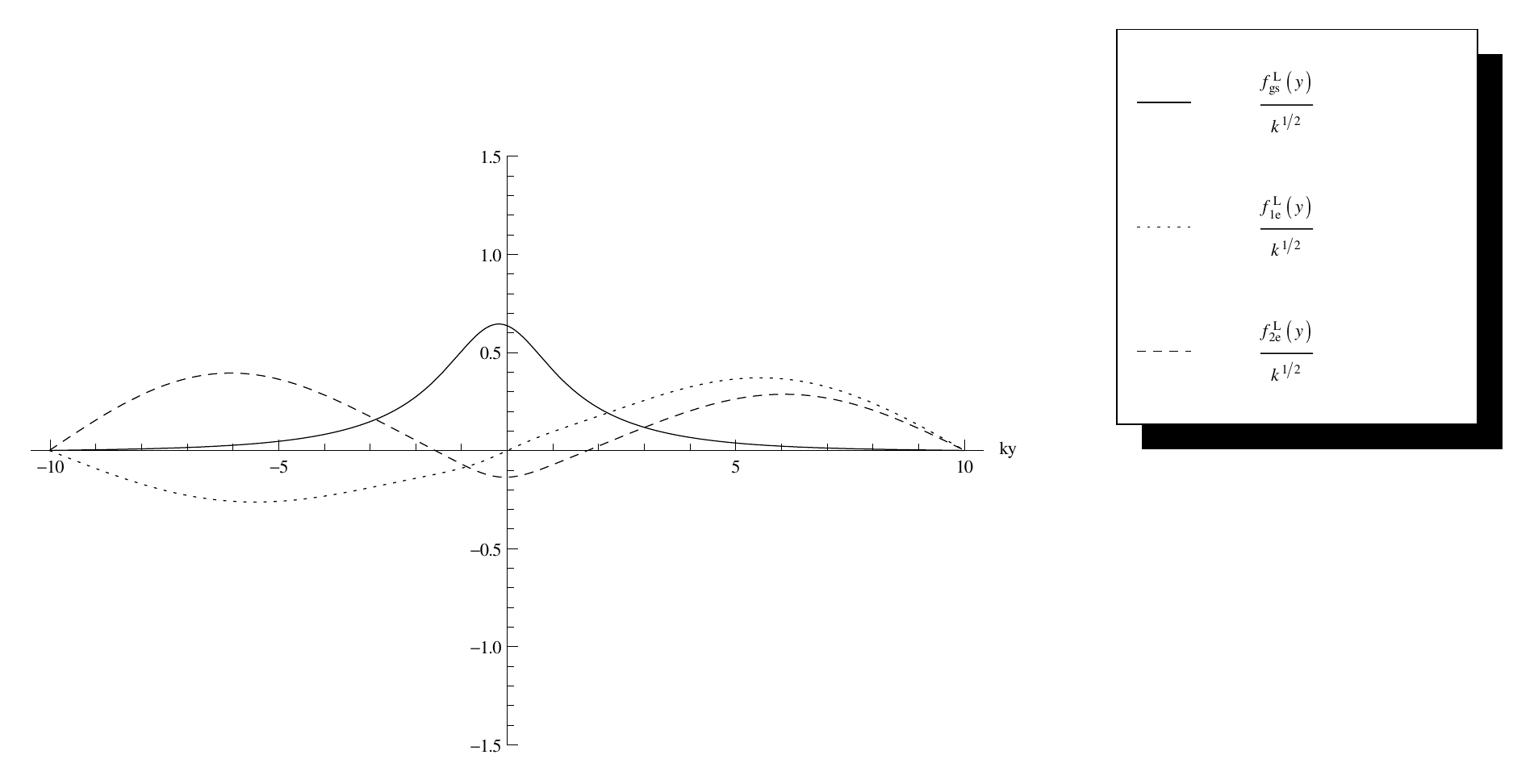}
\caption{A plot of the first three left-chiral (right-chiral) modes of the $\Psi^{--}$ ($\Psi^{++}$) singlet for $\tilde{h}=10$.}
\label{fig:leftchiralsingleth10} 
\end{center}
\end{figure}

\begin{figure}[H]
\begin{center}
\includegraphics[scale=0.9]{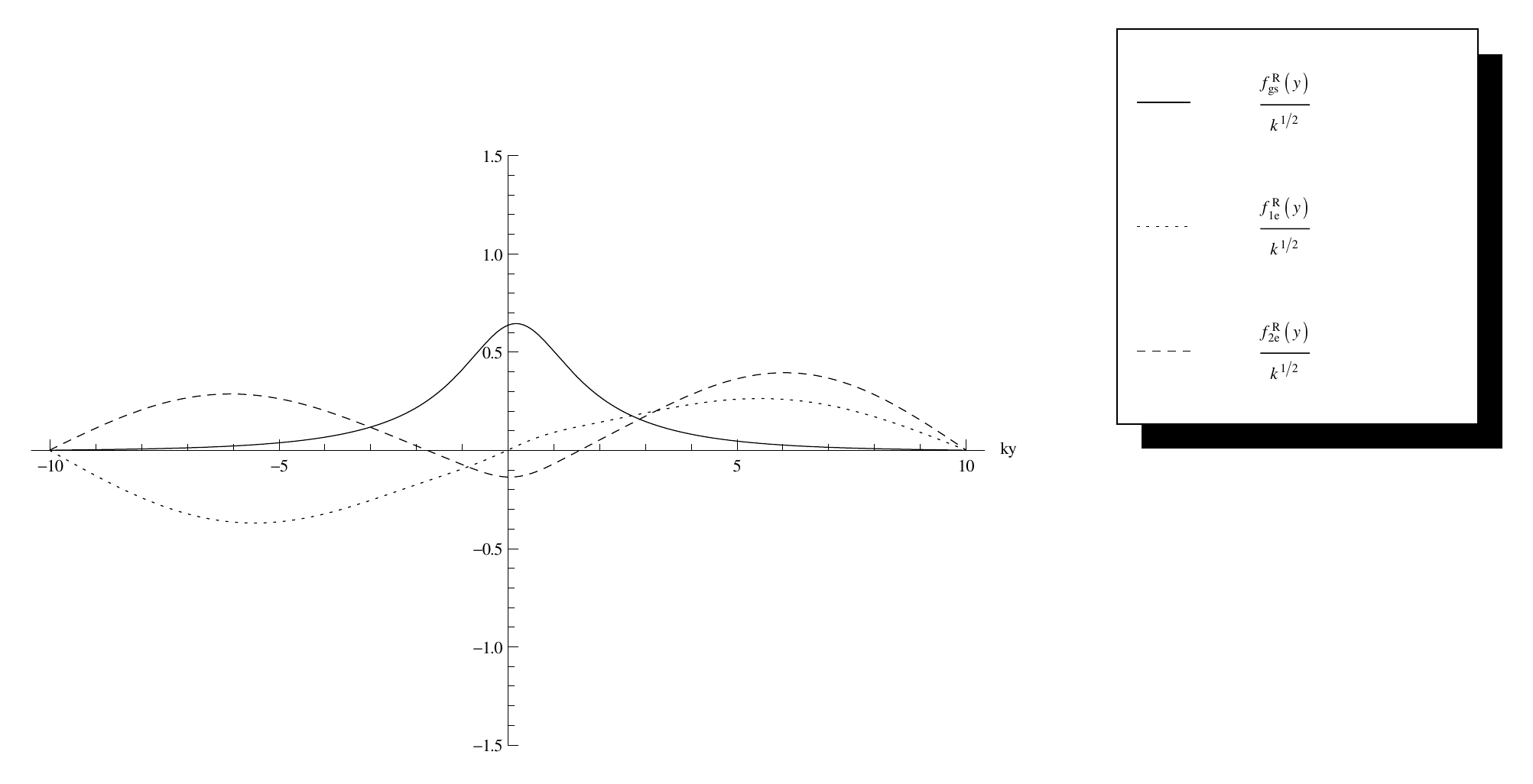}
\caption{A plot of the first three right-chiral (left-chiral) modes of the $\Psi^{--}$ ($\Psi^{++}$) singlet for $\tilde{h}=10$.}
\label{fig:rightchiralsingleth10} 
\end{center}
\end{figure}

For $\tilde{h} = 1000$, the squared masses of the first few localized chiral modes for the singlets are 
\begin{equation}
\begin{aligned}
 \label{eq:masssqrleft12visible}
 m^{2}_{L,gs} &= 66375.7932k^2, \\
 m^{2}_{L,1e} &= 66618.2533k^2, \\
 m^{2}_{L,2e} &= 66858.8267k^2,
 \end{aligned}
\end{equation}
and
\begin{equation}
\begin{aligned}
 \label{eq:masssqrleft12visible}
 m^{2}_{R,gs} &= 66375.7932k^2, \\
 m^{2}_{R,1e} &= 66618.2533k^2, \\
 m^{2}_{R,2e} &= 66858.8267k^2.
 \end{aligned}
\end{equation}
 We show the plots for the first several modes for the choice $\tilde{h}=10$ in Fig.~\ref{fig:leftchiralsingleth10} and Fig.~\ref{fig:rightchiralsingleth10}. The plots for these massive modes for the corresponding choices $\tilde{h}=100$ and $\tilde{h}=1000$ are similar, but more localized.

 We now deal with coupling a fermion, $\Psi_{66}$, in the $66$ representation to the domain wall in the form of the interaction in Eq.~\ref{eq:yukawacoupling1}. For this interaction, we choose for convenience the normalization
 \begin{equation}
  \label{eq:66coupling}
  Y_{66} = 2hTr(\overline{\Psi_{66}}\eta{}\Psi_{66})+2hTr(\overline{\Psi_{66}}\chi{}\Psi_{66}).
 \end{equation}
 To derive the relevant superpotentials for the $SU(5)_{V}\times{}SU(5)_{D}\times{}U(1)_{X}$ components of $\Psi_{66}$, we need to know how to write $\Psi_{66}$ in terms of these components. We may write $\Psi_{66}$ as the matrix, and the correct way to do this in order to attain the appropriate normalizations of the kinetic terms for each component is 
 \begin{equation}
  \label{eq:66matrix}
  \Psi_{66} = \begin{pmatrix}
             \Psi^{10V} && \frac{1}{\sqrt{2}}\Psi^{5V--} && \frac{1}{\sqrt{2}}\Psi^{5V++} && \frac{1}{\sqrt{2}}\Psi^{5V5D} \\
             -\frac{1}{\sqrt{2}}(\Psi^{5V--})^{T} && 0 && \frac{1}{\sqrt{2}}\Psi_{--++} && \frac{1}{\sqrt{2}}\Psi_{5D--} \\
             -\frac{1}{\sqrt{2}}(\Psi^{5V++})^{T} && -\frac{1}{\sqrt{2}}\Psi_{--++} && 0 && \frac{1}{\sqrt{2}}\Psi_{5D++} \\
             -\frac{1}{\sqrt{2}}(\Psi^{5V5D})^{T} && -\frac{1}{\sqrt{2}}(\Psi^{5D--})^{T} && -\frac{1}{\sqrt{2}}(\Psi_{5D++})^{T} && \Psi_{10D}
  \end{pmatrix},
 \end{equation}
where $\Psi^{10V}$ corresponds to the visible $(10, 1, +2, -2, +2)$ decuplet, $\Psi^{10D}$ is the dark decuplet corresponding to the $(1, 10, +2, +2, -2)$, $\Psi^{5V--}$ corresponds to the $(5, 1, -4, -2, 0)$ quintet, $\Psi^{5V++}$ to the $(5, 1, -4, 0, +2)$ quintet, $\Psi^{5D--}$ the $(1, 5, -4, 0, -2)$ quintet, $\Psi^{5D++}$ the $(1, 5, -4, +2, 0)$ quintet, $\Psi^{5V5D}$ is the bi-fundamental $(5, 5, +2, 0, 0)$ component, and $\Psi^{--++}$ is the $(1, 1, -10, 0 , 0)$ singlet component. When one substitutes the matrix representation of Eq.~\ref{eq:66matrix} into the interaction of Eq.~\ref{eq:66coupling}, one derives the superpotentials
\begin{equation}
 \label{eq:visibledecupletsuperpot}
  W^{10V}(\tilde{y}) = 2h[\eta_{-}(\tilde{y})+\chi_{-}(\tilde{y})],
 \end{equation}
for the visible decuplet,
 \begin{equation}
 \label{eq:darkdecupletsuperpot}
  W^{10D}(\tilde{y}) = -2h[\eta_{-}(\tilde{y})+\chi_{-}(\tilde{y})] = -W^{10V}(\tilde{y}),
 \end{equation}
for the dark decuplet, 
\begin{equation}
 \label{eq:5Vmmsuperpot}
  W^{5V--}(\tilde{y}) = h[\eta_{+}(\tilde{y})+\chi_{+}(\tilde{y})-\eta_{-}(\tilde{y})-\chi_{-}(\tilde{y})],
 \end{equation}
for the extra $\Psi^{5V--}$ quintet, 
\begin{equation}
 \label{eq:5Dmmsuperpot}
  W^{5D--}(\tilde{y}) = h[\eta_{+}(\tilde{y})+\chi_{+}(\tilde{y})+\eta_{-}(\tilde{y})+\chi_{-}(\tilde{y})],
 \end{equation}
 for the $\Psi^{5D--}$ quintet,
\begin{equation}
 \label{eq:5Vppsuperpot}
  W^{5V++}(\tilde{y}) = -h[\eta_{+}(\tilde{y})+\chi_{+}(\tilde{y})+\eta_{-}(\tilde{y})+\chi_{-}(\tilde{y})] = -W^{5D--}(\tilde{y})
 \end{equation}
for the $\Psi^{5V++}$ quintet,
\begin{equation}
 \label{eq:5Dppsuperpot}
  W^{5D++}(\tilde{y}) = -h[\eta_{+}(\tilde{y})+\chi_{+}(\tilde{y})-\eta_{-}(\tilde{y})-\chi_{-}(\tilde{y})] = -W^{5V--}(\tilde{y}).
 \end{equation}
for the $\Psi^{5D++}$ quintet, 
\begin{equation}
 \label{eq:biquintetsuperpot}
 W^{5V5D}(\tilde{y}) = 0,
\end{equation}
for the mixed $\Psi^{5V5D}$ bi-fundamental component and
\begin{equation}
 \label{eq:66singletsuperpot}
 W^{--++}(\tilde{y}) = 0,
\end{equation}
for the $\Psi^{--++}$ singlet component.

From the above superpotentials, we can see that the visible and dark decuplet couple to the combination, $\eta_{-}+\chi_{-}$, with equal and opposite strength, in the same way that the quintets from the fundamental did. This means that they will also attain localized chiral zero modes with opposite chiralities, making possible the localization of a left-chiral $(\overline{5}, 1)\oplus{}(10, 1)$ sector embedding the Standard Model fermions together with the localization of a corresponding right-chiral $(1, \overline{5})\oplus{}(1, 10)$ sector embedding a mirror dark fermion sector.

The second important thing to note from the above superpotentials is that the superpotential for the mixed bi-fundamental $\Psi^{5V5D}$ component vanishes. This implies that this component which couples to both the visible and dark $SU(5)$ gauge sectors is completely decoupled from the domain wall, and thus it remains a $4+1D$ fermionic field. Initially, this seems more worrying given that this field will initially be massless, and to have a $4+1D$ massless fermion interacting with a localized $3+1D$ Standard Model sector would be disastrous. However, because $\Psi^{5V5D}$ remains a $4+1D$ fermion, it remains a \emph{Dirac} fermion and will thus be able to form vector-like interactions with any additional scalar fields we later introduce into the theory. This means that when we introduce an additional adjoint scalar field which induces the usual breaking $SU(5)_{V}\rightarrow{}SU(3)_{c}\times{}SU(2)_{I}\times{}U(1)_{Y}$ in the interior of the domain wall, this very component will attain a mass of order the GUT scale in the interior of the domain wall and will thus be removed from the spectrum. Also, the singlet $\Psi^{--++}$ component also experiences a vanishing superpotential and remains delocalized. Given that the singlet has a charge of $-10$ under $U(1)_{X}$, it will attain a mass or, at the very least, become decoupled from the localized sectors when we break $U(1)_{X}$ at a sufficient scale. 

\begin{figure}[H]
\begin{center}
\includegraphics[width=18cm, height=10cm]{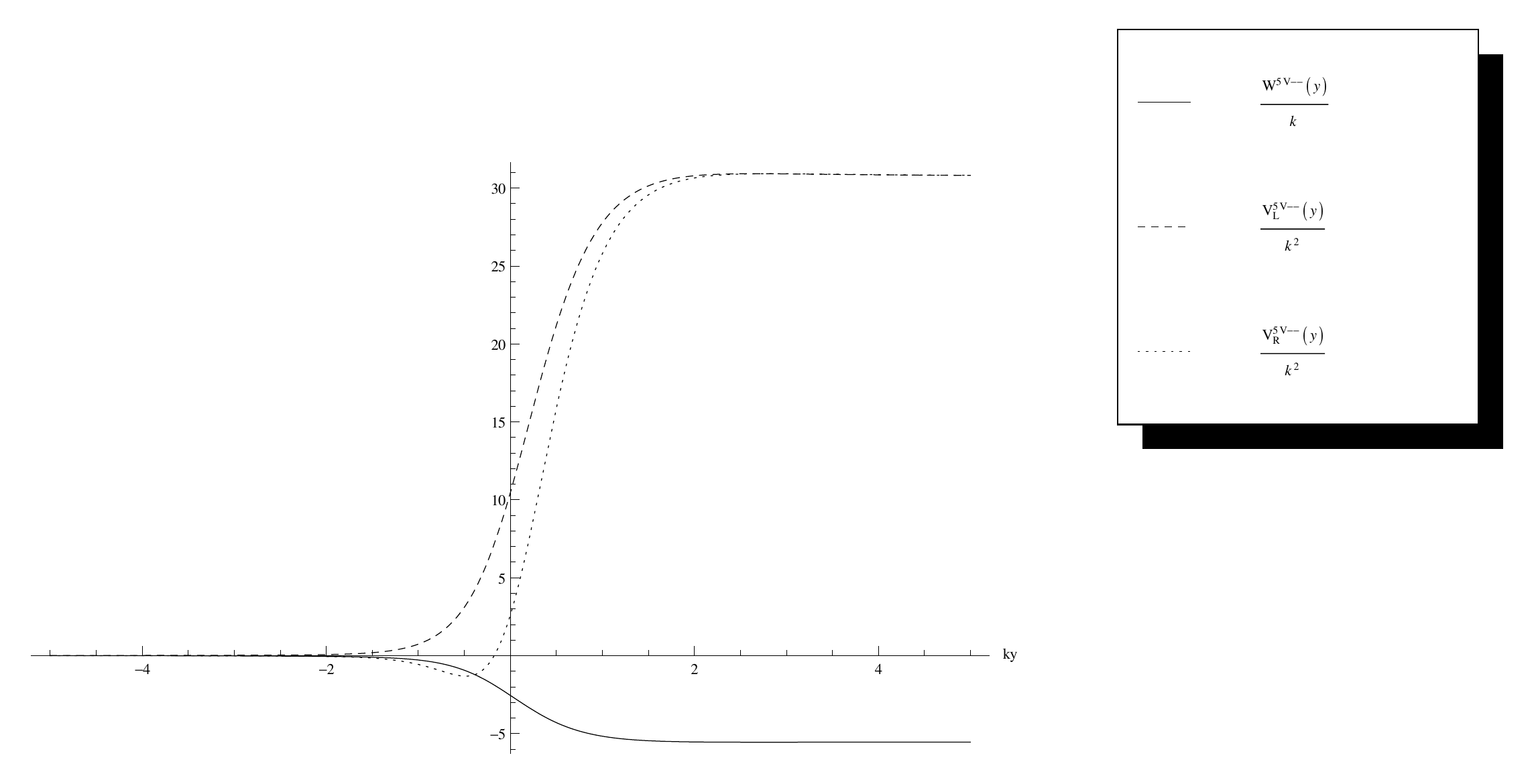}
\caption{A plot of the superpotential $W^{5V--}$ and the resulting left-chiral and right-chiral potentials $V^{5V--}_{L}$ and $V^{5V--}_{R}$ for $\tilde{h}=10$.}
\label{fig:potential5Vmm} 
\end{center}
\end{figure}

\begin{figure}[H]
\begin{center}
\includegraphics[width=18cm, height=10cm]{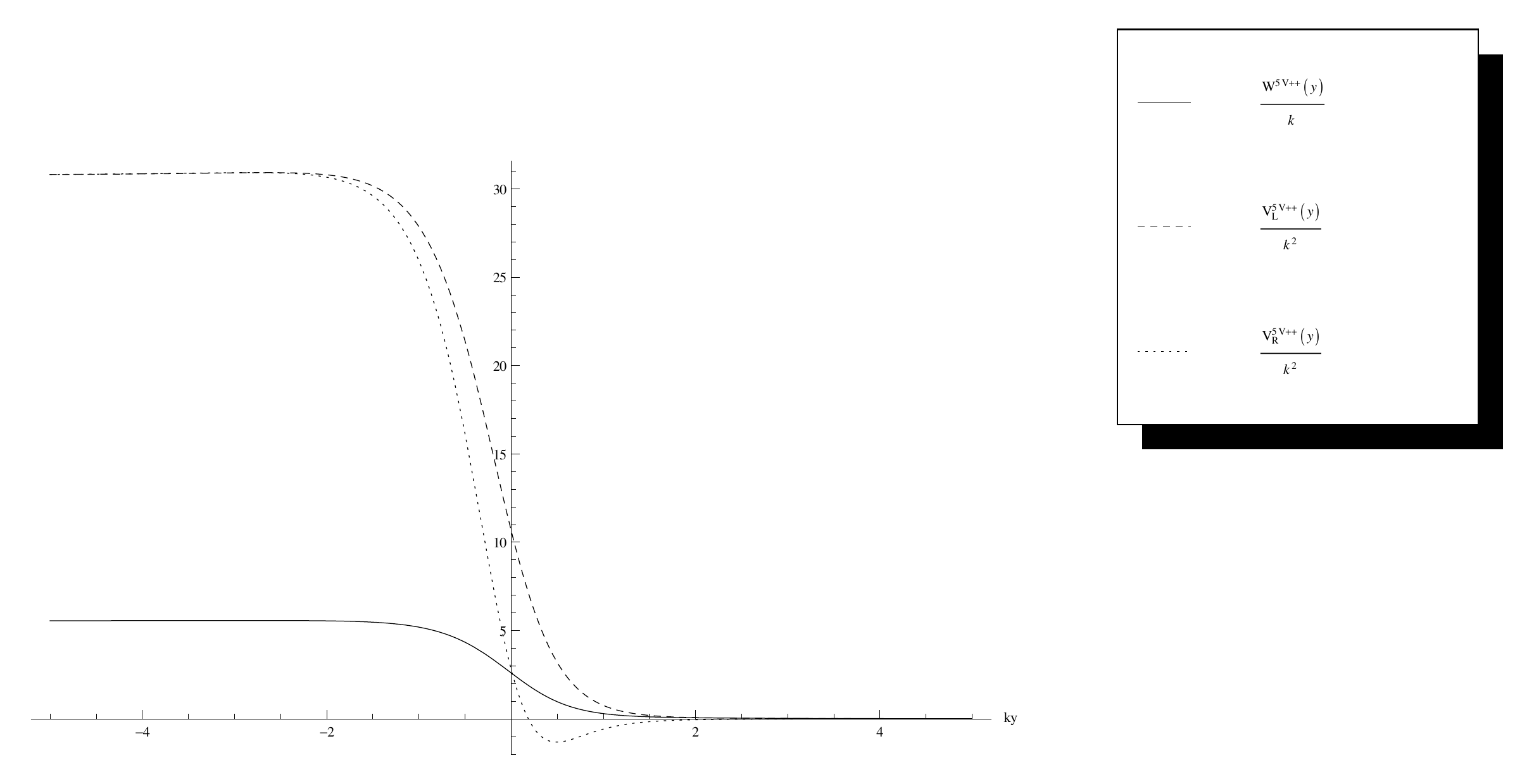}
\caption{A plot of the superpotential $W^{5V++}$ and the resulting left-chiral and right-chiral potentials $V^{5V++}_{L}$ and $V^{5V++}_{R}$ for $\tilde{h}=10$.}
\label{fig:potential5Vpp} 
\end{center}
\end{figure}

\begin{figure}[H]
\begin{center}
\includegraphics[scale=1.5]{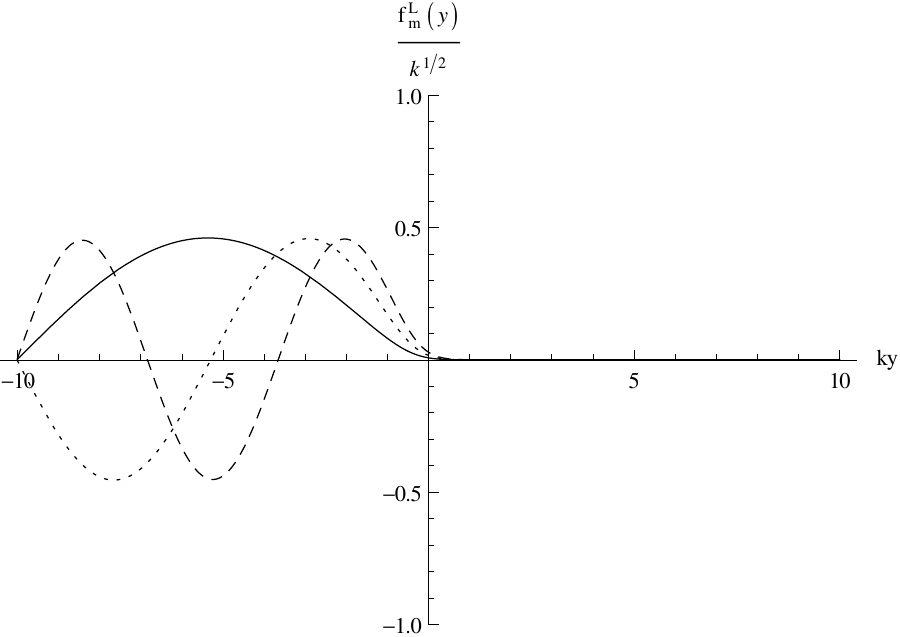}
\caption{A plot of a several left-chiral modes for $\Psi^{5V--}$ for $\tilde{h}=10$.}
\label{fig:L5VIh10modes} 
\end{center}
\end{figure}

\begin{figure}[H]
\begin{center}
\includegraphics[scale=1.5]{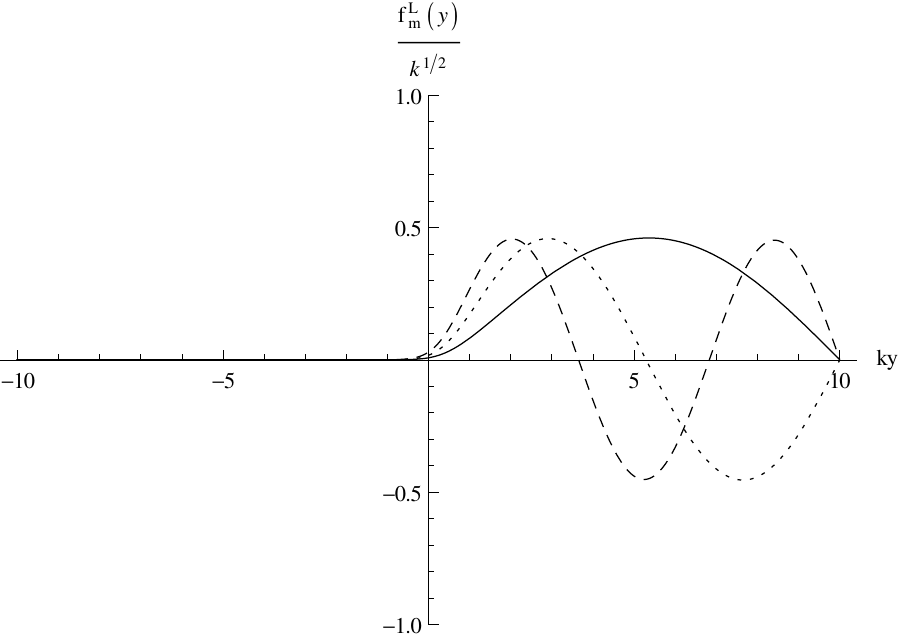}
\caption{A plot of a several left-chiral modes for $\Psi^{5V++}$ for $\tilde{h}=10$.}
\label{fig:L5VIprimeh10modes} 
\end{center}
\end{figure}

\begin{figure}[H]
\begin{center}
\includegraphics[scale=1.5]{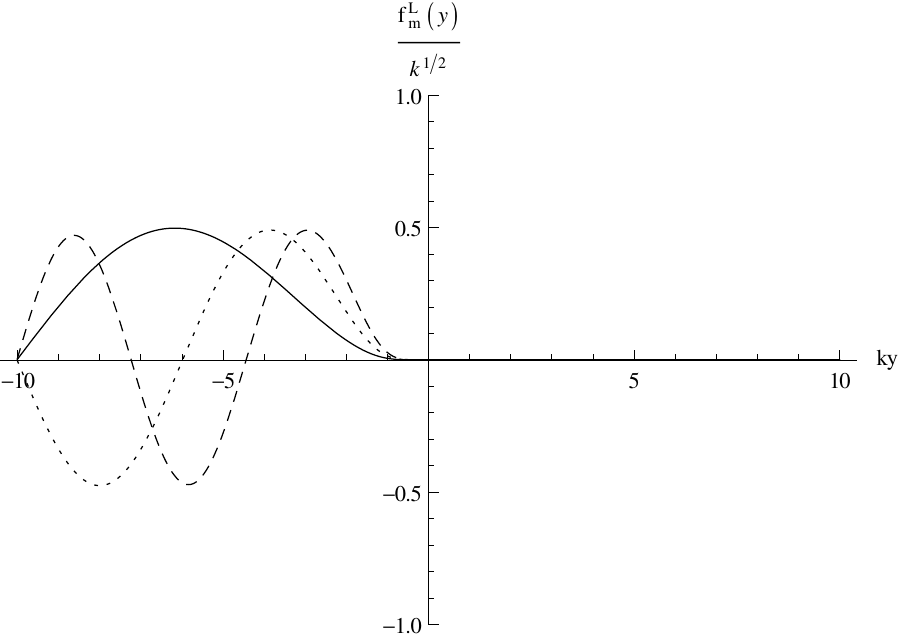}
\caption{A plot of a several left-chiral modes for $\Psi^{5V--}$ for $\tilde{h}=100$.}
\label{fig:L5VIh100modes} 
\end{center}
\end{figure}

\begin{figure}[H]
\begin{center}
\includegraphics[scale=1.5]{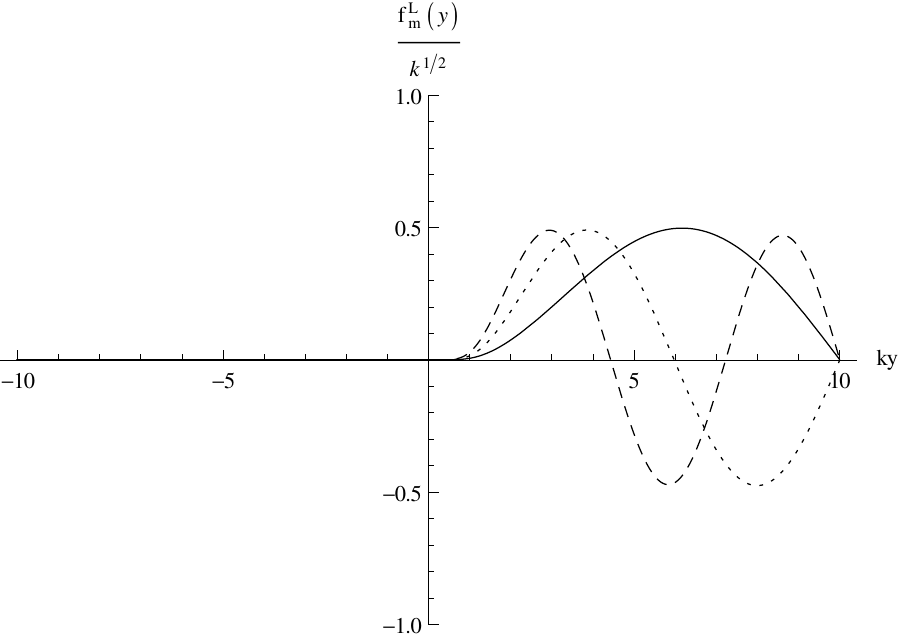}
\caption{A plot of a several left-chiral modes for $\Psi^{5V++}$ for $\tilde{h}=100$.}
\label{fig:L5VIprimeh100modes} 
\end{center}
\end{figure}

\begin{figure}[H]
\begin{center}
\includegraphics[scale=1.5]{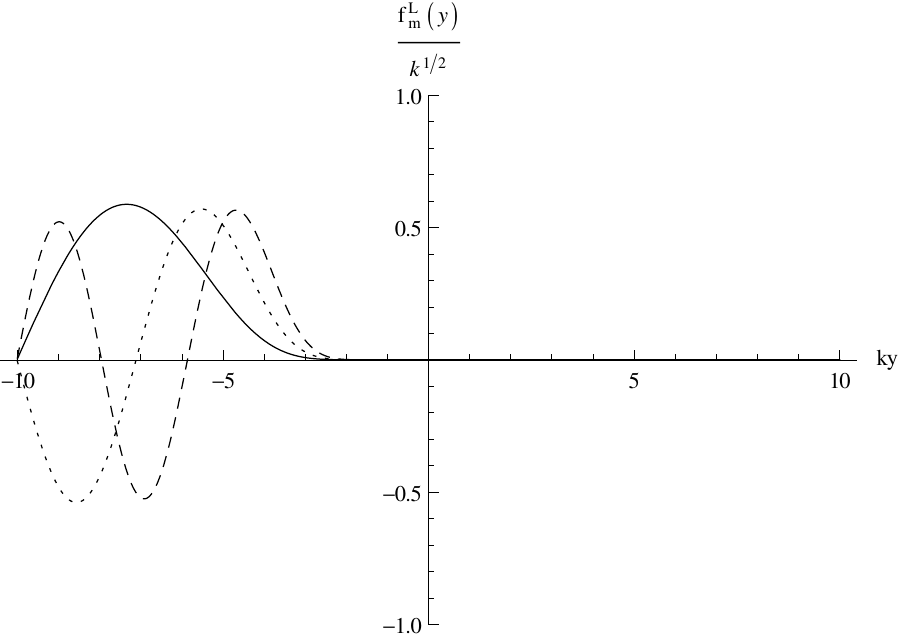}
\caption{A plot of a several left-chiral modes for $\Psi^{5V--}$ for $\tilde{h}=1000$.}
\label{fig:L5VIh1000modes} 
\end{center}
\end{figure}

\begin{figure}[H]
\begin{center}
\includegraphics[scale=1.5]{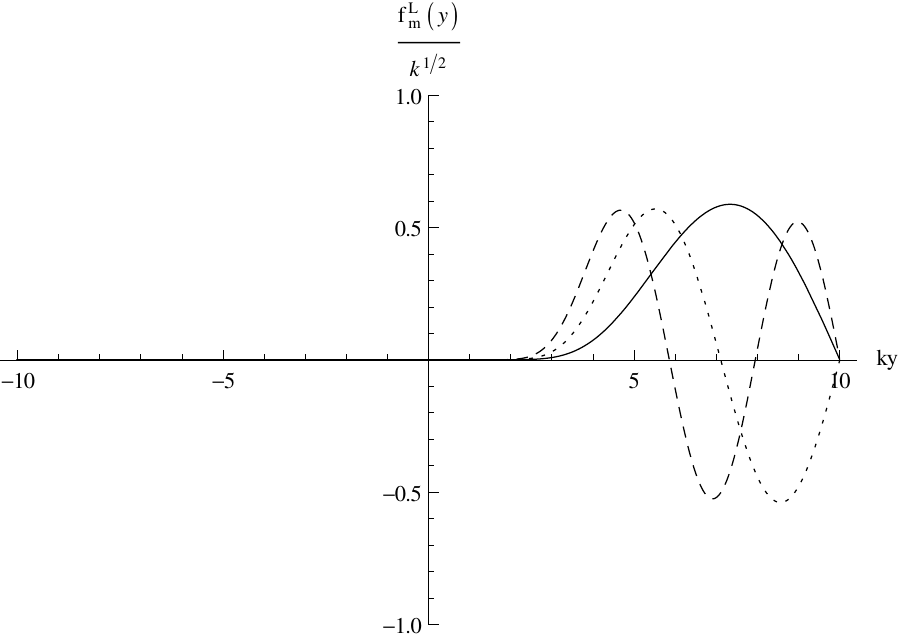}
\caption{A plot of a several left-chiral modes for $\Psi^{5V++}$ for $\tilde{h}=1000$.}
\label{fig:L5VIprimeh1000modes} 
\end{center}
\end{figure}

Finally, there are the additional quintet components from $\Psi^{66}$. When one looks closely at their superpotentials and the resulting potentials, it is clear that they do not attain chiral zero modes. Given that any of the superpotentials in Eqs.~\ref{eq:5Vmmsuperpot}, \ref{eq:5Dmmsuperpot}, \ref{eq:5Vppsuperpot} and \ref{eq:5Dppsuperpot} either contain the combination $\eta_{+}-\eta_{-}$ or $\chi_{+}+\chi_{-}$, these superpotentials interpolate between a non-zero value at spatial infinity at one end (negative or positive) and zero at the other. This means that any potential zero mode would have to be localized at infinity and therefore unphysical. We show plots of the superpotentials $W^{5V--}$ and $W^{5V++}$ along with the resultant left- and right-chiral potentials respectively in Figs.~\ref{fig:potential5Vmm} and \ref{fig:potential5Vpp} for $\tilde{h}=10$. Note that the potentials for the modes of  $\Psi^{5D--}$ and $\Psi^{5D++}$ are the same as those for $\Psi^{5V++}$ and $\Psi^{5V--}$ respectively 
but with the chiralities reversed, again due to a relative minus sign in the superpotentials, so it suffices to analyze the localization properties of $\Psi^{5V++}$ and $\Psi^{5V--}$. In the aforementioned figures, the potentials tend to zero at either negative or positive infinity, and to some positive value on the opposite side. Thus we anticipate that the modes for these fields will exhibit a continuum of massive modes starting from $m=0$, which are delocalized and free to propogate on one side of the wall, but are deeply suppressed on the other. The main concern that we have with these modes is whether they will be able to tunnel sufficiently into the interior of the domain wall and interact with the low-energy localized theory. For simplicity, we just give the plots for several of the left-chiral modes of $\Psi^{5V--}$ and $\Psi^{5V++}$, as there is little qualitative difference between them and the right-chiral modes: we show those for $\tilde{h} = 10$ in in Figs.~\ref{fig:L5VIh10modes} and \ref{fig:L5VIprimeh10modes}, those for $\tilde{h} = 100$ in Figs.~\ref{fig:L5VIh100modes} and \ref{fig:L5VIprimeh100modes} and those for $\tilde{h} = 1000$ in Figs.~\ref{fig:L5VIh1000modes} and \ref{fig:L5VIprimeh1000modes}. For completeness, we give the squared masses of these modes: the masses of the states (which we still label with $gs$, $1e$ and $2e$) in all cases are the same for $\Psi^{5V--}$ and $\Psi^{5V++}$, and for $\tilde{h} = 10$ we have

\begin{equation}
\begin{aligned}
 \label{eq:delocquintetmassesh10}
 m^{2}_{L,gs} &= 0.1156k^2, \\
 m^{2}_{L,1e} &= 0.4450k^2, \\
 m^{2}_{L,2e} &= 0.9793k^2,
 \end{aligned}
\end{equation}
for $\tilde{h} = 100$ we have 
\begin{equation}
\begin{aligned}
 \label{eq:delocquintetmassesh100}
 m^{2}_{L,gs} &= 0.1748k^2, \\
 m^{2}_{L,1e} &= 0.6118k^2, \\
 m^{2}_{L,2e} &= 1.2817k^2,
 \end{aligned}
\end{equation}
and for $\tilde{h} = 1000$ the masses are
\begin{equation}
\begin{aligned}
 \label{eq:delocquintetmassesh1000}
 m^{2}_{L,gs} &= 0.3853k^2, \\
 m^{2}_{L,1e} &= 1.2047k^2, \\
 m^{2}_{L,2e} &= 2.3568k^2,
 \end{aligned}
\end{equation}

As can be seen in the above figures, the stronger the coupling, the less the continuum modes tunnel into the interior of the domain wall. In all three cases the energies of the modes are similar and their exact energies and profiles reflect the fact that we used a program to find them on a truncated mesh. Given the nature of this semi-delocalized potential, the modes are free to propagate in half the domain, which is a length of $L = 10k$. Thus our program finds modes which resemble standing waves with wavelengths of order $L$. Given that the energies of these waves is inversely proportional to $L$, it is no surprise to see that in Eqs.~\ref{eq:delocquintetmassesh10}, \ref{eq:delocquintetmassesh100} and \ref{eq:delocquintetmassesh1000} that the energies of the modes our program found were in the range $0.1k$ to a few $k$. Thus, not surprisingly, modes with roughly the same energies become more suppressed and penetrate less deeply into the interior of the domain wall, as we increase $\tilde{h}$ and thus the height of the energy barriers of their localization potentials from the interior of the wall onwards. Furthermore, the above modes suggest that to get any sort of significant tunneling, even for small $\tilde{h}$, the energy of the modes must be of a scale near $k$. Since $k$ in the parameter region we chose for the scalar fields generating the  domain wall is roughly the inverse width of the wall, $k$ must be at the very least be several $TeV$, hence the interaction of these delocalized modes with the localized modes on the wall is extremely minimal. This is achieved solely with the dynamics of localization of these fields to the wall; there are many other mechanisms which could contribute to the same effect, including further symmetry breaking as well as the addition of a bulk mass, which we will discuss shortly. 

 One may be worried about the fact that in Figs.~\ref{fig:potential5Vmm} and \ref{fig:potential5Vpp}, the potentials $V^{5V--}_{R}$ and $V^{5V++}_{R}$ (and thus, also $V^{5D--}_{L}$ and $V^{5D--}_{L}$) appear become slightly negative near the interior of the wall. This would perhaps suggest the existence of modes localized in these regions, with tachyonic masses. However, no such bound states can exist, since if they did, they would have partners of the opposite chirality, which experience the localization potentials $V^{5V--}_{L}$ and $V^{5V++}_{L}$ (and $V^{5D--}_{R}$ and $V^{5D++}_{R}$ in the dark sector). The potentials $V^{5V--}_{L}$ and $V^{5V++}_{L}$ are positive definite everywhere, and so such opposite chirality partners can only attain positive definite squared masses, and thus cannot possibly be tachyonic. Hence, only modes with squared masses $m^{2} > 0$ exist. Furthermore, as we increase the value of $\tilde{h}$, the well gets pushed further away from the center of the wall, making penetration of (and escape from) the wall by low mass modes negligible. 

 The above results were achieved solely through the localization properties of fermions in particular representations of $SU(12)$ to this domain-wall arrangement. Given that the localized gauge group to this wall is $SU(5)_{V}\times{}SU(5)_{D}\times{}U(1)_{X}$, this setup is not complete, since at the very least we must break the visible GUT $SU(5)_{V}$ to the Standard Model. The most likely way to achieve this breaking is through the introduction of an additional adjoint $143$ field, and choosing parameters such that the $(24, 1)$ component embedded in it condenses in the interior of the domain wall, inducing the breaking in the usual way. In many domain wall models \cite{firstpaper, jayneso10paper}, such a field contributes to the background domain-wall configuration, condensing in the interior of the wall and asymptoting to zero at infinity, leading to a kink-lump background configuration. This lump affects the localization of the different SM components, splitting them according to their hypercharges. Similar physics will happen in the dark sector if we introduce additional fields which break $SU(5)_{D}$ in the interior of the wall. We leave the specific analysis of further symmetry breaking in the interior of the wall to later work.
 
 Note also that the addition of a bulk mass is consistent with the symmetries which underlie the Yukawa interaction in Eq.~\ref{eq:yukawacoupling1}. This is so because under this symmetry, $\Psi_{R}$ and thus $\overline{\Psi_{R}}\Psi_{R}$ is invariant. For the localized quintets and decuplets, the bulk mass will shift their localization centers from $y = 0$ in the usual way, as per the original split fermion mechanism \cite{splitfermions}. In fact, given fields with opposite chiralities have their localization centers shifted in equal amounts in opposite directions for the same bulk mass $M$, a bulk mass term will shift the visible and dark fermions in different directions along the extra dimension, leading to a splitting of the visible and dark sectors. For the mixed $(5, 5)$ fermion and the singlet state from the $66$, a bulk mass simply makes these delocalized $4+1D$ states massive, hence presenting a much easier way to make these fields massive than through symmetry breaking. For the delocalized singlets of the fundamental, the most likely outcome is that their masses get shifted. For the semi-delocalized quintets of the $66$, since their superpotentials can be thought of as approximately of the form $hv(1\pm{}\tanh{(k'y)})/2$, if the bulk mass is opposite that provided by the superpotential, then the resultant mass term will always be less than the maximum of the $\tanh(k'y)$ term, thus making it possible for some of these quintets to attain localized modes. These modes will have the same chirality as the decuplets, hence they would be potentially troublesome, but we can always localize additional quintets of the opposite chirality using the fundamental representation to ensure that these modes attain a GUT scale mass after breaking $SU(5)_{V}$ to the Standard Model if need be.

 We have shown in this section that it is possible to localize a set of 3+1D left-chiral fermions in the set of representations $(\overline{5}, 1)\oplus{}(10, 1)$ of $SU(5)_{V}\times{}SU(5)_{D}$, which contain the visible Standard Model fermions, along with a mirror dark sector of right-chiral fermions in the representations $(1, \overline{5})\oplus{}(1, 10)$ of $SU(5)_{V}\times{}SU(5)_{D}$, by coupling $4+1D$ fermions in the $12$ and $66$ representations of $SU(12)$ to the domain wall. Furthermore, we showed that the troublesome mixed $(5, 5)$ fermion was completely delocalized, implying that it remained a vector-like $4+1D$ Dirac fermion which will attain a GUT scale mass when we add an additional adjoint scalar field to the background configuration to induce the breaking of $SU(5)_{V}$ to the Standard Model. Likewise, the delocalized singlet will attain a mass when we break the additional $U(1)_{X}$. We also showed that the additional unwanted quintet states in the $66$ could be sufficiently suppressed in the interior of the domain-wall brane. The next step is to show that we can localize scalars and that we can therefore localize a Standard Model Higgs field along with a dark mirror Higgs field, opening the possibility of having a fully localized Standard Model and a localized dark mirror sector, which are sufficiently sequestrated to satisfy current experimental limits.

\section{Scalar Localization}
\label{sec:scalarlocalization}

 In this section, we give a simple example of scalar localization to the $m = 1$ domain wall which was described in previous sections and used in the previous section on fermion localization. For simplicity, we solely consider a scalar in the fundamental $12$ representation, which we call $\Phi$. We give a couple of interesting scenarios when considering the localization properties of the individual $SU(5)_{V}\times{}SU(5)_{D}\times{}U(1)_{X}\times{}U(1)_{A}\times{}U(1)_{B}$ components, which we label as $\Phi^{5V}$, $\Phi^{5D}$, $\Phi^{--}$ and $\Phi^{++}$, in correspondence with the labelling we used for the components of the fermionic $\Psi_{12}$ from the previous section. We would like to at the very least be able to give the visible Higgs quintet scalar, $\Psi^{5V}$, a lowest energy localized mode with a tachyonic mass, so that electroweak symmetry breaking can be performed. We would like to also show that there are parameter regions where the singlets $\Phi^{--}$ and $\Phi^{++}$ attain tachyonic masses so that we can break the semi-delocalized $U(1)_{A}$ and $U(1)_{B}$.

 The most general potential which couples $\Phi$ to the domain-wall generating fields  $\eta$ and $\chi$ is
\begin{equation}
 \label{eq:fundamentalscalarlocpotential}
 \begin{aligned}
 V_{loc}(\Phi, \eta, \chi) &= \mu^{2}_{\Phi}\Phi^{\dagger}\Phi+\lambda_{\Phi1}\big(\Phi^{\dagger}\eta{}\Phi{}+\Phi^{\dagger}\chi{}\Phi{}\big)+\lambda_{\Phi2}\big(\Phi^{\dagger}\eta^{2}\Phi{} + \Phi^{\dagger}\chi^{2}\Phi{}\big)     \\
                           &+\lambda_{\Phi3}\big(\Phi^{\dagger}\eta{}\chi{}\Phi{}+\Phi^{\dagger}\chi{}\eta{}\Phi{}\big)+\lambda_{\Phi4}\big(\Phi^{\dagger}\Phi{}Tr[\eta^2]+\Phi^{\dagger}\Phi{}Tr[\chi^2]\big)+\lambda_{\Phi5}\Phi^{\dagger}\Phi{}Tr(\eta{}\chi{}).
 \end{aligned}
\end{equation}

From this, by substituting the form of the $m=1$ solution of Sec.~\ref{sec:solution} and describing the couplings in terms of $\eta_{-}$, $\chi_{-}$, $\eta_{+}$ and $\chi_{+}$, we find that the effective localization potentials for the modes of $\Phi^{5V}$, $\Phi^{5D}$, $\Phi^{--}$ and $\Phi^{++}$ are, respectively,
\begin{equation}
\label{eq:visiblequintetscalarpotential}
\begin{aligned}
V^{5V}_{loc}(\tilde{y}) &= \mu^{2}_{\Phi}+\lambda_{\Phi1}\big(\eta_{-}(\tilde{y})+\chi_{-}(\tilde{y})\big)+\lambda_{\Phi2}\big(\eta^{2}_{-}(\tilde{y})+\chi^{2}_{-}(\tilde{y})\big)+2\lambda_{\Phi3}\eta_{-}(\tilde{y})\chi_{-}(\tilde{y}) \\
                        &+\lambda_{\Phi4}\big(10\eta^{2}_{-}(\tilde{y})+2\eta^{2}_{+}(\tilde{y})+10\chi^{2}_{-}(\tilde{y})+2\chi^{2}_{+}(\tilde{y})\big)+\lambda_{\Phi5}\big(10\eta_{-}(\tilde{y})\chi_{-}(\tilde{y})+2\eta_{+}(\tilde{y})\chi_{+}(\tilde{y})\big),
\end{aligned}
\end{equation}

\begin{equation}
\label{eq:darkquintetscalarpotential}
\begin{aligned}
V^{5D}_{loc}(\tilde{y}) &= \mu^{2}_{\Phi}-\lambda_{\Phi1}\big(\eta_{-}(\tilde{y})+\chi_{-}(\tilde{y})\big)+\lambda_{\Phi2}\big(\eta^{2}_{-}(\tilde{y})+\chi^{2}_{-}(\tilde{y})\big)+2\lambda_{\Phi3}\eta_{-}(\tilde{y})\chi_{-}(\tilde{y}) \\
                        &+\lambda_{\Phi4}\big(10\eta^{2}_{-}(\tilde{y})+2\eta^{2}_{+}(\tilde{y})+10\chi^{2}_{-}(\tilde{y})+2\chi^{2}_{+}(\tilde{y})\big)+\lambda_{\Phi5}\big(10\eta_{-}(\tilde{y})\chi_{-}(\tilde{y})+2\eta_{+}(\tilde{y})\chi_{+}(\tilde{y})\big),
\end{aligned}
\end{equation}

\begin{equation}
\label{eq:mmsingletscalarpotential}
\begin{aligned}
V^{--}_{loc}(\tilde{y}) &= \mu^{2}_{\Phi}+\lambda_{\Phi1}\big(\eta_{+}(\tilde{y})+\chi_{+}(\tilde{y})\big)+\lambda_{\Phi2}\big(\eta^{2}_{+}(\tilde{y})+\chi^{2}_{+}(\tilde{y})\big)+2\lambda_{\Phi3}\eta_{+}(\tilde{y})\chi_{+}(\tilde{y}) \\
                        &+\lambda_{\Phi4}\big(10\eta^{2}_{-}(\tilde{y})+2\eta^{2}_{+}(\tilde{y})+10\chi^{2}_{-}(\tilde{y})+2\chi^{2}_{+}(\tilde{y})\big)+\lambda_{\Phi5}\big(10\eta_{-}(\tilde{y})\chi_{-}(\tilde{y})+2\eta_{+}(\tilde{y})\chi_{+}(\tilde{y})\big),
\end{aligned}
\end{equation}
and
\begin{equation}
\label{eq:ppsingletscalarpotential}
\begin{aligned}
V^{++}_{loc}(\tilde{y}) &= \mu^{2}_{\Phi}-\lambda_{\Phi1}\big(\eta_{+}(\tilde{y})+\chi_{+}(\tilde{y})\big)+\lambda_{\Phi2}\big(\eta^{2}_{+}(\tilde{y})+\chi^{2}_{+}(\tilde{y})\big)+2\lambda_{\Phi3}\eta_{+}(\tilde{y})\chi_{+}(\tilde{y}) \\
                        &+\lambda_{\Phi4}\big(10\eta^{2}_{-}(\tilde{y})+2\eta^{2}_{+}(\tilde{y})+10\chi^{2}_{-}(\tilde{y})+2\chi^{2}_{+}(\tilde{y})\big)+\lambda_{\Phi5}\big(10\eta_{-}(\tilde{y})\chi_{-}(\tilde{y})+2\eta_{+}(\tilde{y})\chi_{+}(\tilde{y})\big).
\end{aligned}
\end{equation}

 To find the localized modes of these potentials, we first perform a mode expansion in the usual way, representing a given $SU(5)_{V}\times{}SU(5)_{D}\times{}U(1)_{X}$ component $\Phi^{R}$ in the form
 \begin{equation}
  \label{eq:scalarmodeexpansion}
  \Phi^{R}(x, y) = \sum_{m} p^{R}_{m}(y)\phi^{R}_{m}(x),
 \end{equation}
where again $m$ stands for the mass of the mode $\phi^{R}_{m}$. When we substitute this mode expansion into the $4+1D$ Klein-Gordon equation, noting that $\Box_{3+1D}\phi^{R}_{m} = -m^2\phi^{R}_{m}$, we find that profiles $p^{R}_{m}(y)$ satisfy the Schr\"{o}dinger equations
\begin{equation}
 \label{eq:scalarmodeSE}
 \big[-\frac{d}{dy^2} + V^{R}_{loc}(y)\big]p^{R}_{m}(y) = m^2p^{R}_{m}(y).
\end{equation}
We solve for the three lowest energy modes for the above set of equations in the same way that we did for the corresponding equations for the fermions in the previous section, by finding the eigenvectors and eigenvalues of the Hamiltonian acting on the $2001$-dimensional space spanned by the values of the eigenmodes at each of the lattice points. We do this for two parameter choices.

\begin{figure}[H]
\begin{center}
\includegraphics[scale=0.85]{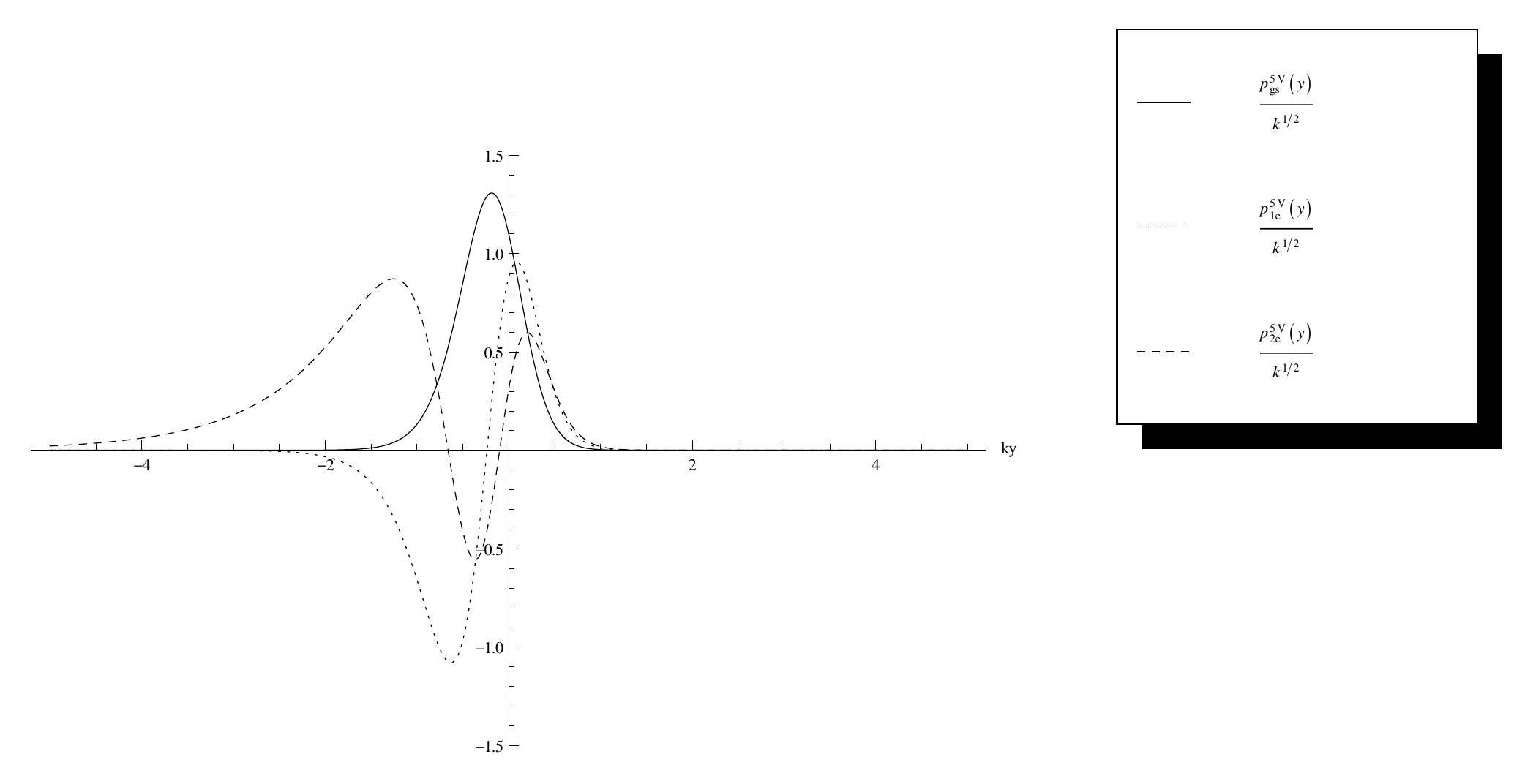}
\caption{A plot of the first three modes for the visible scalar quintet $\Phi^{5V}$ for the parameter choice in Eq.~\ref{eq:firstscalarlocchoice}.}
\label{fig:visiblehiggsquintet1} 
\end{center}
\end{figure}

\begin{figure}[H]
\begin{center}
\includegraphics[scale=0.9]{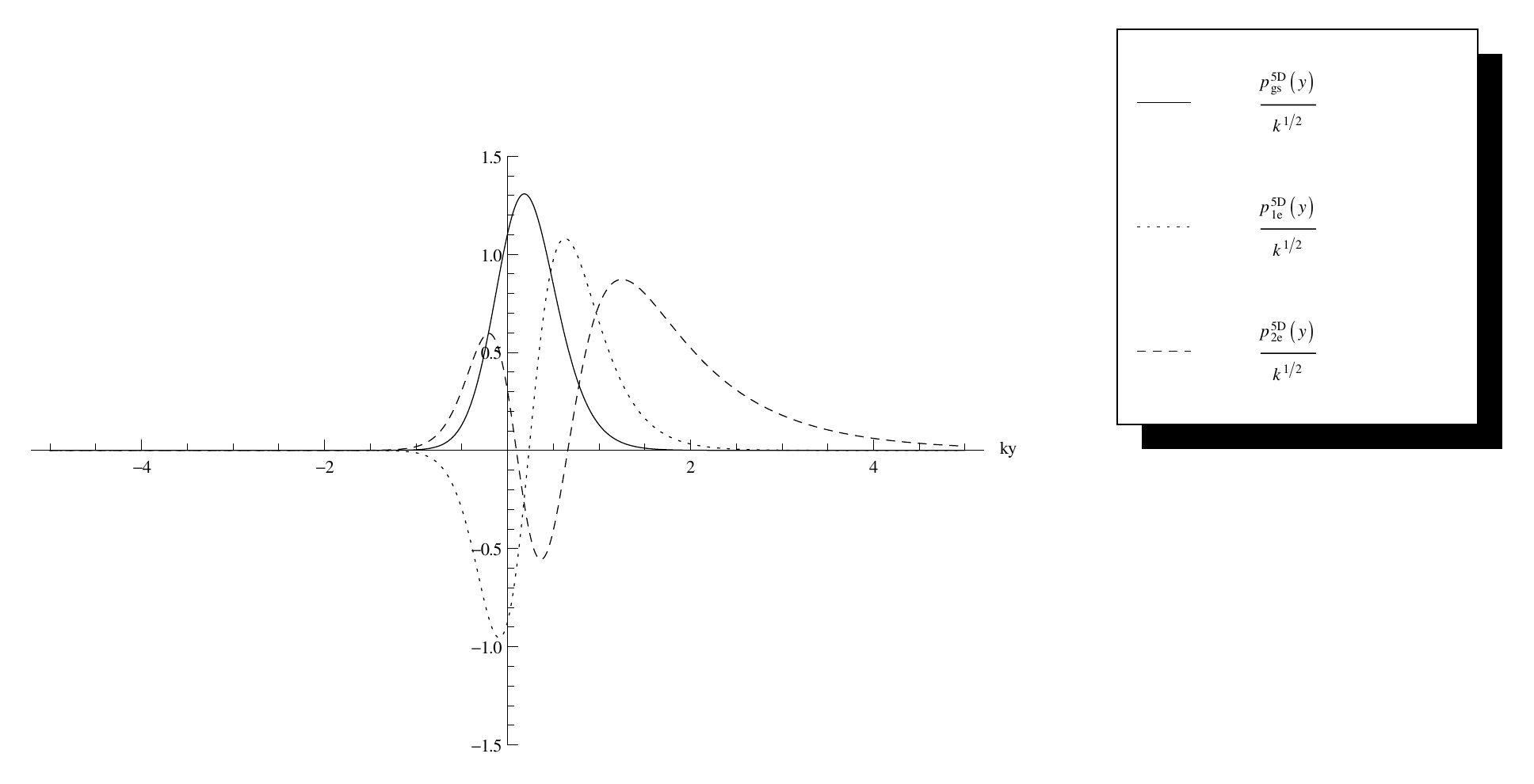}
\caption{A plot of the first three modes for the dark scalar quintet $\Phi^{5D}$ for the parameter choice in Eq.~\ref{eq:firstscalarlocchoice}.}
\label{fig:darkhiggsquintet1} 
\end{center}
\end{figure}

\begin{figure}[H]
\begin{center}
\includegraphics[scale=0.7]{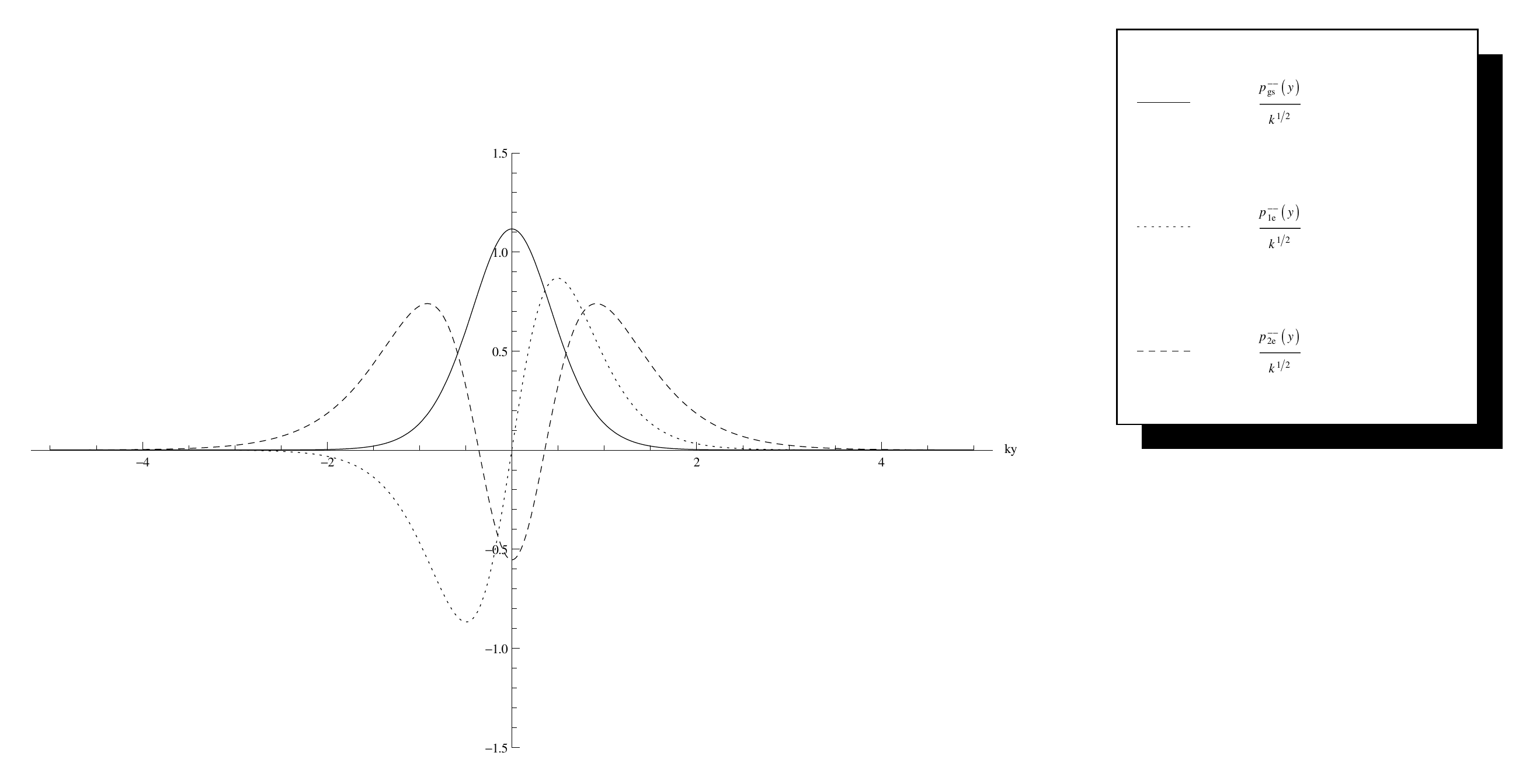}
\caption{A plot of the first three modes for the singlet $\Phi^{--}$ for the parameter choice in Eq.~\ref{eq:firstscalarlocchoice}.}
\label{fig:singletmmhiggsquintet1} 
\end{center}
\end{figure}

\begin{figure}[H]
\begin{center}
\includegraphics[scale=0.75]{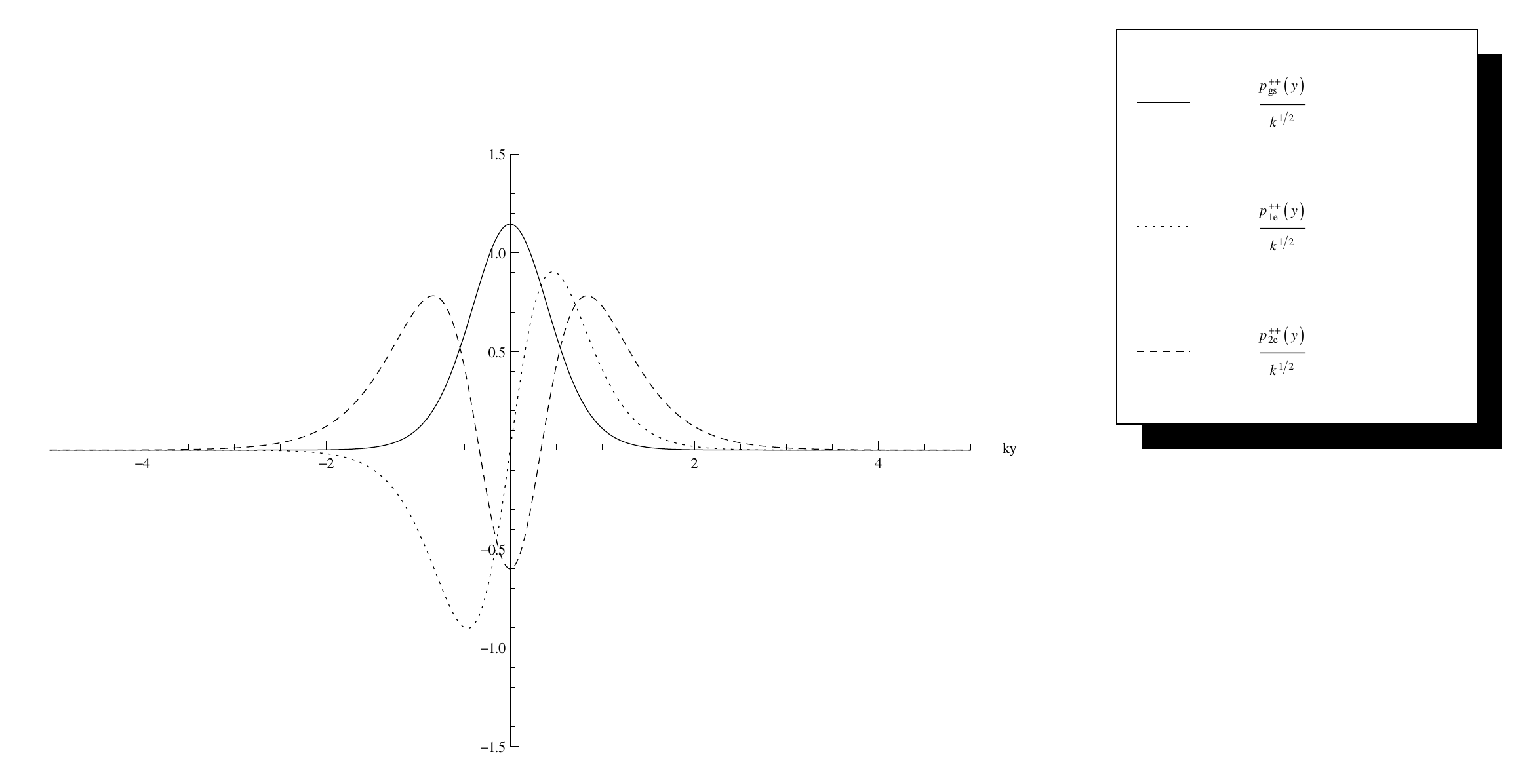}
\caption{A plot of the first three modes for the singlet $\Phi^{++}$ for the parameter choice in Eq.~\ref{eq:firstscalarlocchoice}.}
\label{fig:singletpphiggsquintet1} 
\end{center}
\end{figure}

 For the first parameter choice, we choose
\begin{equation}
 \label{eq:firstscalarlocchoice}
 \begin{aligned}
 \mu^{2}_{\Phi} &= 5.0k^2, \\
 \lambda_{\Phi1} &= \frac{100.0}{k}, \\
 \lambda_{\Phi2} &= \frac{-600.0}{k}, \\
 \lambda_{\Phi3} &= \frac{600.0}{k}, \\
 \lambda_{\Phi4} &= \frac{100.0}{k}, \\
 \lambda_{\Phi5} &= \frac{150.0}{k},
 \end{aligned}
\end{equation}
and we find the masses of the three lightest modes, which label again with the subscripts $gs$, $1e$ and $2e$, are 
\begin{equation}
\begin{aligned}
 \label{eq:visiblequintetscalarmasses1}
 m^{2}_{5V,gs} &= -4.3809k^2, \\
 m^{2}_{5V,1e} &= 12.2846k^2, \\
 m^{2}_{5V,2e} &= 22.3332k^2,
 \end{aligned}
\end{equation}
for the visible Higgs quintet $\Phi^{5V}$,
\begin{equation}
\begin{aligned}
 \label{eq:darkquintetscalarmasses1}
 m^{2}_{5D,gs} &= -4.3809k^2, \\
 m^{2}_{5D,1e} &= 12.2846k^2, \\
 m^{2}_{5D,2e} &= 22.3332k^2,
 \end{aligned}
\end{equation}
for the dark Higgs quintet $\Phi^{5D}$,
\begin{equation}
\begin{aligned}
 \label{eq:mmsingletscalarmasses1}
 m^{2}_{--,gs} &= 3.4731k^2, \\
 m^{2}_{--,1e} &= 12.3468k^2, \\
 m^{2}_{--,2e} &= 18.1113k^2,
 \end{aligned}
\end{equation}
for the singlet Higgs scalar $\Phi^{--}$, and
\begin{equation}
\begin{aligned}
 \label{eq:mmsingletscalarmasses1}
 m^{2}_{++,gs} &= 55.4509k^2, \\
 m^{2}_{++,1e} &= 65.4019k^2, \\
 m^{2}_{++,2e} &= 72.3116k^2,
 \end{aligned}
\end{equation}
for the singlet Higgs scalar $\Phi^{++}$. We show plots of the profiles for $\Phi^{5V}$, $\Phi^{5D}$, $\Phi^{--}$ and $\Phi^{++}$ respectively in Figs.~\ref{fig:visiblehiggsquintet1}, \ref{fig:darkhiggsquintet1}, \ref{fig:singletmmhiggsquintet1} and \ref{fig:singletpphiggsquintet1}.

From this we see that the lowest energy modes of both the visible and dark quintets are localized and attain tachyonic masses, while all the modes for the singlet states have positive squared masses. This means that these quintets can go on to induce symmetry breaking in the visible and dark sectors, while the semi-delocalized Abelian groups $U(1)_{A}$ and $U(1)_{B}$ are left unbroken. Notice that the profiles for these lowest energy states of $\Phi^{5V}$ and $\Phi^{5D}$ are split and their masses are degenerate. The splitting is due to the cubic interaction corresponding to the coupling constant $\lambda_{\Phi1}$; this term introduces a contribution to the potential proportional to the combination $\eta_{-}+\chi_{-}$, which is kink-like, thus shifting the localization centers from zero. Given that the visible and dark quintets experience this term equally but with the opposite sign, they experience shifts in opposite directions from $\tilde{y}=0$. In fact, one can deduce that $V^{5D}_{loc}(\tilde{y}) = V^{
5V}_{loc}(-\tilde{y})$, so that the potential of the dark quintet is a mirror image of the one for the visible quintet, explaining the degeneracy of the masses for their respective modes. 

The same cubic interaction does something very different for the singlet modes. This interaction leads to terms in the potentials for $\Phi^{--}$ and $\Phi^{++}$ which are proportional to $\eta_{+}+\chi_{+}$. As discussed previously, $\eta_{+}+\chi_{+}$ is an even function and behaves, for the most part, as mass-like rather than kink-like. This means that this term will either raise or lower the masses of the localized modes and, given that $\Phi^{--}$ and $\Phi^{++}$ experience this interaction equally but with a relative minus sign, the masses of one of them will be lowered while those for the other will be raised. This is why the masses of the modes of $\Phi^{--}$ and $\Phi^{++}$ are not degenerate.

 We have shown that symmetric breaking of the $SU(5)_{V}$ and $SU(5)_{D}$ is possible. Given that the singlet Higgs scalars are charged under $U(1)_{A}$ and $U(1)_{B}$, they can be potentially used to break one of these symmetries, so we are thus also interested in whether these components can attain tachyonic masses. In the second parameter choice, we will show that this is possible. If one of these components attains a tachyonic mass, $U(1)_{A}\times{}U(1)_{B}$ is broken to $U(1)_{A-B}$. In a realistic model, we would need to add a scalar in another representation to localize a singlet component which can break $U(1)_{A-B}$.

\begin{figure}[H]
\begin{center}
\includegraphics[scale=0.9]{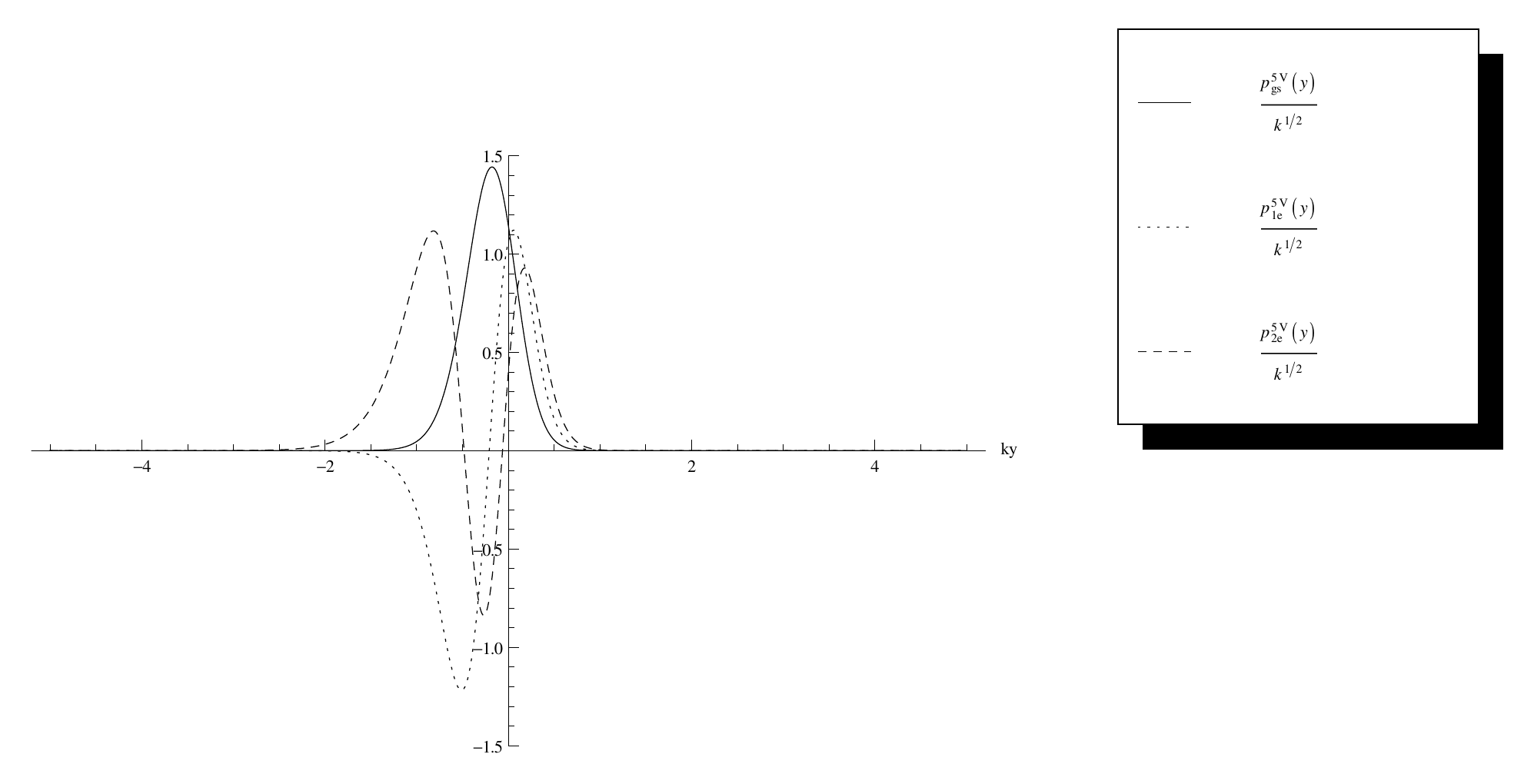}
\caption{A plot of the first three modes for the visible scalar quintet $\Phi^{5V}$ for the parameter choice in Eq.~\ref{eq:secondscalarlocchoice}.}
\label{fig:visiblehiggsquintet2} 
\end{center}
\end{figure}

\begin{figure}[H]
\begin{center}
\includegraphics[scale=0.75]{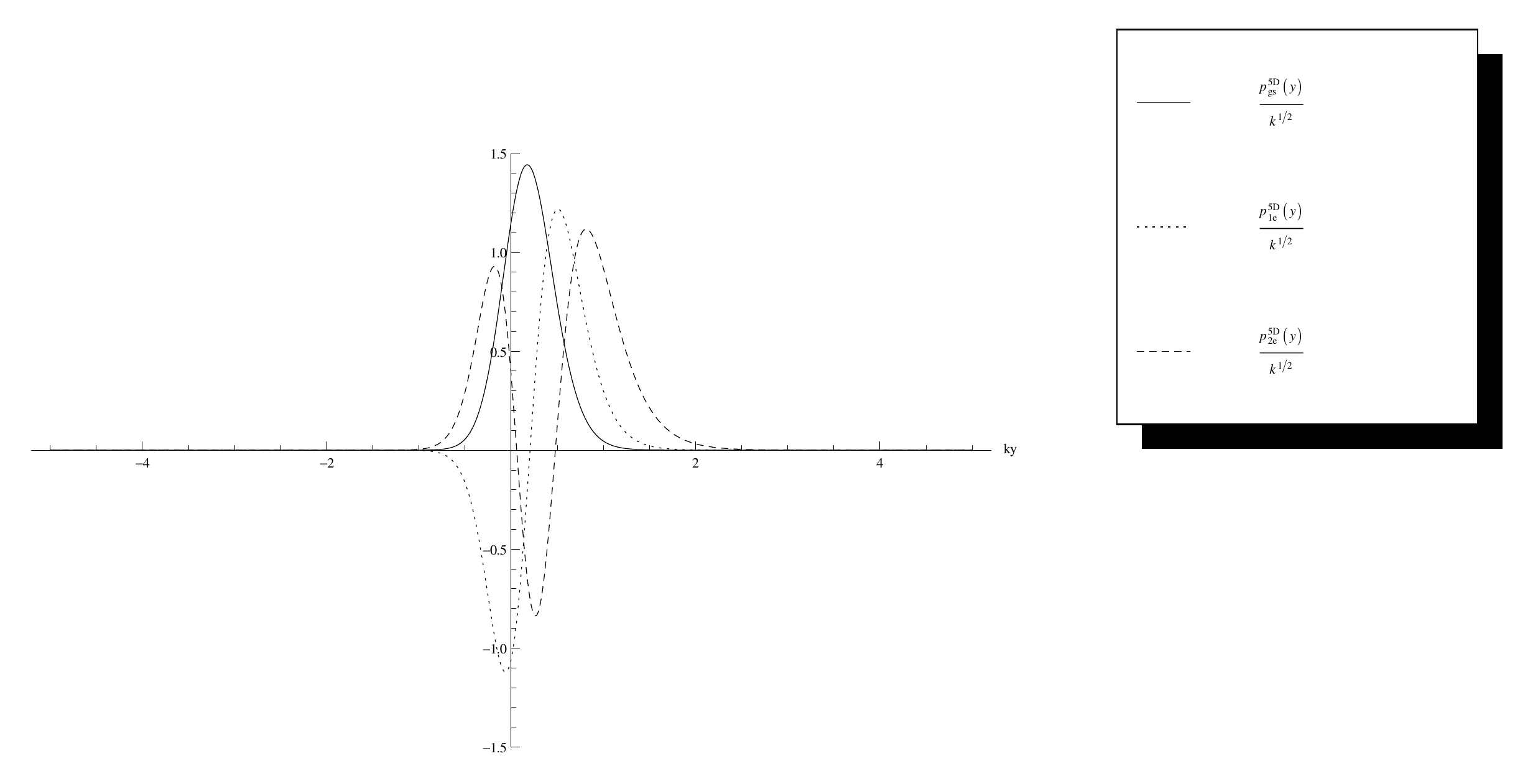}
\caption{A plot of the first three modes for the dark scalar quintet $\Phi^{5D}$ for the parameter choice in Eq.~\ref{eq:secondscalarlocchoice}.}
\label{fig:darkhiggsquintet2} 
\end{center}
\end{figure}

\begin{figure}[H]
\begin{center}
\includegraphics[scale=0.75]{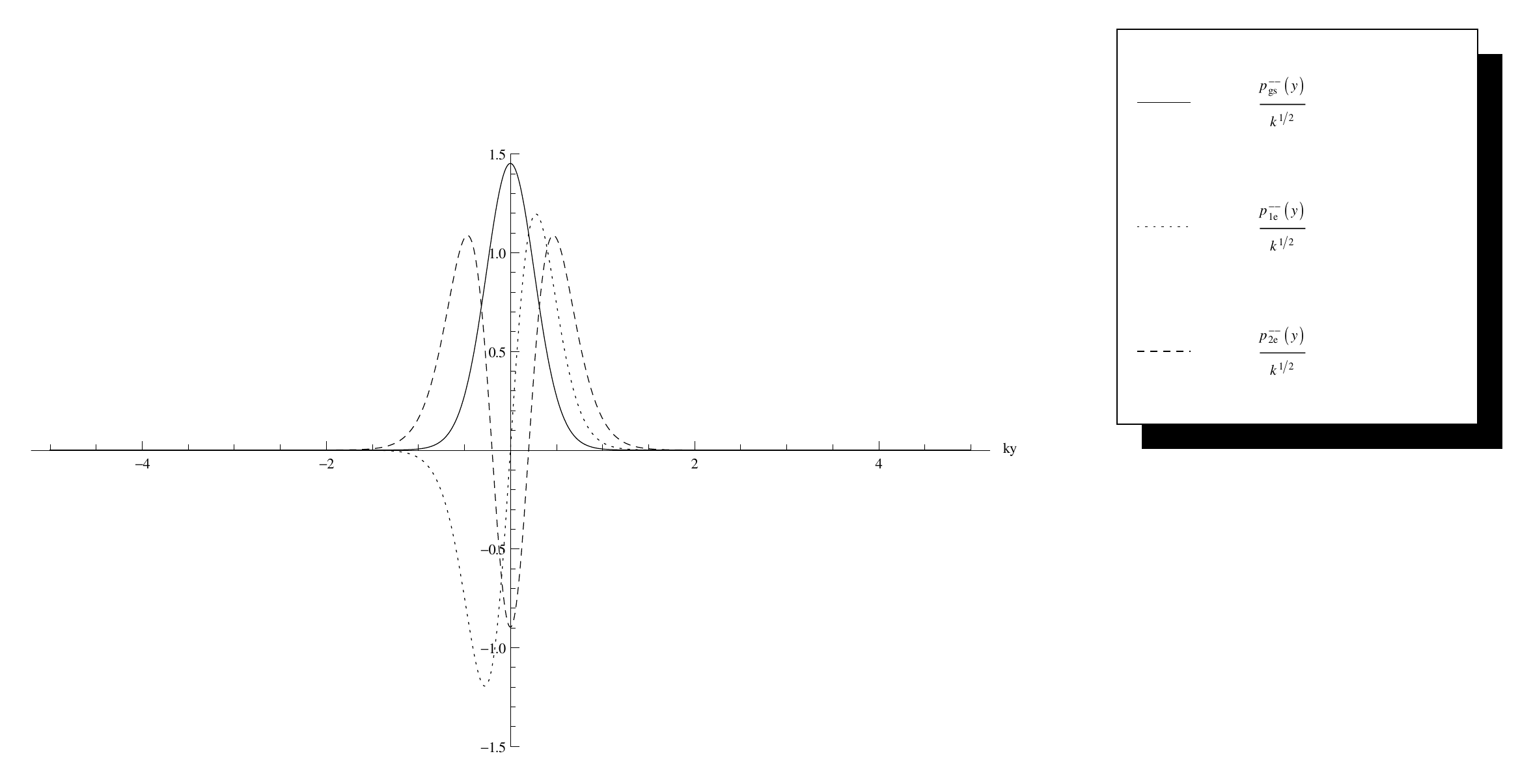}
\caption{A plot of the first three modes for the singlet $\Phi^{--}$ for the parameter choice in Eq.~\ref{eq:secondscalarlocchoice}.}
\label{fig:singletmmhiggsquintet2} 
\end{center}
\end{figure}

\begin{figure}[H]
\begin{center}
\includegraphics[scale=0.75]{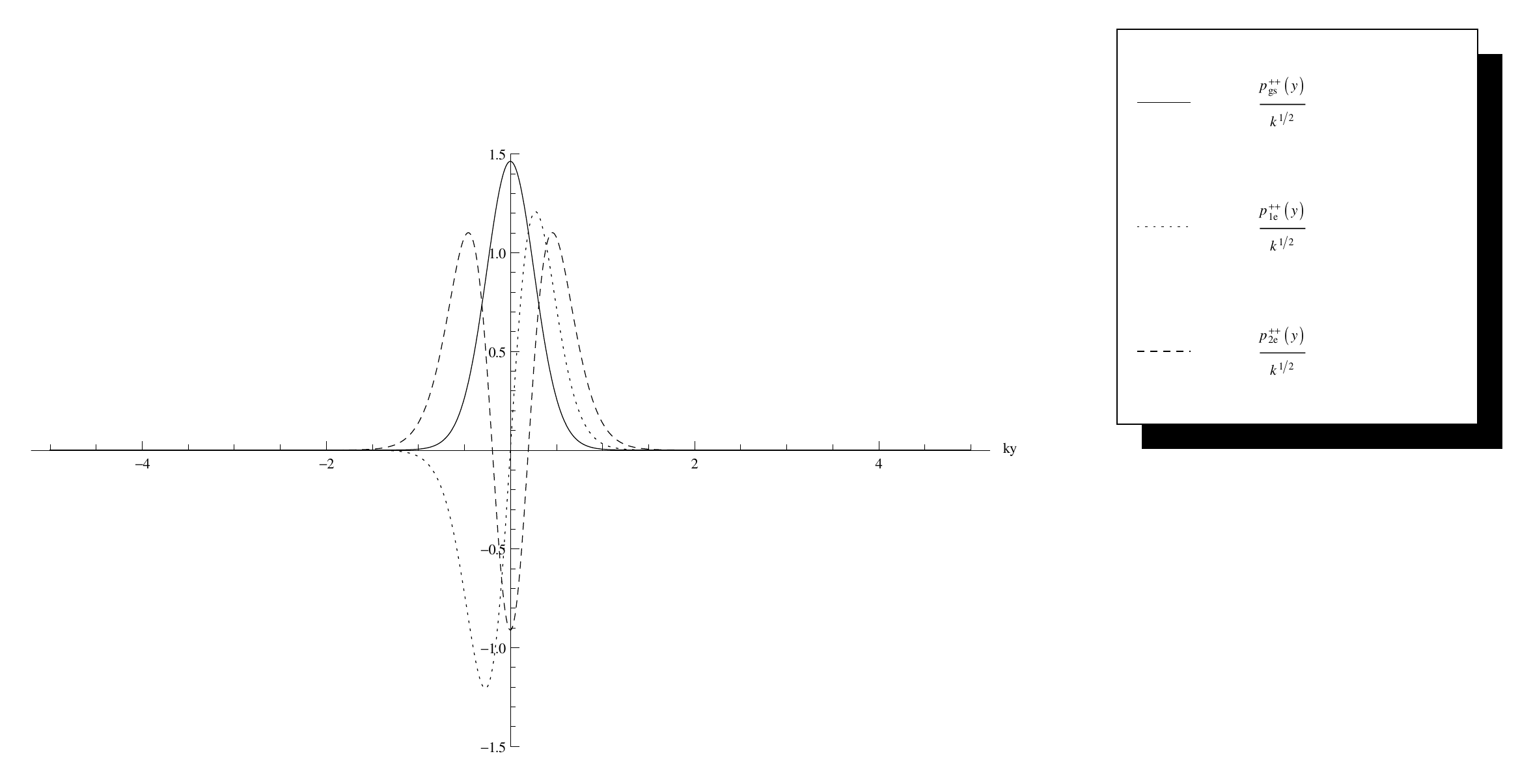}
\caption{A plot of the first three modes for the singlet $\Phi^{++}$ for the parameter choice in Eq.~\ref{eq:secondscalarlocchoice}.}
\label{fig:singletpphiggsquintet2} 
\end{center}
\end{figure}

For the second parameter choice, we choose
\begin{equation}
 \label{eq:secondscalarlocchoice}
 \begin{aligned}
 \mu^{2}_{\Phi} &= 32.0k^2, \\
 \lambda_{\Phi1} &= \frac{200.0}{k}, \\
 \lambda_{\Phi2} &= \frac{100.0}{k}, \\
 \lambda_{\Phi3} &= \frac{100.0}{k}, \\
 \lambda_{\Phi4} &= \frac{100.0}{k}, \\
 \lambda_{\Phi5} &= \frac{550.0}{k},
 \end{aligned}
\end{equation}
and we find the masses of the three lightest modes this time are
\begin{equation}
\begin{aligned}
 \label{eq:visiblequintetscalarmasses2}
 m^{2}_{5V,gs} &= 13.8579k^2, \\
 m^{2}_{5V,1e} &= 39.5516k^2, \\
 m^{2}_{5V,2e} &= 59.6312k^2,
 \end{aligned}
\end{equation}
for the visible Higgs quintet $\Phi^{5V}$,
\begin{equation}
\begin{aligned}
 \label{eq:darkquintetscalarmasses2}
 m^{2}_{5D,gs} &= 13.8579k^2, \\
 m^{2}_{5D,1e} &= 39.5516k^2, \\
 m^{2}_{5D,2e} &= 59.6312k^2,
 \end{aligned}
\end{equation}
for the dark Higgs quintet $\Phi^{5D}$,
\begin{equation}
\begin{aligned}
 \label{eq:mmsingletscalarmasses2}
 m^{2}_{--,gs} &= -24.8537k^2, \\
 m^{2}_{--,1e} &= 1.9428k^2, \\
 m^{2}_{--,2e} &= 24.4997k^2,
 \end{aligned}
\end{equation}
for the singlet Higgs scalar $\Phi^{--}$, and
\begin{equation}
\begin{aligned}
 \label{eq:mmsingletscalarmasses2}
 m^{2}_{++,gs} &= 78.5005k^2, \\
 m^{2}_{++,1e} &= 106.1446k^2, \\
 m^{2}_{++,2e} &= 129.6727k^2,
 \end{aligned}
\end{equation}
for the singlet Higgs scalar $\Phi^{++}$. For this parameter choice, we show plots of the profiles for $\Phi^{5V}$, $\Phi^{5D}$, $\Phi^{--}$ and $\Phi^{++}$ respectively in Figs.~\ref{fig:visiblehiggsquintet2}, \ref{fig:darkhiggsquintet2}, \ref{fig:singletmmhiggsquintet2} and \ref{fig:singletpphiggsquintet2}.

 From the above equations for the squared masses, we can clearly see that for the second parameter choice, the lowest energy localized modes for the visible and dark quintets as well as the singlet $\Phi^{++}$ have positive definite squared masses, while the lowest energy localized mode for $\Phi^{--}$ attains a tachyonic mass. This will lead to the breaking $U(1)_{A}\times{}U(1)_{B}\rightarrow{}U(1)_{A-B}$

 We have used this section to give an example of scalar localization for the various components of a fundamental 4+1D scalar field coupled to the $m=1$ CoS solution. Any realistic model will have to include additional scalar fields, including at least one adjoint which breaks the visible $SU(5)_{V}$ to the Standard Model. Given the magnitude of the GUT scale, such a field is expected to have a significant back-reaction on the kink-generating scalar fields, hence an additional scalar inducing the breaking of $SU(5)_{V}$ is expected to form part of the background scalar field configuration, rather than merely just being localized. Coupling a fundamental scalar to this field would then lead to splitting of different SM-covariant components, and would also lead to a breaking of degeneracy in the masses between these components. Thus, it should be possible to choose parameters such that the electroweak Higgs embedded in the visible quintet attains a tachyonic mass, while the colored Higgs component of $\Phi^{5V}$ attains a positive squared mass, maintaining $SU(3)_{c}$. 
 
 In later models, we could also introduce symmetry breaking in the dark $SU(5)_{D}$ sector via scalar fields in the background. In the dark sector, we obviously have a lot more freedom in how we break $SU(5)_{D}$. If we were to break $SU(5)_{V}$ and $SU(5)_{D}$ symmetrically, that is to break $SU(5)_{D}$ to a mirror SM gauge group at the same scale is the corresponding breaking for $SU(5)_{V}$, we would end up with a localized mirror matter model on the domain wall. Another interesting possibility would to break $SU(5)_{V}$ and $SU(5)_{D}$ asymmetrically, as per the models in Ref.~\cite{lonsdalegrandunifieddm, lonsdaleso10xso10adm}. The potential to do asymmetric symmetry breaking generally in the context of this model is very rich; we could do this through the background scalar field configuration or, alternatively, through the localization of a set of scalar fields which experience an asymmetric symmetry breaking potential in the interior of the wall. We leave an analysis of these symmetry breaking scenarios to later work. 
 
 In this section along with the two previous ones, we have outlined the construction of a domain-wall brane model based on the Clash-of-Symmetries mechanism in which gauge bosons corresponding to the gauge group $SU(5)_{V}\times{}SU(5)_{D}\times{}U(1)_{X}$ are localized, along with fermions and scalars. In particular, fermion localization in this model has some special properties, with only the components which transform solely under either $SU(5)_{V}$ or $SU(5)_{D}$ localized to the wall; the troublesome $(5, 5)$ mediator is completely delocalized and will attain a GUT scalar mass when we perform the required breaking of $SU(5)_{V}$. Higgs scalars can be localized with tachyonic masses and can hence induce further symmetry breaking on the wall as required. Hence, we have been successful in constructing a prototype model in an extra-dimensional field theory in which a visible gauge sector along with a dark, non-Abelian gauge sector arises from a unified $SU(12)$ theory. Given that it would be interesting to construct other viable models of this sort, before we conclude this paper, we give an overview of several other interesting alternatives which could lead to the same dynamics in the next section.

\section{Some Alternative Models}
\label{sec:alternativemodels}

\subsection{Another interesting model: $SU(9)$}
\label{subsec:su9model}

 In this section, we briefly outline how to construct a potentially realistic model using the group $SU(9)$. In this case, we claim that the CoS mechanism can be used to generate an $SU(5)_{V}\times{}SU(2)_{D}\times{}U(1)_{X}$ invariant theory on the wall after breaking $SU(9)$ to two differently embedded copies of $SU(6)\times{}SU(3)\times{}U(1)$. Surprisingly, it turns out that the visible SM fermions as well as dark $SU(2)_{D}$ quarks can be acceptably embedded in a combination of the $9$ and $84$ representations, with all fermionic mediators and other unwanted states either being completely decoupled from the wall or attaining a coupling potential which does not permit chiral zero mode solutions. 

  Just as we did before with the $SU(12)$ model described previously, we generate the CoS domain wall generating our desired theory using two scalar fields charged under the adjoint representation, which in $SU(9)$ is 80-dimensional, transforming under a discrete $\mathbb{Z}_{2}$ interchange symmetry. Unlike for $SU(12)$, the discrete reflection transformation $\eta{}\rightarrow{}-\eta$ is outside $SU(9)$, however, the required breakings to subgroups isomorphic to $SU(6)\times{}SU(3)\times{}U(1)$ at positive and negative spatial infinity requires the cubic invariant to make the VEV pattern generating $SU(6)\times{}SU(3)\times{}U(1)$ globally minimal. 

  Just like for the $SU(12)$ model, the scalar potential may be written as 
\begin{equation}
\begin{aligned}
\label{eq:twoadjointpotentialpartssu9}
V(\eta) &= -\frac{1}{2}Tr[\eta^2]-\frac{1}{3}cTr[\eta^3]+\lambda_{1}(Tr[\eta^2])^2+\lambda_{2}Tr[\eta^4], \\
V(\chi) &= -\frac{1}{2}Tr[\chi^2]-\frac{1}{3}cTr[\chi^3]+\lambda_{1}(Tr[\chi^2])^2+\lambda_{2}Tr[\chi^4], \\
I(\eta, \chi) &= 2\delta^2Tr[\eta{}\chi]+dTr[\eta^2\chi]+dTr[\eta{}\chi^2]+l_{1}Tr[\eta^2]Tr[\chi^2]+l_{2}Tr[\eta^2\chi^2]+l_{3}(Tr[\eta{}\chi])^2 \\
              &+l_{4}Tr[\eta{}\chi{}\eta{}\chi{}]+l_{5}Tr[\eta^2]Tr[\eta{}\chi]+l_{5}Tr[\eta{}\chi]Tr[\chi^2]+l_{6}Tr[\eta^3\chi]+l_{6}Tr[\eta{}\chi^3].
\end{aligned}
\end{equation}

 We will not go into the specifics of ensuring the desired minima and whether the desired CoS solution can be made the most stable, although like the $SU(12)$ model, we expect that this can be done given the generic features of the potential in Eq.~\ref{eq:twoadjointpotentialpartssu9}. We only give the minima required and the localization properties for fermions and scalars which follow. 

 We wish to choose parameters such that the minima are of the form $\eta{}\neq{}0$, $\chi=0$ and $\eta=0$, $\chi{}\neq{}0$, to which the CoS domain wall solution asymptotes to at spatial infinity. Without loss of generality, let the minimum at $y=-\infty$ be of the form $\eta{}\neq{}0$, $\chi=0$, with $\eta$ proportional to
\begin{equation}
\label{eq:su9neginftyminimum} 
\eta(y=-\infty) \propto{} A = diag(-1, -1, -1, -1, -1, -1, +2, +2, +2).
\end{equation}
 We will again choose $l_{2}>l_{4}$ so that the paths with $[\eta, \chi] = 0$ are minimal, and so that $\chi$ is thus simultaneously diagonalizable with $\eta$. Then the desired VEV pattern, of the form $\eta=0$, $\chi{}\neq{}0$, at positive infinity to localize an $SU(5)\times{}SU(2)\times{}U(1)_{X}$ gauge group is obviously one in which $\chi$ is proportional to (up to trivial gauge rotations connecting to other diagonal forms)
\begin{equation}
\label{eq:su9posinftyminimum} 
\chi(y=+\infty) \propto{} B = diag(+1, +1, +1, +1, +1, -2, +1, -2, -2).
\end{equation}

 At negative infinity, the breaking induced is $SU(9)\rightarrow{}SU(6)_{1}\times{}SU(3)_{1}\times{}U(1)_{A}$, and at positive infinity, the induced breaking is $SU(9)\rightarrow{}SU(6)_{2}\times{}SU(3)_{2}\times{}U(1)_{B}$. On the domain-wall brane, there is further breaking to the overlap of these two subgroups and, clearly, $SU(6)_{1}\cap{}SU(6)_{2}\supset{}SU(5)_{V}$ and $SU(3)_{1}\cap{}SU(3)_{2}\supset{}SU(2)_{D}$. To determine the form for the localized Abelian generator $X$, we need to look at the leftover generators from the $SU(6)$ and $SU(3)$ subgroups on each side of the wall. 

 From $SU(6)_{1}\times{}SU(3)_{1}$, the leftover generators are respectively 
 \begin{equation}
 \label{eq:t1su9}
  T_{1} = diag(+1, +1, +1, +1, +1, -5, 0, 0, 0),
  \end{equation}
  and 
 \begin{equation}
 \label{eq:t2su9}
 T_{2} = diag(0, 0, 0, 0, 0, 0, -2, +1, +1). 
 \end{equation}
 For $SU(6)_{2}\times{}SU(3)_{2}$, the leftover generators are 
 \begin{equation}
 \label{eq:t1primesu9}
 T'_{1} = diag(+1, +1, +1, +1, +1, 0, -5, 0, 0)
\end{equation} and 
\begin{equation}
\label{eq:t2primesu9}
 T'_{2} = diag(0, 0, 0, 0, 0, -2, 0, +1, +1).
\end{equation}
One sees immediately that the generator 
\begin{equation}
\label{eq:xgeneratorlocsu9}
 X = 2T_{1}+5T_{2} = 2T'_{1}+5T'_{2} = diag(+2, +2, +2, +2, +2, -10, -10, +5, +5),
\end{equation}
is preserved on the domain wall interior and its corresponding photon is localized. 

 Thus the full symmetry preserved on the domain wall can be written $SU(5)_{V}\times{}SU(2)_{D}\times{}U(1)_{X}\times{}U(1)_{A}\times{}U(1)_{B}$, with the $SU(5)_{V}\times{}SU(2)_{D}\times{}U(1)_{X}$ subgroup completely dynamically localized. The photons corresponding to the generators $A$ and $B$ are semi-delocalized. 

 The next step is to write the simplest representations of $SU(9)$ in terms of the representations of $SU(5)_{V}\times{}SU(2)_{D}\times{}U(1)_{X}\times{}U(1)_{A}\times{}U(1)_{B}$ so that we can embed fermions and scalars and determine their localization properties. Under $SU(5)_{V}\times{}SU(2)_{D}\times{}U(1)_{X}\times{}U(1)_{A}\times{}U(1)_{B}$, the fundamental $9$ representation, the rank 2 antisymmetric $36$ representation, and the rank 3 totally antisymmetric $84$ representation decompose respectively as 
\begin{equation}
\label{eq:su9fundamental} 
 9 = (5, 1, +2, -1, +1)\oplus{}(1, 1, -10, -1, -2)\oplus{}(1, 2, +5, +2, -2)\oplus{}(1, 1, -10 , +2, +1),
\end{equation}
\begin{equation}
\begin{aligned}
\label{eq:su9rank2antisym}
 36 &= (10, 1, +4, -2, +2)\oplus{}(5, 1, -8, -2, -1)\oplus{}(5, 2, +7, +1, -1)\oplus{}(5, 1, -8, +1, +2) \\
    &\oplus{}(1, 2, -5, +1, -4)\oplus{}(1, 1, -20, +1, -1)\oplus{}(1, 1, +10, +4, -4), 
\end{aligned}
\end{equation}
and
\begin{equation}
\begin{aligned}
\label{eq:su9rank3totantisym}
 84 &= (\overline{10}, 1, +6, -3, +3)\oplus{}(10, 1, -6, -3, 0)\oplus{}(10, 2, +9, 0, 0)\oplus{}(10, 1, -6, 0, +3)\oplus{}(5, 2, -3, 0, -3)\oplus{}(5, 1, -18, 0, 0) \\
    &\oplus{}(5, 1, +12, +3, -3)\oplus{}(5, 2, -3, +3, 0)\oplus{}(1, 1, 0, +3, -6)\oplus{}(1, 2, -15, +3, -3)\oplus{}(1, 1, 0, +6, -3). 
\end{aligned}
\end{equation}

 Given the methods we developed in the previous section for analyzing the localization properties of the various components,  we can see that for this type of solution, assuming that the fermions are coupled to the background fields by the type of coupling described by Eq.~\ref{eq:yukawacoupling1}, that for the $9$ of $SU(9)$, the $(5, 1, +2, -1, +1)$ and $(1, 2, +5, +2, -2)$ components experience a kink and anti-kink respectively (or vice versa, depending on the sign of the coupling $h$), and thus develop chiral zero modes, while the $(1, 1, -10, -1, -2)$ and $(1, 1, -10 , +2, +1)$ components experience bulk masses which do not lead to chiral zero modes. 
 
 Looking at the $36$ representation, it at first seems unlikely that we can reproduce a realistic model, given that it contains an undesirable $(5, 2, +7, +1, -1)$ component which will attain a chiral zero mode given the type of coupling in Eq.~\ref{eq:yukawacoupling1}. However, if we look at the $84$ representation, one of the fermionic mediator components, the $(10, 2, +9, 0, 0)$ component, is uncharged under both $A$ and $B$ and is thus completely decoupled from the domain wall and, therefore, delocalized. The other two fermionic mediator components, the $(5, 2, -3, 0, -3)$ and $(5, 2, -3, +3, 0)$ components, are uncharged under one of $A$ or $B$ and thus also do not attain chiral zero modes and are semi-delocalized, analogously to the additional quintet components of the $66$ in the $SU(12)$ model discussed previously. Similarly, the two decuplets $(10, 1, -6, -3, 0)$ and $(10, 1, -6, 0, +3)$ are also semi-delocalized. The quintet $(5, 1, -18, 0, 0)$ is fully decoupled from the wall and will attain a mass of order $M_{GUT}$ when we break $SU(5)_{V}$. This leaves only the anti-decuplet $(\overline{10}, 1, +6, -3, +3)$ along with a quintet $(5, 1, +12, +3, -3)$, an $SU(2)_{D}$ doublet $(1, 2, -15, +3, -3)$ and two $SU(5)_{V}\times{}SU(2)_{D}\times{}U(1)_{X}$-singlet components $(1, 1, 0, +3, -6)$ and $(1, 1, 0, +6, -3)$, which develop chiral zero modes. If we choose the coupling constant such that the $(\overline{10}, 1, +6, -3, +3)$ component develops a right-chiral zero mode, then the $(5, 1, +12, +3, -3)$ component develops a left-chiral mode, which means that $SU(5)_{V}$ chiral multiplets arising from the $84$ are equivalent to a left-chiral combination of $5\oplus{}10$. The required combination is $\overline{5}\oplus{}10$, so from the $84$ we are guaranteed an additional $5$ which we must make massive, which can be done by localizing another left-chiral $\overline{5}$, which is readily done by embedding it in a second $9$ representation. Hence, the minimal required fermionic particle content to contain three generations of the SM fermions in this model is three copies of the combination 
 \begin{equation}
 \label{eq:su9fermioncontent} 
 9\oplus{}9\oplus{}84. 
 \end{equation}

 After forming this solution, we will have to include another adjoint Higgs field which will contain a component charged under the adjoint of $SU(5)_{V}$, in order to break $SU(5)_{V}$ to the Standard Model. As might be expected, we can embed an $SU(5)_{V}$ quintet Higgs in the $(5, 1, +2, -1, +1)$ component of the fundamental, and we can use any number of combinations of the various $SU(5)_{V}\times{}SU(2)_{D}$-singlet components embedded in the $9$, $36$ and $84$ to break the $U(1)$ subgroups associated with $X$, $A$ and $B$.

\subsection{Models based on Dvali-Shifman localization on non-Clash-of-Symmetries Domain Walls}
\label{subsec:noncossolution}

 In this section, we briefly outline how we can alternatively use one of the non-Clash-of-Symmetries solutions, the $m=0$ or $m=6$ solutions from Sec.~\ref{sec:solution}, as a basis to construct a realistic model. One of the immediate benefits of this is that it is very easy to ensure that one of these solutions is the most stable, since we can impose the additional $\eta{}\rightarrow{}\chi{}$, $\chi{}\rightarrow{}-\eta$ and $\eta{}\rightarrow{}-\eta{}$, $\chi{}\rightarrow{}-\chi{}$ symmetries, and then choose the coupling constant of the $Tr[\eta{}\chi{}]^2$ interaction to be negative. In a setup based on these solutions, we have to utilize the original Dvali-Shifman mechanism to localize $SU(5)_{V}$ and $SU(5)_{D}$, since the $SU(6)_{V}\times{}SU(6)_{D}$ group which is respected in the interior of the wall at the level of symmetries is otherwise delocalized. We simply achieve this with the addition of extra scalar fields which induce the breakings $SU(6)_{V,D}\rightarrow{}SU(5)_{V,D}$ in the interior of the wall. Thus, one of the costs to using the non-CoS walls is the addition of extra fields to the background scalar field configuration. 
 
 For the non-CoS solutions, the symmetry respected in the interior of the wall is $SU(6)_{V}\times{}SU(6)_{D}\times{}U(1)_{A}$. Consider the $m=0$ solution, where $B=-A$. The $12$ and $66$ representations decompose under $SU(6)_{V}\times{}SU(6)_{D}\times{}U(1)_{A}$ as

\begin{equation}
 \label{eq:fundamentalrepnoncossol}
 12 = (6, 1, -1)\oplus{}(1, 6, +1),
 \end{equation}
 and
\begin{equation}
 \label{eq:fundamentalrepnoncossol}
 66 = (15, 1, -2)\oplus{}(6, 6, 0)\oplus{}(1, 15, +2).
 \end{equation}
Since $B = -A$ for the $m=0$ solution, if we couple a fermionic field in the $12$ representation to $\eta$ and $\chi$ according to the type of interaction in Eq.~\ref{eq:yukawacoupling1}, if the Yukawa coupling $h$ is positive, the $(6, 1, -1)$ component will experience a kink-like interaction and develop left-chiral zero mode and the $(1, 6, +1)$ will experience an anti-kink interaction, attaining a right-chiral zero mode, and vice verse if $h$ is negative. For the same reason, if $h$ for a fermion in the $66$ representation is positive, the $(15, 1, -2)$ component attains a left-chiral zero mode and $(1, 15, -2)$ component attains a right-chiral zero mode, and vice versa if $h$ is negative. Once again, the mixed bi-fundamental $(6, 6, 0)$ is completely decoupled from the domain wall and is completely delocalized, since for this component $A=B=0$. This component will then attain a mass at least as large as the GUT scale if we break $SU(6)_{V}\rightarrow{}SU(5)_{V}\times{}U(1)$ by introducing an additional adjoint scalar which forms a lump in the interior of the domain wall. That leaves the $SU(6)_V$ and $SU(6)_{D}$ components charged under the respective $6$ and $15$ representations to attain localized chiral zero modes. 

 Given that the visible $SU(6)_V$ $15$ component will contain an additional $SU(5)_V$ quintet component with the same chirality as the the decuplet, we need to include two fundamentals, so that a localized quintet from one of them can form a mass term with the unwanted quintet from the visible $15$. Given that under $SU(6)_{V}\rightarrow{}SU(5)_{V}$, $6 = 5\oplus{}1$ and $15 = 10\oplus{}5$, in choosing the combination $12\oplus{}12\oplus{}66$, and choosing the background couplings such that the $(6, 1, -1)$ components in the two $12$ fermions attain right-chiral zero modes, and the $(15, 1, -2)$ component of the $66$ attains a left-chiral zero mode, the localized visible content will consist of left-chiral $\overline{5}$ fermion and a left-chiral $10$ fermion as required, along with two right-chiral (sterile) neutrinos, and a Dirac $5$ fermion, which will have a mass of order $M_{GUT}$. 
 
 In the dark sector, the $(1, 6, +1)$ components will attain left-chiral modes, and the $(1, 15, +2)$ components will attain right-chiral modes. To localize gauge groups in the dark sector, we must again utilize the ordinary Dvali-Shifman by spontaneously breaking $SU(6)_{D}$ to a subgroup. Unlike the visible sector, we have a great deal of freedom in what we break $SU(6)_{D}$ down to: we could break it symmetrically to $SU(3)\times{}SU(2)\times{}U(1)$, yielding a mirror matter scenario, or asymmetrically to something else entirely. Hence, what the localized $(1, 6, +1)$ and $(1, 15, +2)$ components break down to depends on how we break $SU(6)_{D}$.

  Using non-CoS domain walls, we can actually produce a model with the simpler gauge group $SU(10)$, in which the breaking in the visible sector leads directly to the Standard Model. If we break $SU(10)$ to the same $SU(5)_{V}\times{}SU(5)_{D}\times{}U(1)$ subgroup on both sides of the wall, we can then localize the Standard Model gauge group by introducing an additional scalar field which induces the usual breaking $SU(5)_{V}\rightarrow{}SU(3)\times{}SU(2)\times{}U(1)$. To localize gauge groups in the dark sector, we need to have an additional scalar field which breaks $SU(5)_{D}$, and we have the same freedom in choosing what subgroup we break it to as before. 

  Under $SU(10)\rightarrow{}SU(5)_{V}\times{}SU(5)_{D}\times{}U(1)$, the fundamental and rank two antisymmetric representations decompose as
 \begin{equation}
 \label{eq:decupletnoncos} 
 \overline{10} = (\overline{5}, 1, +1)\oplus{}(1, \overline{5}, -1),
 \end{equation}
 and
 \begin{equation}
  \label{eq:45noncos}
  45 = (10, 1, -2)\oplus{}(5, 5, 0)\oplus{}(1, 10, +2).
 \end{equation}
 Again, we see that the mixed component of the $45$, the $(5, 5, 0)$ component, is uncharged under the generator which induces the breaking $SU(10)\rightarrow{}SU(5)_{V}\times{}SU(5)_{D}\times{}U(1)$, and is thus decoupled from the wall, and, for reasons stated previously, removed from the low-energy spectrum. Hence, picking the background Yukawa couplings appropriately, if we choose the combination $\overline{10}\oplus{}45$ for our fermionic particle content, the visible SM fermions embedded in the $(\overline{5}, 1, +1)$ and $(10, 1, -2)$ components will attain left-chiral zero modes, leading to the required visible content, and the dark matter fermions embedded in the $(1, \overline{5}, -1)$ and $(1, 10, +2)$ components will attain right-chiral zero modes.

 In this subsection, we have shown that realistic models can also be constructed from the non-CoS solutions. We showed that the non-CoS solutions from the $SU(12)$ model discussed through the majority of this paper can lead to a realistic model, and we showed that this scenario could be further refined and simplified by using theory based on an $SU(10)$ gauge group, which leads directly to a domain-wall brane localized SM in the visible sector. The price for using the non-CoS solutions is that we must revert to the ordinary Dvali-Shifman mechanism, rather than the Clash-of-Symmetries mechanism, for localizing the gauge fields, and this in turn requires additional background scalar fields, greatly increasing the number of parameters of the scalar field theory generating the background field configuration.

\section{Conclusion}
\label{sec:conclusion}

 In this paper, we have shown how to construct a $4+1D$ theory based on domain-wall branes in which a realistic $SU(5)_{V}\times{}SU(5)_{D}\times{}U(1)_{X}$ was localized to a Clash-of-Symmetries domain wall, starting from a grand unified theory based on $SU(12)$. To motivate the addition of a higher dimension, we first argued that $3+1D$ grand unified theories based on $SU(N)\rightarrow{}SU(5)_{V}\times{}SU(N-5)_{D}\times{}U(1)$ were highly difficult to construct due the existence of chiral fermions charged under representations of both the visible and dark gauge groups. We then constructed a scalar field theory based on $SU(12)$ in $4+1D$ with two adjoint scalar fields transforming under a discrete $\mathbb{Z}_{2}$ interchange symmetry. We chose parameters such that the theory had two disconnected vacuum manifolds with the topology of $SU(12)/SU(6)\times{}SU(6)\times{}U(1)$, which meant we could construct clash-of-symmetries domain-wall solutions which break $SU(12)$ to differently embedded copies of $SU(6)\times{}SU(6)\times{}U(1)$, leading to a further breaking to the overlap of these differently embedded groups in the interior of the domain wall. Furthermore, we then showed it was possible to choose parameters such that one of the domain-wall solutions which lead to a localized $SU(5)_{V}\times{}SU(5)_{D}\times{}U(1)_{X}$ was made the most stable. 
 
 We then demonstrated that fermions could be localized to this $SU(5)_{V}\times{}SU(5)_{D}\times{}U(1)_{X}$-respecting wall in a phenomenologically interesting and acceptable way, showing that it was possible to localize left-chiral $\overline{5}\oplus{}10$ combination in the visible $SU(5)_{V}$ sector along with a right-chiral $\overline{5}\oplus{}10$ combination in the dark $SU(5)_{D}$ sector. Furthermore, we showed that the potentially troublesome $(5, 5)$ component charged under both $SU(5)_{V}$ and $SU(5)_{D}$ was completely decoupled from the wall and remained a vector-like $4+1D$ Dirac fermion; this means that this fermionic mediator will attain a GUT scale mass in the interior of the wall and be removed from the spectrum when we include an additional background adjoint scalar field that performs the required breaking $SU(5)_{V}\rightarrow{}SU(3)_{c}\times{}SU(2)_{I}\times{}U(1)_{Y}$. We also showed that other undesirable components did not attain localized modes and could also be removed from the localized theory on the wall. This means that we have a localized theory on the wall which has a visible sector containing the Standard Model particles along with a hidden, dark sector which is completely sequestrated from it at low energies. 
 
 We showed that scalars could be localized to the wall, and we demonstrated that the parameters controlling the coupling of a fundamental scalar to domain wall could be chosen to make certain $SU(5)_{V}\times{}SU(5)_{D}\times{}U(1)_{X}$-covariant components have either tachyonic or positive definite squared masses. In particular, we showed that it was possible to choose parameters such that the visible and dark quintets could be made tachyonic, initiating symmetry breaking  in the $SU(5)_{V}$ and $SU(5)_{D}$ sectors. Alternatively, we can make the singlet components, which are charged under the semi-delocalized $U(1)_{A}$ and $U(1)_{B}$ subgroups, tachyonic in order to break these troublesome Abelian gauge symmetries on the wall. Further analysis of the localization of scalars and spontaneous symmetry breaking in this model is left to later work; this work would include an analysis which takes into account the breaking $SU(5)_{V}$ to the Standard Model, as well an analysis of symmetric and asymmetric symmetry breaking scenarios in the dark sector.
 
 Given that a desirable goal would be to extend the above work and find other interesting scenarios leading to realistic models based on breaking a grand unified group $G$ to $G_{V}\times{}G_{D}$, we then outlined several other interesting potential models. We showed that another interesting model based on $SU(9)$ could lead to a localized theory with the gauge group $SU(5)_{V}\times{}SU(2)_{D}\times{}U(1)_{X}$, and we gave a set of representations for the fermions which could lead to a realistic theory without fermionic mediators. We also showed that the non-CoS domain walls in the $SU(12)$ model, in which $SU(5)_{V}$ and $SU(5)_{D}$ are localized by utilizing the original Dvali-Shifman mechanism, could be constructed, and, that this scenario could be further refined to generate localized $SU(3)\times{}SU(2)\times{}U(1)$ groups in the visible and dark sectors by utilizing the corresponding non-CoS solutions for $SU(10)$, which first induce the breaking $SU(10)\rightarrow{}SU(5)\times{}SU(5)\times{}U(1)$.
 
 The work in this paper represents the first step for using the Clash-of-Symmetries mechanism in particular to generate dark matter gauge groups and, potentially, asymmetric dark matter scenarios on a domain-wall brane. The next steps are to explore both symmetric and asymmetric breaking scenarios in this model with the introduction of additional background fields which break $SU(5)_{V}$ and $SU(5)_{D}$, and to explore the phenomenology in the visible and dark sectors in these scenarios. Another step would be doing a detailed calculation which checks, given the small energy differences between the different CoS solutions, that the stability of one of the $SU(5)_{V}\times{}SU(5)_{D}\times{}U(1)_{X}$ generating solutions can be preserved under quantum corrections for some parameter choice. Such a calculation would perhaps have to be done first in a lower dimensional toy model. Another interesting further work with the Clash-of-Symmetries mechanism could be to investigate whether it could be alternatively used to generate a gauge flavor symmetry instead of, or in addition to, a dark matter gauge symmetry.

\subsection*{Acknowledgments}
 
 This work was supported in part by the Australian Research Council and the Commonwealth of Australia. BCD would like to thank Raymond Volkas for useful discussion and advice. BDC would also like to thank Stephen Lonsdale, Claudia Hagedorn and Michael Schmidt for some useful discussions.

\newpage

\bibliographystyle{ieeetr}
\bibliography{bibliography2.bib}

\end{document}